TIINA LIIMETS

Nebulosities and jets from
outbursting evolved stars

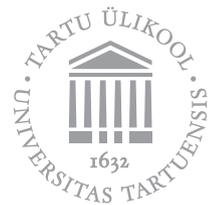

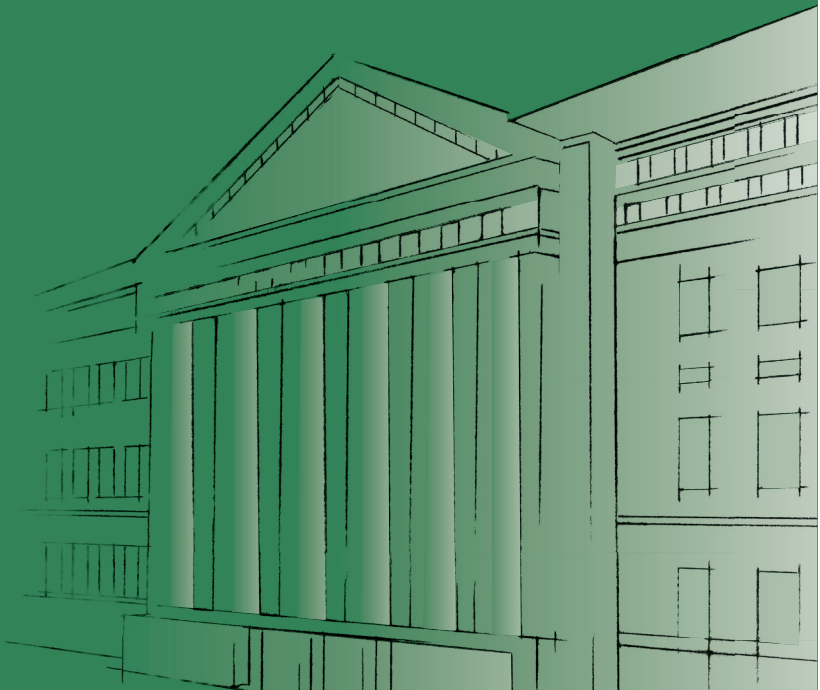

# TIINA LIIMETS

## Nebulosities and jets from outbursting evolved stars

This study was carried out at the Tartu Observatory, University of Tartu, Estonia.

The dissertation was admitted on June 20, 2019, in partial fulfilment of the requirements for the degree of Doctor of Philosophy in physics, and allowed for defence by the Council of the Institute of Physics, University of Tartu.

Supervisors:     PhD Romano L. M. Corradi,
                 Instituto de Astrophísica de Canarias,
                 GRANTECAN,
                 Spain

                 PhD Indrek Kolka,
                 University of Tartu,
                 Estonia

Opponent:        Dr Eric Lagadec,
                 Observatoire de la Côte d'Azur,
                 Laboratoire Lagrange,
                 Université Côte d'Azur,
                 France

Defence:         October 3, 2019, University of Tartu, Estonia

# CONTENTS

# LIST OF ORIGINAL PUBLICATIONS

**This thesis consists of a review of the following original papers:**

I **T. Liimets**, R. L. M. Corradi, M. Santander–García, E. Villaver, P. Rodríguez-Gil, K. Verro, & I. Kolka 2012, *A 3D view of the remnant of Nova Persei 1901 (GK Per)*, The Astrophysical Journal, 761, 34

II **T. Liimets**, R. L. M. Corradi, D. Jones, K. Verro, M. Santander–García, I. Kolka, M. Sidonio, E. Kankare, J. Kankare, T. Pursimo, & P. A. Wilson 2018, *New insights into the outflows from R Aquarii*, Astronomy & Astrophysics, 612, 118.

III M. Kraus, **T. Liimets**, C. E. Cappa, L. S. Cidale, D. H. Nickeler, N. U. Duronea, M. L. Arias, D. S. Gunawan, M. E. Oksala, M. Borges Fernandes, G. Maravelias, M. Curé, & M. Santander–García 2017, *Resolving the Circumstellar Environment of the Galactic B[e] Supergiant Star MWC 137 from Large to Small Scales*, The Astronomical Journal, 154, 186.

**Related conference publications:**

**T. Liimets**, R. L. M. Corradi, B. Balick, M. Santander–García: *The growth of the outflows from evolved stars: A live poster*. IAU Symposium, 2012, 283, 420.

**T. Liimets**, R. L. M. Corradi, M. Santander–García, et al.: *A dynamical study of the nova remnant of GK Per*. Astronomical Society of the Pacific, 2014, 490, 109.

M. Kraus, L. S. Cidale, **T. Liimets**, et al.: *Clumpy Molecular Structures Revolving the B[e] Supergiant MWC 137*. Astronomical Society of the Pacific, 2017, 508, 381.

**Author's contribution to the publications**

Author's research has given an essential contribution to all these publications. Here the author's contribution to the original publications is indicated. The Roman numerals correspond to those in the list of publications.

**Publication I.** The author reduced, measured, and analysed most of the data. Performed part of the observations. Wrote most of the text for the paper and prepared most of the figures.

**Publication II.** The author reduced, measured, and analysed most of the data. Performed part of the observations. Wrote most of the text for the paper and prepared all the figures.

**Publication III.** The author contributed significantly to the preparation of the observing proposal to obtain the optical images and spectra. Reduced and analysed mentioned data as well as wrote the text for the paper and prepared the related figures.

# ABSTRACT

In this PhD Thesis we study the occurrence of powerful stellar outbursts in the late phases of evolution of stars. Mass loss is a fundamental ingredient to understand the fate of stars, but the related physical mechanisms are still not well understood. This is for instance illustrated by the extraordinary variety of shapes revealed by observations of these stellar outflows, which are in principle not expected considering that stars are basically spherical all over their lives. Three intriguing stellar outflows are studied: the classical nova remnant GK Persei, the nebulosity surrounding the symbiotic binary R Aquarii, and the nebula of the massive B[e] supergiant star MWC 137.

The stellar ejecta that we study are related to eruptive events, and therefore occur and develop on relatively short time scales, in some cases few years or even months. Combined with the fact that ejection velocities are relatively large, and that these sources are relatively close, we have the opportunity to monitor the evolution of these outflows "in real-time", which is a rare opportunity in Astrophysics! To this aim, we have gathered over the years a unique collection of multi epoch images, from which "movies" can be produced that reveal the expansion of the nebulae as projected in the plane of the sky. Images are complemented by spectra that measure the line-of-sight velocities, also required to build 3D models of the dynamical evolution of the outflows, a privileged information for theoretical modelling.

We find that the expansion of the nova remnant GK Persei is rather homogeneous, contrary to previous predictions. The thick spherical shell is expanding with a broad range of velocities, showing only modest deceleration since the nova explosion that occurred more than a century ago. The R Aquarii nebula has a more irregular and fascinating behaviour. Additionally to its generally ordered global expansion pattern, we detect rapidly changing features that apparently move laterally in the jet. This most likely reflects changes in the illumination and ionisation properties rather than real motions. In addition, radial velocities show a blue/red-shift pattern opposite to the one found in previous observations, further demonstrating the complexity of the object. For MWC 137's nebula, we derive a double cone model, which with the current data available is the most plausible explanation for its shape.

Even though these stellar outflows are studied in detail in the current Thesis, several unanswered questions in their nature remain, providing us a plentiful of work for the future.

*"We Are All Made Of Star Stuff."*

Carl Sagan

# CHAPTER 1
# INTRODUCTION

Probably the most famous "nebula" of all time is the Andromeda Galaxy. It was the Persian astronomer Abd al-Rahman al-Sufi, in his *Book of Fixed stars* at $\sim$ AD 964 (translated by Hafez 2010), who mentioned the *Andromeda Nebula* for the very first time. Al-Sufi referred to Andromeda as a "nebulous smear". It took scientists nearly a millennium to disentangle the true distance to Andromeda. In 1922 the Estonian astronomer Ernst Julius Öpik derived a distance of 450 kpc (Öpik 1922). In 1925 Edwin Hubble calculated the distance to Andromeda to be 275 kpc (Hubble 1929). With this distance estimate, not only for Andromeda but for many other such nebulae, it was finally accepted that many of the beforehand known nebulosities are actually galaxies being significantly further away than any other object within the Milky Way. Nevertheless, many nebulosities still exist in the skies, which are not connected to galaxies but to stars in diverse evolutionary phases. And nowadays, the terminology *nebula* is exclusively used for a cloud of gas and/or dust, often concentrated around a stellar object which provides the energy to excite the nebula. To understand the multitude of nebulae, we first need to take a close look of the evolution of stars.

## 1.1   Stellar evolution in brief

Stellar evolution starts with a nebula as well. Stars are born from clouds of dust and gas contracting under their own gravity. After the gravitational collapse, the evolution of individual stars depends on the initial stellar mass. Those, which have lower mass, live a steady and long life, trillions of years. While those with higher mass go through their life-cycles in a hectic and dramatic way in just a few millions of years.

Substellar objects, such as brown dwarfs, with a mass M < 0.08 $M_\odot$, do not reach high enough temperatures in their cores to start the hydrogen (H) fusion. These stars gain their energy predominately from gravitational contraction. They slowly cool until they are no longer observable.

The majority of stars, M = 0.08 − 0.8$M_\odot$, are called very-low mass stars. They have sufficient energy and temperature, T = $15 \times 10^6$ K, to ignite the H to burn into helium (He) in their cores. Steady H burning phase in stars lives is referred to as the main sequence (MS), a diagonal track from low-temperature, low-luminosity to the high-temperature, high-luminosity region in the so called Hertzsprung-Russell diagram (HRD) in Figure 1.1. After the MS, when H is exhausted, gravity compresses the He-rich core into a white dwarf (WD) with a size similar to the Earth. The white dwarf will slowly cool until it is too faint to be detected.

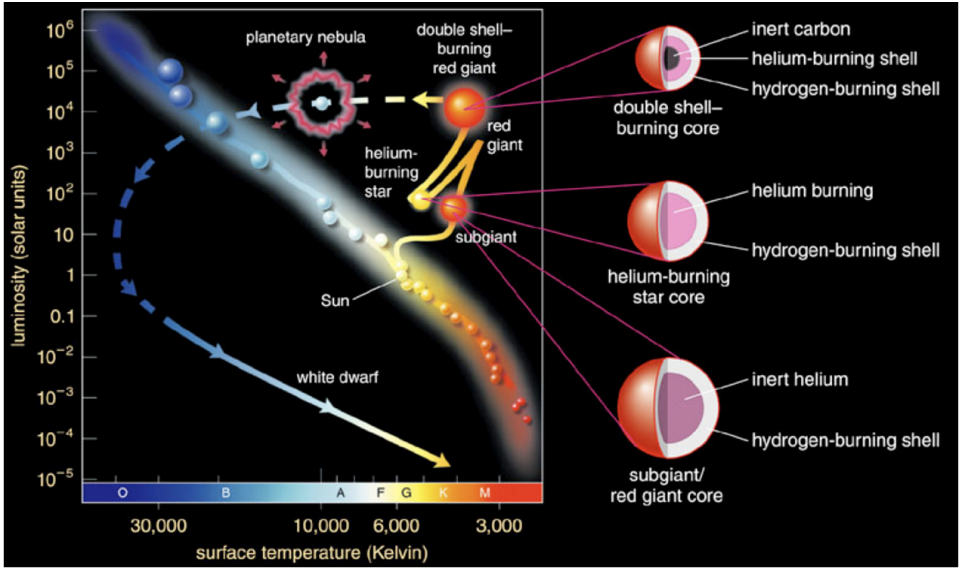

Figure 1.1: Hertzsprung-Russell diagram. Shown are the main sequence for low-mass up to very high mass stars and the evolutionary path of a solar mass type star. Image credit: https://sites.google.com/site/thestarthatshouldnotexist/low-mass-stars

Low- and intermediate-mass stars, $0.08 < M < 8\ M_\odot$, also spend most of their lives on the MS with steady H-burning. When the H in the core is all used up the star has a He core surrounded by H shell. The temperature in the core is too low to start the He fusion and too low to ignite the H in the outer layers. The star has no source of energy to balance the gravitational force. Therefore the core starts to collapse which causes it to heat up. When temperature is high enough the H burning starts in a shell around the He core. The He core continues to contract and heats up, while the outer H layers expand and cool. The star swells and cools at nearly constant luminosity moving towards the red giant branch and then expands at constant temperatures moving to higher luminosities in the HR diagram. Eventually, when the core gets hot enough – T = $10^8$ K for stars with a mass lower than ∼2.5-3 $M_\odot$ (low-mass stars) – the degenerate He core is ignited in an explosive reaction called He-flash. The luminosity of the star decreases, and the outer layers shrink, while the surface temperature of the star increases. Helium is now fused steadily in the core and star is located in the HRD in the so-called

Horizontal Branch (in Figure 1.1 referred to as a He-burning star). Intermediate-mass stars, with masses larger than 3 $M_\odot$, have non-degenerate He core. They quietly start their burning.

Eventually all the He in the core will be depleted and a C/O degenerate core is developed. Thermonuclear fusion continues now only in the outer layers, with a He burning shell surrounding the C/O core, and a H burning shell embracing the He burning shell. The star expands, once again, entering to the asymptotic giant branch (AGB) on the HRD, moving towards the lower surface temperature and higher luminosity. During this phase, instabilities in the He-burning shell may occur causing so called thermal pulses. Stars may loose 50-60% of their mass via stellar winds and ejection of the outer layers during the AGB phase (Vassiliadis & Wood 1993). The mass-loss rate can be as high as $10^{-3}$ $M_\odot$ $yr^{-1}$. This is the reason why low- and intermediate-mass stars never exceed the Chandrasekhar mass limit, $M_{Ch} = 1.44$ $M_\odot$, to end their lives as supernovae. The mass loss during the AGB phase ultimately leads to the complete ejection of the stellar envelope, which is heated and ionised by the hot central stellar remnant on its journey toward the WD phase. These bright nebulae are called planetary nebulae (PN).

High-mass stars are those with initial masses greater than 8 $M_\odot$. The more massive the star, the faster it burns its H content in the core shortening its MS lifetime (hundred thousand to a few million years). Owing to their higher mass, they reach a higher core temperature after the He core burning finishes. These stars can hence continue with fusion processes of heavier elements. Follow-up core burning phases encompass C, Ne, O, and Si burning, and these phases are all very short, ranging from 100 years for C burning down to 1 day for Si burning. At the same time, the previous fusion processes all continue in shells surrounding the core. The heaviest element to burn in the core is silicon. Once all silicon is transformed into iron, the fusion process stops as thermonuclear reactions are not exothermic anymore. The star rapidly collapses under its own gravity causing a shock wave which tears apart the whole star in an energetic explosion called supernova (SN). A tremendous amount of energy of about $10^{51}$ erg is released in SN explosions. Their brightness with up to $10^{10}$ $L_\odot$ can even exceed the one of their host galaxy. Supernovae might leave behind a super dense core such as a neutron star or a black hole, or only the expanding remnant. Supernova explosions are the only way to enrich the Universe with heavier elements, the "star stuff", which most of the Earth and life as we currently know is made of.

Schematically, the interior of a massive star in its late evolutionary phase would look like an onion with all the individual shell burning phases. However, the radiation from massive stars can drive strong winds with mass-loss rates of $10^{-8}$ to $10^{-6}$ $M_\odot$ $yr^{-1}$, which peel off the outer layers already during the MS evolution, and even more with mass-loss rates up to $10^{-4}$ to $10^{-3}$ $M_\odot$ yr$^{-1}$ in certain, sometimes eruptive late evolutionary stages such as the red supergiant (RSG), the B[e] supergiant, the luminous blue variable (LBV) or the Wolf-Rayet (WR) phase. The amount of mass lost during individual evolutionary phases and hence their fate strongly depend on several physical parameters, such as the initial mass of the star, its chemical composition (metallicity), and rotation speed (e.g. Ekström et al. 2012; Georgy et al. 2012, 2013a,b). In addition, mixing processes in the stellar interior transport matter from the core and the shells to the surface, enriching it with chemically processed material. Hence it is not surprising that the observed post-MS stages of massive stars do not match the theoretical phases related to the various core burning stages, and the evolution of massive stars beyond the MS still bears many uncertainties.

The next evolutionary step of a massive star after the MS is conceptually similar to the lower mass star evolution, but the exact behaviour as a function of the main stellar parameters is not precisely known. When the H in the core is depleted, He fusion starts, and the star expands and cools. Evolutionary models predict that stars in the mass range 10-25 $M_\odot$ reach the red, low-temperature regime of the HR diagram which is populated by the red supergiants (Levesque 2010). The red supergiant (RSG) phase might be considered as the high-luminosity counterpart to the red giant phase in low mass stars. Red supergiants are the largest stars in the Universe. Their radius can reach up to 1500 $R_\odot$ , while their temperatures stay below 4100 K (Levesque et al. 2005). Stars with masses below 20 $M_\odot$ will end their life in the RSG phase, where they will explode as type II SN. Stars in the mass range 20-25 $M_\odot$ might evolve back to the blue, hot side of the HRD after the RSG phase. These stars first appear as Yellow Hypergiants (de Jager 1998). When reaching a temperature of about 7000 K, their atmospheres become dynamically unstable. The stars shed their outer layers in a series of outbursts (e.g., Kraus et al. 2019) before they continue their lives as blue supergiants (BSG) or eventually as B[e] supergiants (B[e]SGs), depending on their physical parameters. BSGs are hot O or B type evolved stars, not as large as RSG, but with much higher temperatures, up to about 50 000 K. B[e]SGs are B-type stars with dense winds and a Keplerian rotating molecular and dusty circumstellar disk or rings (e.g., Kraus et al. 2016; Maravelias et al. 2018), possible remnant of previous mass ejection phases.

Stars more massive than 30 $M_\odot$ are not expected to reach the RSG stage but to return back to the blue, hot side much earlier. They may eventually pass through a B[e]SG phase and/or a luminous blue variable (LBV) phase. LBVs are massive unstable blue supergiants which show different types of photometric and spectroscopic variability. They experience periodic outbursts as well as occasional very massive explosions, almost as energetic as SN. The ejected vast amounts of material create huge nebulae (e.g., Nota et al. 1995; Gvaramadze et al. 2010). LBVs belong to the most luminous stars in the Universe, with values exceeding $10^6$ L$_\odot$. The last stage before the supernova explosion in these very massive stars is the Wolf-Rayet (WR) phase. However, during the evolution, massive stars may turn into LBVs and WRs more than once (see, e.g., Meynet et al. 2011). The WR phase can be paralleled as a PN phase of low mass stars, when a star ejects its outer layers via massive stellar winds creating magnificent nebulae. The difference with a PN central star is that the WR central star has a C/O core with continues fusion into heavier and heavier elements until it ends its life in a core collapse supernova.

All the above considerations apply to isolated, single stars. But stars are typically born in double or multiple systems (Sana et al. 2012), adding further complexity to the evolution of low and high-mass stars. If the binary separation is small enough, effects such as gravitational interactions and common envelope evolution (e.g. Ivanova et al. 2013) can strongly affect the evolution of the individual stars in the system, dramatically modifying the evolutionary timescales, stellar fates, and leading to powerful eruptions such as novae and Type-Ia SN. These latter are explosions produced in a close binary system, when a carbon-oxygen WD has reached the Chandrasekhar mass. While in a low- to intermediate-mass stars the thermonuclear runaway occurs only on the surface of the WD, in SN Ia the thermonuclear runaway occurs within the WD itself. The SN Ia can be produced by recurrent novae, symbiotic stars, supersoft X-ray sources, double WD binaries, and WDs accreting material from red-giant companions (e.g Giovannelli & Sabau-Graziati 2015).

The dramatic eruptions and explosions that occur in the later stages of stellar evolution result in different type of ejecta which produce magnificent stellar outflows with a variety of shapes and sizes: from spherical to unipolar- and/or bipolar nebulosities, irregular nebulae, filamentary structures, and highly collimated jets (Figure 1.2). The material which was once used to create a star is now enriched with heavier elements synthesised in the stellar interiors and blown back to the interstellar medium (ISM) to start the stellar life-cycle all over again.

## 1.2   Nebulae

The leftover stellar remnants studied in this Thesis are gaseous ionised nebulae. Nebulae are mostly glowing due to their excitation by the ultraviolet radiation of their hot central stars. However, shock excitation plays an important role as well, specially on the fast outflows like nova remnants (around $10^3$ km s$^{-1}$) and supernova remnants (up to $10^4$ km s$^{-1}$).

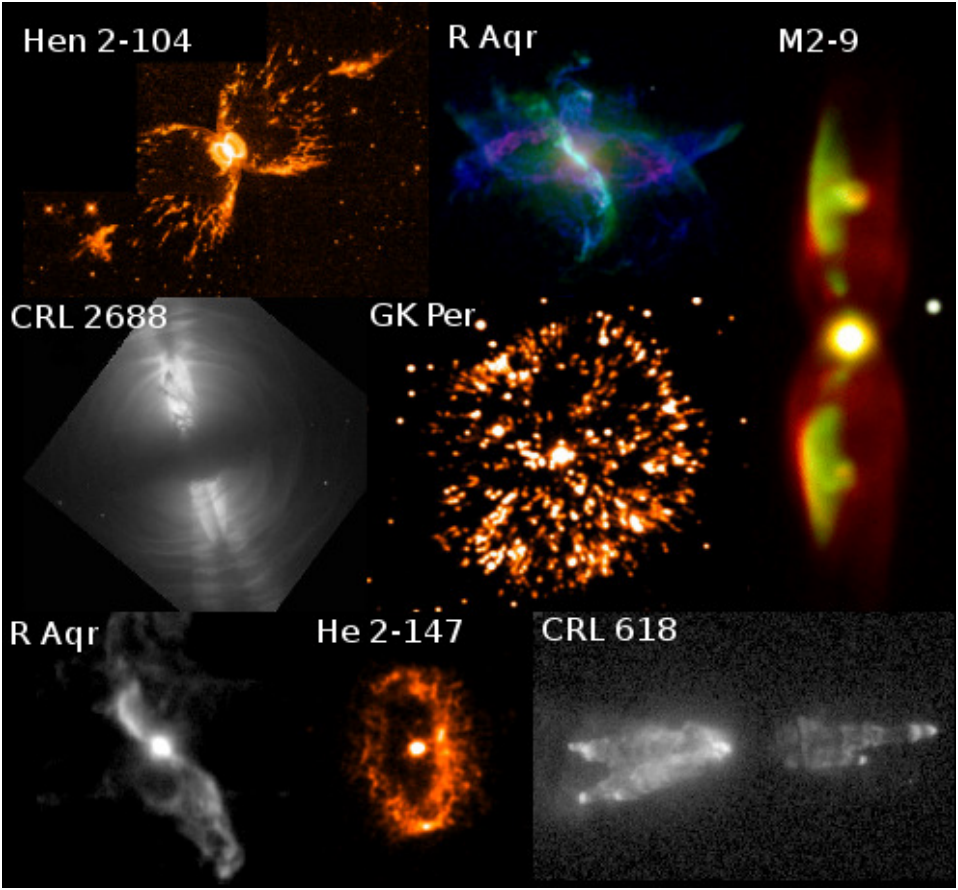

Figure 1.2: Various shapes and types of stellar outflows. Figure is a reprint from Liimets et al. (2012).

When the surface temperatures of the central stars are T $\gtrsim$ 30 000°K, they emit substantial amounts of ultraviolet photons which photoionise the gas. Due to the large amount of H present in astronomical objects the photoionisation of H is the main energy supply. Photons with energy larger than the ionisation threshold of H, 13.6 eV, are absorbed. The leftover energy of each absorbed photon over the ionisation threshold is transformed into kinetic energy of the ejected photoelectrons. This energy is redistributed by collisions between free electrons and between the electrons and ions. In this way, electrons and ions reach a Maxwell-Boltzmann velocity distribution with a kinetic temperature in the range of a typical nebulae, 5000°K to 20 000°K. The collisions between ions and thermal electrons excite ions in their electronic ground state into so called meta-stable stages. Downwards radiation transitions from meta-stable levels have very low probability. However, as the collisional de-excitation has even smaller probability at the low densities of these nebulae ($N_e \lesssim 10^4$ cm$^{-3}$), most of the excitations result in emitting a photon, in the so called forbidden transitions. Most of the forbidden transitions are collisionally excited. Depending on the temperature of the central star as well as on the temperature of the nebula, different atoms in the nebula at various ionisation states get excited and therefore numerous forbidden emission lines are detected. The lower energy forbidden emission lines often observed in nebulae are e.g. [O I] $\lambda\lambda$6300,6364, [O II] $\lambda\lambda$3726, 3729, [N II] $\lambda\lambda$6548,6583, and [S II] $\lambda\lambda$6716,6731 doublets. If the central star has a very high temperature, the amount of high-energy photons is also larger and as a result the atoms in the nebula are as well in a higher ionised state. Then excitation lines, e.g. of [O III] $\lambda\lambda$4959,5007 are observed.

In addition to forbidden lines, permitted recombination lines are also present in nebulae. Depending on the kinetic energy of the free electron and the excitation energy of the ions, collision between the free electrons and ions can also result in the capture of the electron by the ion, the so-called recombination. Typically, the captured electron will populate a high-excitation level, from which it will then cascade down to the lowest so-called stable energy level, emitting a series of photons with well defined energy forming the recombination lines. As Hydrogen is the most common element in the Universe, and hence also in nebulae, its recombination lines are also most frequently detected. In the optical wavelength the most usual H recombination lines are called the Balmer series, with the strongest emission line being H$\alpha$ $\lambda$6563.

When the ejecta have large velocities (e.g. novae and supernovae remnants, jets), kinetic energy may become comparable with the UV radiation from the central stars, and shock-excitation may play a dominant role in the excitation of the nebular plasma. Shocks are created due to the fact that the ejected matter moves faster than the speed of sound into the environment colliding with and compressing the ISM. In cases of shocks the gas is almost instantaneously heated and ionised and then cools by radiation. The relative contribution of photo- and shock-ionisation strongly varies depending on the kind of objects and their exact evolutionary status.

Astrophysical nebulae display a large variety of shapes, which also depend on the wavelength at which they are observed, as different regions and physico-chemical conditions within the multi-component nebula are mapped. (e.g. Crab Nebula by Dubner et al. 2017). In cases such as planetary nebulae the underlying shaping mechanisms are not completely understood. In spite of the fact that their progenitors are spherical stars with low angular momentum, ejecta can be strongly aspherical and show multiple kinds of symmetries. It is widely believed that in this respect binarity plays a decisive role.

To understand the shaping mechanisms, following the evolution of the outflows in real time provides valuable information. In fact, evolutionary timescales in astrophysics are generally so long that it is not possible to see appreciable changes during human life time. However, selected stellar outflows which are close and bright enough, and have large enough expansion velocities, give us the opportunity to study their evolution on timescales of few years, or even in few months. This is a rare opportunity in astrophysics. And, if the monitoring of the evolution of a nebula in the plane of the sky is combined with the measurement of Doppler shift velocities, a nearly fully 3D view of the evolution of the stellar outflow can be gained, which helps to better understand their formation mechanism.

For this reason, starting from 1991, our group has been monitoring three outstanding stellar outflows: the classical nova remnant Nova Persei 1901 (GK Persei or GK Per), the bipolar nebula and a jet of the symbiotic star R Aquarii, and the nebula and jet of the massive B[e] supergiant MWC 137.

### 1.2.1 Classical Nova GK Persei

Classical nova GK Persei (GK Per) is a close binary system belonging to the class of cataclysmic variables (CVs). CVs consist of a primary WD and a secondary late type low mass main sequence star. The WD is accreting matter, usually rich in Hydrogen, from the donor star via Roche lobe overflow and an accretion disk is formed around the primary. Accretion is to be blamed for the several phenomena occurring in these systems: irregular brightness changes, ejection of material via different eruption events (e.g. jets, novae, dwarf novae), high velocity winds up to 5000 $km\,s^{-1}$ (Froning 2005). Based on optical outburst characteristics, it is possible to divide CVs into four groups: classical novae, recurrent novae, dwarf novae, nova-like objects (e.g. Giovannelli & Sabau-Graziati 2015 and references therein). The eruption time scales vary from tens of seconds (oscillations in dwarf novae at outburst) to years in a case of larger outbursts. The mass transfer rates range from $10^{-11}$ to $10^{-8}$ $M_\odot\,yr^{-1}$ which are correlated with orbital periods (Patterson 1984).

A classical nova is occurring when the temperature and the pressure in the accreted matter around the WD is high enough to ignite thermonuclear reaction. On average energy of $10^{45}$ erg is released ejecting $10^{-4}$ to $10^{-5}$ $M_\odot$ over a time interval of a year. During the outburst the system brightens by more than 12 mag (Schaefer 2010). Probably the most famous classical nova is Nova Persei 1901 (GK Per), which created a bright and spectacular knotty remnant (Figure 1.3). GK Per also exhibits dwarf nova outbursts (maximum $\sim$ 3 mag) which are in-stabilities in the accretion disk itself. As neither of these events, nova and dwarf nova eruption, destroy the star completely, it is debated that all novae might be recurrent. The recurrence time depends on the accretion rate. From the models in Iben (1982) it is found that an accretion rate of $10^{-10}$ M$_\odot$ yr$^{-1}$ would produce a nova every tens of thousands of years but for a much higher accretion rate, $10^{-7}$ M$_\odot$ yr$^{-1}$, the recurrence time is less than 40 years. This is why some novae have only one recorded outburst event, while others have had several nova explosions during the last centuries (Schaefer 2010).

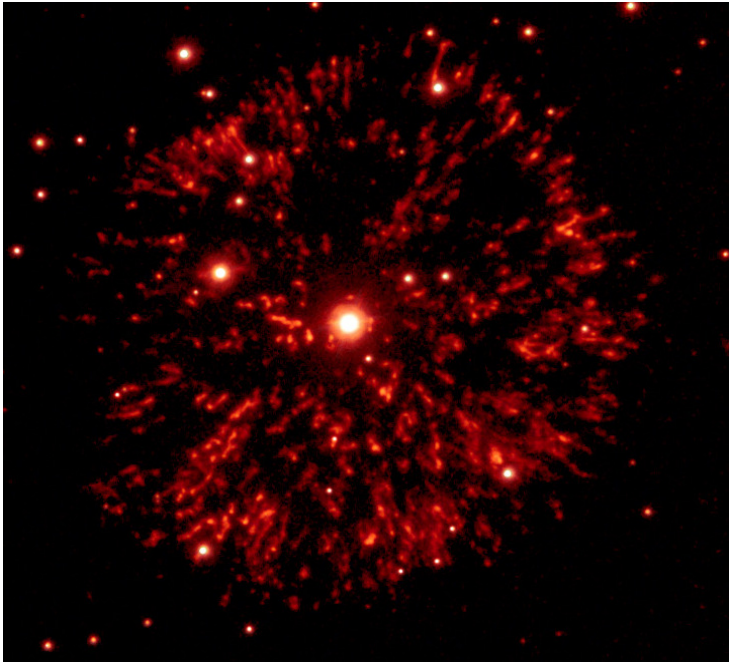

Figure 1.3: Classical nova remnant GK Persei. An overall slightly inhomogeneous circular shape consisting of numerous filaments is apparent. Figure is a reprint from Liimets & Corradi (2009).

Nova remnants are difficult to observe because they are short-lived, on scales of stellar evolution, and they are relatively faint. In that respect GK Per is unique due to its larger surface brightness and closeness, 400 pc, offering a superb possibility to monitor its evolution on time-scales of years, even months.

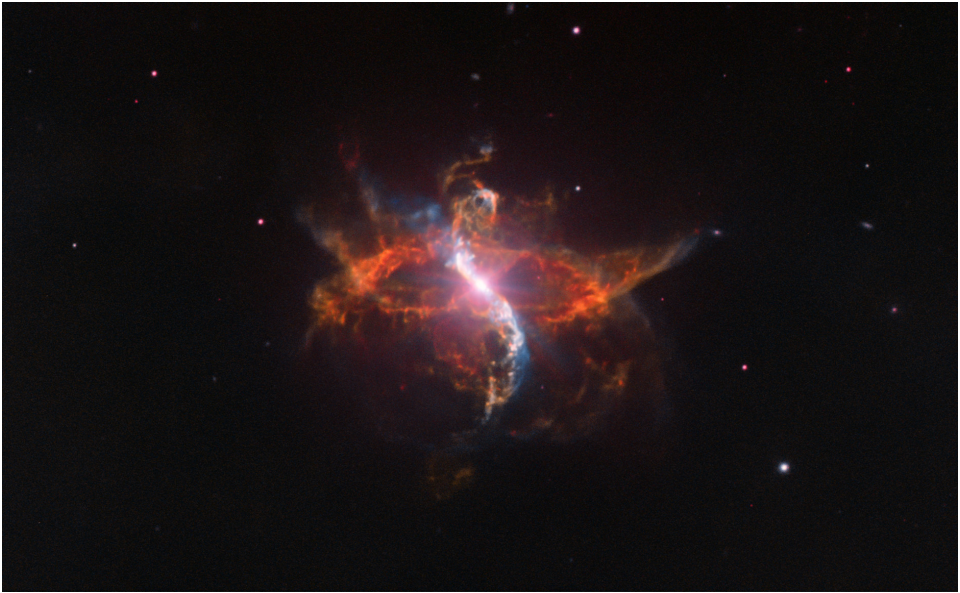

Figure 1.4: The hourglass nebula and jet of the symbiotic binary R Aquarii. Credit: ESO

### 1.2.2 Outflows from the symbiotic star R Aquarii

Another type of evolved binary stars which produce magnificent outflows are symbiotic stars. They are interacting binaries, composed of a hot component, usually a WD or a main sequence star, and a mass losing red giant, mostly a Mira giant. They are historically called symbiotic because their spectra simultaneously show features from a red giant, such as TiO absorption bands, as well as from the nebula, which is ionised due to the hot primary.

Symbiotic Miras are pulsating red giants on the AGB with a brightness variability larger than 2.5 magnitudes and periods longer than 100 days. Pulsations are caused by the star expanding and contracting in the He-shell burning phase. They have a large mass-loss rate, $10^{-8}$ to $10^{-4}\,M_\odot\,\mathrm{yr}^{-1}$ via relatively slow winds, up to 30 $\mathrm{km\,s}^{-1}$ (e.g. Goldman et al. 2017), which is accreted by the WD. At the same time the WD has a low mass-loss rate of $10^{-9}$ to $10^{-8}\,M_\odot\,\mathrm{yr}^{-1}$ (e.g. Kudritzki et al. 1997 and references therein) but the velocities can be up to several thousands of $\mathrm{km\,s}^{-1}$. The winds of both stars as well as the occurrence of nova-like explosions, produce a rich circumstellar environment around these systems. Various shapes of nebulae can be combined with collimated jets, just like in the case of R Aquarii (Figure 1.4). Jets are common phenomenon in astrophysics.

Highly collimated jets are produced in various stages of stellar evolution before and after the main sequence as well as on galactic scales (e.g. active galactic nucleus, quasars). Jets have an important role in evolution by carrying mass, energy, and momentum. They sustain vast amount of energy and enriched material in the ISM. Although the stage of stellar evolution that produces jets in certain types of objects may be short-lived, the consequence on the ISM can be long-lasting. Launching from magnetised accretion discs is the most common explanation for the formation of jets in most astrophysical objects (Livio 2011). However, the exact mechanism, in particular the collimation, is still under debate. In addition, there are observational constraints, mainly because resolving the jet launching regions is beyond the capability of present instrumentation. In spite of that, it is possible to observationaly study jets from bright stellar systems, to investigate the nature of the stars and the associated phenomena (accretion, magnetic fields, jets evolution, etc.). Symbiotic stars provide such an opportunity, as some of them show powerful jets. However, the distance and the relatively dense CSM around these eruptive stars makes them usually difficult to observe and study. With a distance of about 200 pc and a low local extinction, R Aquarii is one of the best cases giving a unique opportunity to study in detail the evolution of its complicated jets and nebula.

### 1.2.3 Double-cone nebula and jet of the massive B[e] supergiant MWC 137

B[e] stars are emission line B type stars, including forbidden emission, with a strong infrared excess emission due to the hot (500 to 1000 K) circumstellar dust (e.g. Lamers et al. 1998) related to the steady or eruptive mass loss. The physical mechanism creating such mass ejection is not well known nor the evolutionary status of their central stars. They can be found to have low to high mass and to be in different evolution phases: from pre-main sequence to post-main sequence. The following classification of B[e] stars was proposed by Lamers et al. (1998): B[e] supergiants, pre-MS B[e], compact PN B[e], symbiotic B[e], and unclassified B[e]. Despite the fact that the central stars of B[e] stars can be very different the physical conditions of creating the emission lines and the dust emitting region are quite similar. Therefore the B[e] status refers just to a specific physical and spectral characteristics but not to an evolutionary status.

B[e] supergiants (B[e]SGs) are evolved massive stars with dense dusty discs (Kraus 2017). It is not clear if they are on pre- or post-RSG stage. B[e]SGs luminosities exceed $10^4$ L$_\odot$. They experience small photometric variability, have a significant amount of heavier elements indicating an evolved nature, show composite spectrum of a narrow low-excitation emission lines of low-ionised metals (e.g. [O I], Fe II) and a broad absorption features of high-excitation lines, and suffer mass loss (Lamers et al. 1998).

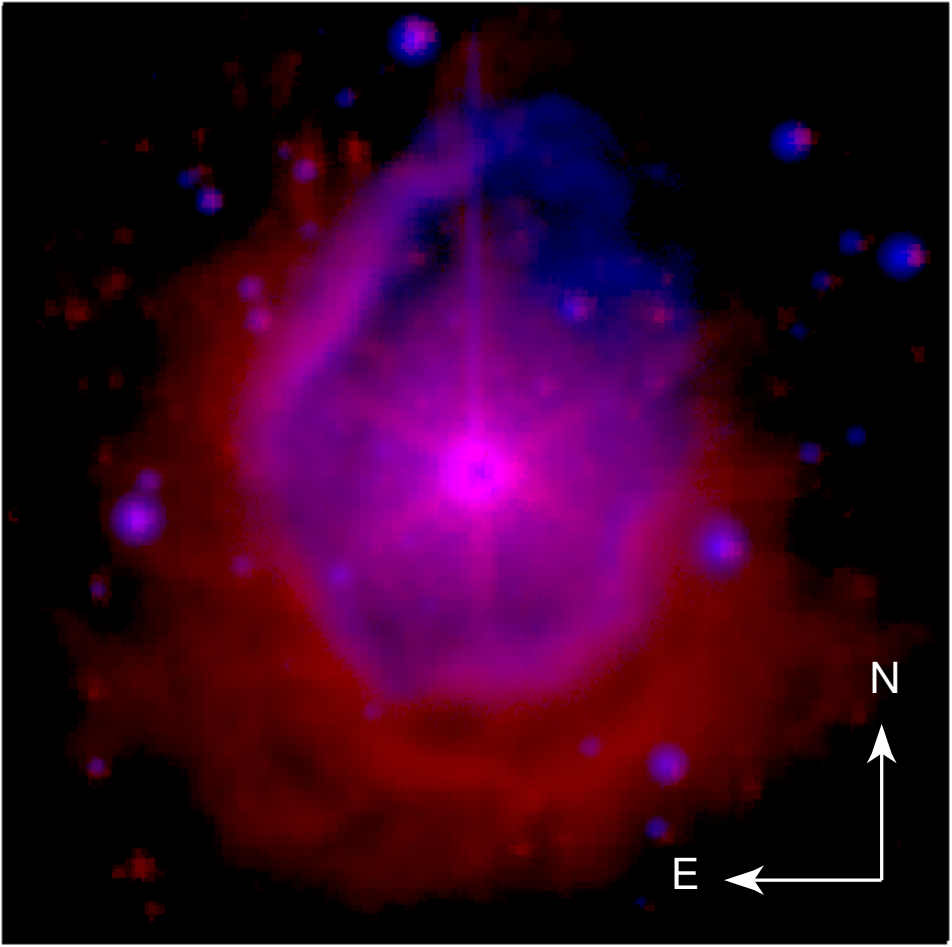

Figure 1.5: Nebula of a massive B[e] supergiant MWC 137. Figure is a reprint from Paper III (Kraus et al. 2017).

In the current Thesis we investigate the massive B[e]SG MWC 137 which has an evolved nature (Muratore et al. 2015). It is surrounded by a double-cone nebula (Figure 1.5) and a jet. MWC 137 is the only known B[e]SG, so far, possessing a jet! The mass of MWC 137 is 10-15 M$_\odot$ (Mehner et al. 2016) and the discovered jet is an indication of a binary nature. Hence, possibly, MWC 137 may explode as a Type Ia supernova.

## 1.3 Motivation for this Thesis

Bright stellar outflows are an opportunity to shed light on the controversial mass ejection mechanisms in evolved stars, and the return of gas to the ISM. To provide constraints to theoretical modelling we chose 3 stellar outflows of different type and studied their evolution in the plane of the sky as well as along the line of sight. Two of the selected objects, GK Persei and R Aquarii, are close enough to detect their evolution in the plane of the sky in few years, even in a few months. Adding radial velocities to imaging data gave us a complete 3D view of these ejecta. The third object of the current Thesis, MWC 137, is further away but nevertheless its nebulosities provide a superb possibility for detailed dynamical and kinematical investigation.

All the three targets of this Thesis posses peculiar characteristics and provide therefore a fruitful base for a valuable investigation. In particular, the seemingly spherical classical nova remnant was predicted by Bode et al. (2004) to have asymmetric expansion. This scenario is investigated in Chapter 4. Additionally, we wanted to disentangle the true shape of the remnant, to calculate the possible deceleration rate of the outwards expanding features in the remnant, the kinematical distance to it, and investigate its brightness variation.

The collimated jet of the symbiotic binary R Aquarii has raised several questions: creation of the northern and southern jet, origin of the collimation, kinematical behaviour? Besides the investigation of the jet, using the expansion of the outer nebula we wanted to refine the distance to R Aquarii and calculate the age of its different type of ejecta. The results are presented and discussed in Chapter 5.

Chapter 6 is dedicated to B[e] supergiant MWC 137. The main question to answer in this Thesis about the MWC 137 was if the visible double cone appearance of its nebula is supported by the spectral observations.

# CHAPTER 2
# DATA AND REDUCTION PROCEDURES

## 2.1 Optical imaging

Optical imaging data for this dissertation was collected over more than two decades at various observatories. In the case of R Aquarii, the first images were obtained in 1991 with the New Technology Telescope (NTT) at ESO Chile equipped with ESO Multi-Mode Instrument (EMMI, $0''.35$ pix$^{-1}$ and a FOV of $6'.2 \times 6'.2$, Dekker et al. 1986). Since 1997, the monitoring continued with the 2.6m Nordic Optical Telescope (NOT) at La Palma with a set of images taken approximately after 5 years, occasionally more frequently. At the NOT the Andalucia Faint Object Spectrograph and Camera (ALFOSC) was used. At those times ALFOSC had a pixel scale $0''.19$ pix$^{-1}$ and a field of view (FOV) of $6'.4 \times 6'.4$. The most recent images, in 2012, were obtained with the Very Large Telescopes (VLT) and its Focal Reducer/low dispersion Spectrograph 2 (FORS2, Appenzeller et al. 1998). The Standard Resolution collimator of FORS2 was used resulting in a pixel scale of $0''.25$ pix$^{-1}$ and a FOV of $6'.8 \times 6'.8$. Several narrow band filters were used (H$\alpha$+[N II], [N II], [O I], [O II], and [O III]), covering the following emission lines: H$\alpha$ $\lambda6562.82$; [N II] $\lambda\lambda6548.03,6583.41$; [O I] $\lambda6300.30$; [O II] $\lambda\lambda3726.03,3728.82$; and [O III] $\lambda\lambda4958.91,5006.84$, respectively. Due to the moderate radial velocities present in the R Aquarii jet and bipolar nebula, all filters include all the light from corresponding emission lines, except the 2002 [N II] filter which only includes radial velocities larger than $+45$ km s$^{-1}$ considering the [N II] $\lambda6583$ rest wavelength. Filters centred at H$\alpha$ also include emission from the [N II] $\lambda\lambda6548,6583$ doublet. Details of the observations are listed in Table 2.1. In the Table the first column lists the date at the start of the observing night in the format yyyy-mm-dd, the second column JD is the Julian Date at mid point of the observations, column 3 lists the telescope and instrument used, column 4 contains the nebular lines included and the central wavelength together with the FWHM of the filter, column 5 is the total exposure time, column 6 is the number of frames added together, and column 7 provides the seeing. As can be seen from the table, most of the data were taken with the NOT+ALFOSC, under good seeing conditions. The airmass mostly never exceeded 1.5. This allows sufficiently accurate result when comparing individual images among this homogeneous set.

An additional set of deep observations of R Aquarii were obtained on 2016 October 2, 8, and 2006 November 2, 3 in Terroux Observatory, Canberra, Aus-

Table 2.1: Log of the imaging data of R Aquarii. See text for more details.

| Date | JD | Telescope+ instrument | Filter CW/FWHM (Å) | Total exp. time (sec) | Nr. of frames | Seeing ($''$) |
|------|-----|------|------|------|------|------|
| 1991-07-06[a] | 2448443.79 | NTT+EMMI | Hα+[N II] 6568/73 | 300, 30, 1 | 1, 1, 1 | 1.1 |
| 1998-09-04 | 2451061.54 | NOT+ALFOSC | Hα+[N II] 6577/180 | 100 | 1 | 1.3 |
| 2009-07-08* | 2455021.68 | NOT+ALFOSC | Hα+[N II] 6577/180 | 100, 2 | 1, 1 | 1.3, 1.2 |
| 2012-09-05 | 2456175.78 | VLT+FORS2 | Hα+[N II] 6563/61 | 90 | 3 | 0.7 |
| | | | | | | |
| 1997-07-13[a] | 2450643.77 | NOT+ALFOSC | [N II] 6584/10 | 120, 20, 5 | 1, 1, 1 | 1.1 |
| 2002-06-26 | 2452452.73 | NOT+ALFOSC | [N II] 6588/9 | 600 | 1 | 0.8 |
| | | | | | | |
| 2002-06-27 | 2452453.77 | NOT+ALFOSC | [O III] 5008/30 | 60, 30, 10 | 1, 1, 1 | 0.9 |
| 2007-09-05* | 2454349.62 | NOT+ALFOSC | [O III] 5007/30 | 400, 30 | 1, 1 | 0.7, 0.6 |
| 2009-07-08 | 2455021.68 | NOT+ALFOSC | [O III] 5007/30 | 60 | 1 | 1.2 |
| 2009-08-24 | 2455068.65 | NOT+ALFOSC | [O III] 5007/30 | 600 | 2 | 0.6 |
| 2011-09-05 | 2455810.57 | NOT+ALFOSC | [O III] 5007/30 | 180, 40 | 1, 1 | 0.9, 0.8 |
| 2012-09-05 | 2456175.78 | VLT+FORS2 | [O III] 5001/57 | 300, 1 | 3, 1 | 0.7 |
| | | | | | | |
| 1997-07-15[a] | 2450645.74 | NOT+ALFOSC | [O II] 3727/30 | 900 | 3 | 1.1 |
| 1998-09-04[b] | 2451061.54 | NOT+ALFOSC | [O II] 3727/30 | 300 | 1 | 1.3 |
| 2002-06-27 | 2452453.77 | NOT+ALFOSC | [O II] 3725/50 | 900 | 1 | 1.3 |
| 2007-09-05* | 2454349.62 | NOT+ALFOSC | [O II] 3726/51 | 600 | 1 | 0.7 |
| 2009-08-24 | 2455068.65 | NOT+ALFSOC | [O II] 3726/51 | 1200 | 2 | 0.8 |
| 2011-09-05 | 2455810.57 | NOT+ALFOSC | [O II] 3726/51 | 600 | 1 | 0.9 |
| 2012-09-05 | 2456175.78 | VLT+FORS2 | [O II] 3717/73 | 540, 10 | 3, 1 | 0.8 |
| | | | | | | |
| 2007-09-05* | 2454349.62 | NOT+ALFOSC | [O I] 6308/29 | 120, 30 | 1, 1 | 0.6 |
| 2009-08-24 | 2455068.65 | NOT+ALFSOC | [O I] 6300/30 | 600 | 2 | 0.6 |

[a] Published in Navarro et al. (2003) and Gonçalves et al. (2003).

[b] Published in Corradi (2003).

* Indicate the reference epoch for the pixel by pixel matching for a given filter (see Section 2.1.1).

tralia using a 30cm f3.8 Newtonian telescope with a CCD camera and narrow band [O III] (5010 Å, FWHM=60 Å) and Hα (6560 Å, FWHM=60 Å) filters. The pixel scale was $0''.84$ pix$^{-1}$. 34 frames for a total exposure time of 7.6h were obtained in [O III], while 32 frames adding up to 6.8 hours were taken in Hα.

Monitoring of the nova remnant GK Per was performed starting from 2004 until the end of 2011. It was mostly carried out with the Isaac Newton Telescope (INT) at La Palma equipped with the Wide Field Camera (WFC). WFC has a pixel scale of $0''.33$ pix$^{-1}$. A narrow band Hα+[N II] filter with a central wavelength 6568 Å and a FWHM 95 Å was used. Few extra observations were observed with the NOT+ALFOSC with a wider Hα+[N II] (see Table 2.2). During that period of time 19 good quality data points, evenly distributed in time, were obtained.

In order to expand our baseline, archival data from 1987, 1989, 1995, 1997, and 1999 were used, which were observed with different telescopes and fil-

Table 2.2: Log of the imaging data of GK Persei. See text for more details.

| Date | JD | Telescope+ Instrument | Filter CW/FWHM (Å) | Total exp. time (sec) | Nr. of frames | Seeing (″) |
|------|-----|------|------|------|------|------|
| 1953-01-10 | 2434387.62 | Pal48 | phot. plate | 2400 | 1 | 5.0 |
| 1987-07-27[a] | 2447003.69 | INT+PFCU | R | 300 | 1 | 2.8 |
| 1989-10-05 | 2447804.87 | Pal48 | phot. plate | 5100 | 1 | 4.9 |
| 1995-11-08[b] | 2450030.42 | HST+WFPC2 | 6590/22 | 1400 | 2 | 0.2 |
| 1997-01-08[b] | 2450457.18 | HST+WFPC2 | 6590/22 | 1200 | 2 | 0.2 |
| 1999-11-30*[c] | 2451513.56 | INT+WFC | 6568/95 | 7026 | 3 | 2.3 |
| **2004-01-29** * | 2453034.37 | INT+WFC | 6568/95 | 900 | 3 | 1.2 |
| 2004-08-04* | 2453222.66 | INT+WFC | 6568/95 | 1500 | 5 | 1.2 |
| 2004-09-03* | 2453252.67 | INT+WFC | 6568/95 | 1500 | 5 | 1.1 |
| 2004-12-30* | 2453370.43 | INT+WFC | 6568/95 | 1200 | 4 | 1.1 |
| 2005-09-16* | 2453630.72 | INT+WFC | 6568/95 | 900 | 3 | 1.0 |
| 2006-10-12* | 2454021.53 | INT+WFC | 6568/95 | 1200 | 4 | 1.0 |
| 2007-01-04* | 2454105.49 | INT+WFC | 6568/95 | 300 | 1 | 1.0 |
| 2007-03-01* | 2454161.42 | INT+WFC | 6568/95 | 2100 | 7 | 1.2 |
| 2007-09-03 | 2454347.68 | NOT+ALFOSC | 6577/180 | 1800 | 3 | 0.6 |
| 2007-09-05 | 2454349.71 | NOT+ALFOSC | 6577/180 | 1800 | 3 | 0.5 |
| 2007-09-07* | 2454351.74 | INT+WFC | 6568/95 | 1200 | 4 | 1.1 |
| 2008-01-08* | 2454474.42 | INT+WFC | 6568/95 | 1200 | 4 | 0.8 |
| 2008-08-19* | 2454698.69 | INT+WFC | 6568/95 | 4650 | 31 | 1.1 |
| 2008-09-12 | 2454722.70 | INT+WFC | 6568/95 | 1200 | 4 | 1.3 |
| 2008-11-14* | 2454785.49 | INT+WFC | 6568/95 | 1500 | 5 | 1.2 |
| 2009-02-11 | 2454874.43 | NOT+ALFOSC | 6577/180 | 1200 | 2 | 0.8 |
| 2009-12-08 | 2455174.48 | INT+WFC | 6568/95 | 300 | 1 | 1.3 |
| 2010-02-09* | 2455237.37 | INT+WFC | 6568/95 | 1500 | 5 | 0.9 |
| 2011-12-13* | 2455909.41 | INT+WFC | 6568/95 | 2700 | 9 | 0.8 |

[a] Published in Seaquist et al. (1989)

[b] Published in Shara et al. (2012)

[c] Published in Bode et al. (2004)

**2004-01-29** was used as reference frame for the astrometry and flux match of all images, dates marked with * were flux and PSF matched with the reference image (see Section 2.1.1).

ters/photographic plates, apart of the 1999 data, which was from the INT+WFC. A very early image from 1953 was used for the visual comparison but not for the actual measurements. The log and details of all the imaging observations of the GK Per are presented in Table 2.2 which has the same format as Table 2.1.

MWC 137 imaging data was collected in the period 2016-11-06 with the NOT+ALFOSC using the same narrow band H$\alpha$+[N II] 6577/180 filter as for the other two objects. One long (600s) and one short (10s) exposure was obtained. For these observations the pixel scale of ALFOSC was $0''.21$ pix$^{-1}$.

### 2.1.1 Image processing

Optical imaging data were pre-processed using the standard tasks in the package `ccdred` in IRAF[1]. The following steps were performed: bias and flat field correction, combination of sub-exposures, and cosmic rays removal. The additional images of R Aqr from the Terroux Observatory were reduced by the co-author of the paper II, M Sidonio. Individual frames were flat field corrected and median combined.

After the pre-processing the R Aqr and GK Per data were further prepared for the analyses. Firstly, an astrometric correction for each combined frame was applied. For GK Per frames stars in the same FOV, in a case of R Aqr frames other well populated fields in the corresponding filters were used. Secondly, pixel by pixel matching was performed using the stars in the FOV.

For R Aqr and GK Per data also seeing and flux matching were needed. For that firstly the FOV stars were used to match the seeing of the worst frame. The flux matching was performed for GK Per frames using the FOV stars, because the telescope+instrument+filter was mostly the same. In the case of R Aqr, as several telescopes with a different filter bandpasses, as well as several different filters were utilized, part of the nebula was used to match the flux. The latter is suitable, as the nebula is emitting in certain emission lines which, considering the slow velocities present in R Aqr, all fit into the corresponding filter bandpasses. While, the stars have more complex spectra, which in slightly different filter bandpasses can create a large brightness difference. For the nova remnant GK Per the seeing and flux matching were needed to perform accurate brightness measurements, whereas for R Aqr these were required for the magnification method which is described in Section 3.3.

Finally, since the monitoring of R Aqr was over a long period, correction for the proper motion of the start was applied to these images.

For more specific details of the imaging processing of each object the reader is referred to the original papers of the current Thesis.

## 2.2 Spectroscopy

The accompanying spectroscopic data were mostly collected on a single date per object, or close in time dates. The medium-resolution, long-slit spectra of GK Per were obtained with the INT and NOT respectively on 2007 January 13-14 and September 3-5. At the INT the Intermediate Dispersion Spectrograph (IDS) with grating R1200Y and a $1''.5$ slit was used. This configuration covers a spectral range from 5730 Å to 7610 Å with a dispersion of 0.47 Å pix$^{-1}$. At the NOT the ALFOSC with grism #17 and $0''.5$ slit was used, giving a spectral range from 6350 Å to 6850 Å, and a dispersion of 0.26 Å pix$^{-1}$. Twelve different position angles (PA) were obtained distributed across the nova remnant to cover as many knots of the remnant as possible (see Figure 2.1). In the Figure the slits positions obtained with the NOT are presented with blue colour and the spectra observed with the INT with red. The boxes indicate the real width of the slits: $0''.5$ at NOT, $1''.5$ at INT. One spectrum per telescope, was observed at a similar position angle ($48°$ and $49°$) to exclude any systematic deviations. In Table 2.3 we present the details of the observations. PA is measured from North to East. These data were reduced and wavelength calibrated using the standard routines in the package `longslit` in IRAF.

---

[1]IRAF is distributed by the National Optical Astronomy Observatory, which is operated by the Association of Universities for Research in Astronomy (AURA) under cooperative agreement with the National Science Foundation.

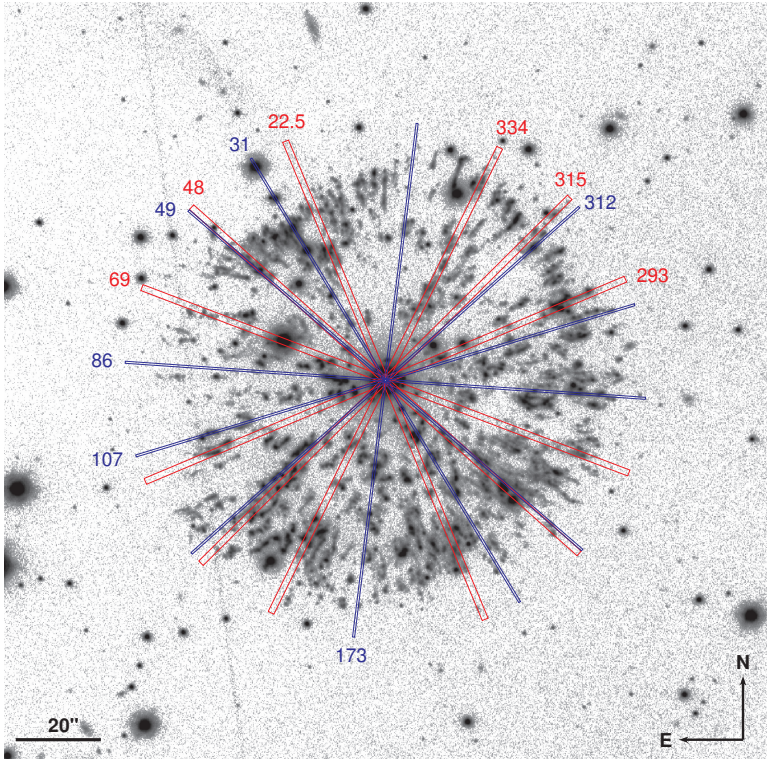

Figure 2.1: The slit positions of GK Persei observations overlaid on a NOT 2007 Hα+[N II] image. Blue: NOT slits, red: INT slits. FOV $3' \times 3'$. See text for more details.

Table 2.3: Log of the spectroscopic observations of GK Persei. See text for more details.

| Date | JD | Telescope | PA (°) | Resolution (Å) | Total exp. time (sec) | Nr. of frames |
|------|-----|-----------|--------|----------------|----------------------|---------------|
| 2007-01-13 | 2454114.448490 | INT | 22.5 | 1.3 | 3600 | 2 |
| | 2454114.555786 | | 48 | | 3600 | 2 |
| | 2454114.486704 | | 315 | | 1800 | 1 |
| 2007-01-14 | 2454115.567185 | | 69 | 1.3 | 1800 | 1 |
| | 2454115.529594 | | 293 | | 3600 | 2 |
| | 2454115.482216 | | 334 | | 3600 | 2 |
| 2007-09-03 | 2454347.735000 | NOT | 31 | 0.7 | 3600 | 2 |
| 2007-09-04 | 2454348.649340 | | 86 | 0.7 | 1800 | 1 |
| | 2454348.730359 | | 107 | | 1500 | 1 |
| | 2454348.677106 | | 312 | | 1800 | 1 |
| 2007-09-05 | 2454349.742280 | | 49 | 0.7 | 1800 | 1 |
| | 2454349.674919 | | 173 | | 1800 | 1 |

The spectral data of the jet of R Aqr were obtained with the VLT on October 3, 2012. The GIRAFFE/ARGUS mode of the Fibre Large Array Element Spectrograph FLAMES (Pasquini et al. 2002) was used, which Integral Field Unit (IFU) provides a continuous spectral coverage for a $11''.4 \times 7''.3$ field of view. Each spaxel in the IFU has a size $0''.52 \times 0''.52$, giving in total of 300 spectra in one exposure. Four telescope pointings covering most of the jet (see Figures 2.2 and 5.2) were obtained, therefore in total 1200 spectra. In Table 2.4 we present the details of those spectra. The first column is the start of the observing night, the second column is the Julian date at mid point of all exposures from the same pointing, the third column refers to the telescope pointing, the fourth column shows the total integration time and the last column the number of frames combined together. The FLAMES data were reduced using the ESO GIRAFFE pipeline version 2.9.2 by the co-author of paper II D. Jones. A standard spectral calibration was performed.

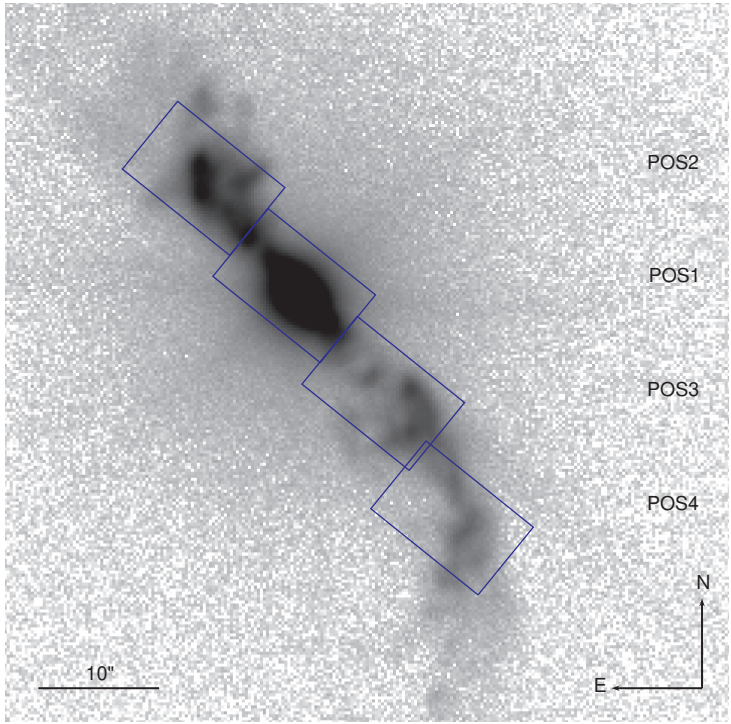

Figure 2.2: VLT+ARGUS pointings overlaid on VLT 2012 [O III] 10s image. FOV $1' \times 1'$.

Table 2.4: Log of the VLT+FLAMES spectroscopic observations of R Aquarii. See text for more details.

| Date | JD | POS | Total exp. time (sec) | Nr. of frames |
|------|-----|-----|-----------------------|---------------|
| 2012-10-03 | 2456203.683866 | 1 | 30; 100; 570 | 1, 1, 2 |
| | 2456203.699699 | 2 | 1535 | 5 |
| | 2456203.769827 | 3 | 1228 | 4 |
| | 2456203.786574 | 4 | 1228 | 4 |

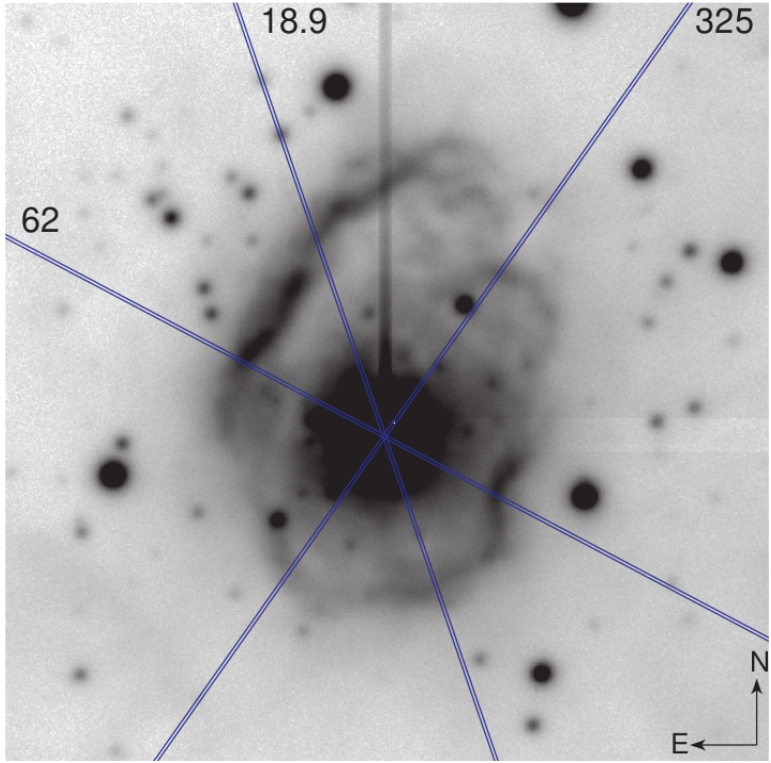

Figure 2.3: The slit positions of MWC 137 overlaid on ALFOSC 2016 Hα+[N II] 600s image. FOV $2' \times 2'$.

As for GK Per, the spectra of MWC 137 were obtained with the NOT+ALFOSC using grism #17 and $0''.5$ slit. Spectral coverage was from 6315 Å to 6760 Å with a dispersion of 0.29 Å pix$^{-1}$. Two slit positions were centred on the star with PAs of 62° and 325°. For each a 30 minutes exposure was obtained. Additional spectra, aligned with the jet components (but slightly off the central star), were obtained at the PA=18°.9. Due to the faintness of the jet, two exposures with 30 minutes each were taken on that PA and combined. IRAF standard routines in the package `longslit` were used to pre-process and wavelength calibrate these data.

# CHAPTER 3
# ANALYSIS METHODS

In this Chapter we present the mathematical background together with the designations used in this Thesis.

## 3.1 Designations

Throughout the Thesis the designations presented in Figure 3.1 are used. In the left panel of the Figure, a view in the plane of the sky is drawn. The $X$ axis points towards the West (W) and $Y$ towards the North (N). The line of sight, $Z$ axis, is perpendicular and pointing out of the drawing plane. The black dots 1 and 2 refer to a knot moving in the plane of the sky between two different epochs with a proper motion $\mu$. PA is the position angle of the knot with respect to the central star (marked with a star) at the first epoch, while $\alpha$ is the angle of the proper motion. Both angles are measured from North to East counterclockwise. The apparent distance from the central star of the knot is $d$. S is South, E is East.

The right panel of Figure 3.1 presents a plane perpendicular to the plane of the sky passing through a given knot (black dot). The plane is defined by the central star, the knot, and the observer. Therefore, for every knot this plane is unique. Every knot has a spatial distance, $R$, from the central star. The apparent distance $d$ is also indicated as well as the distance along the line of sight direction, $Z$. The vector of the expansion velocity is $v_{exp}$, its components in the plane of the sky (tangential velocity) and along the line of sight (radial velocity) are $v_{sky}$ and $v_{rad}$, respectively. Angles $\theta_1$ and $\theta_2$ are both measured from the line of sight toward the expansion velocity vector. If the expansion is purely radial, $\theta_1 = \theta_2$.

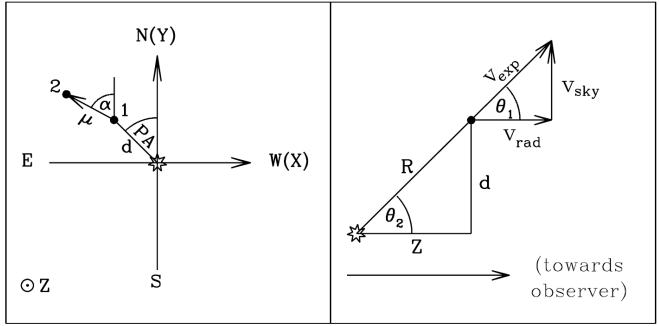

Figure 3.1: Designations used in this Thesis. Left panel: sketch of an example knot (black dot) as seen in the plane of the sky moving between epochs 1 and 2. Right panel: view in the perpendicular plane including the central star and the knot. See text for more details.

## 3.2 Direct measuring

In a case of a well distinguished circular/spherical knot, the $XY$ measure of its centroid for each epoch was used. Then the proper motion, $\mu$, simply follows from

$$\mu \left[ '' \, yr^{-1} \right] = \frac{\sqrt{(x_j - x_i)^2 + (y_j - y_i)^2} \, ['']}{\Delta t_{j-i} \, [yr]}, \tag{3.1}$$

where $(x_i, y_i)$ and $(x_j, y_j)$ refer to the central coordinates of a given knot at the first and second epoch, respectively, and $\Delta t_{j-i}$ is the time span between the two epochs. This method was applied to the data of the nova remnant GK Per.

## 3.3 Magnification method

In the case of extended objects, where no precise central coordinates can be determined, the so-called *magnification method* (see Reed et al. 1999; Santander-García et al. 2007) is used. This method is based on finding the magnification factor, $M$, to match the size of the extended object between the first and second epoch. The image from the first epoch is magnified with a steadily increasing magnification factor. Then the image from the second epoch is subtracted from these various magnified frames. The difference frame that has the smallest residuals is assumed to have the correct magnification factor. In Figure 3.2 an example is presented. The Figure shows the eastern part of the R Aqr bipolar nebula. Black colours represent the features of the nebula on the first epoch, the image which is magnified to match the extent of the features on the second epoch. White colours represent the features on the second epoch. White arrows show the general expansion direction. On the left frame the applied magnification factor, $M = 1.003$, is too small, because the black is seen before the white features when considering the expansion direction. In the middle frame, the difference image with the $M = 1.030$ has the minimal residuals (the features have almost disappeared into the background, black features have been magnified with the proper value to coincide with the white features). In the right frame $M$ is too large, because the black features are ahead of the white ones.

The magnification method was used to investigate the bipolar nebula and the jet of R Aqr. The method assumes homologous expansion of the investigated

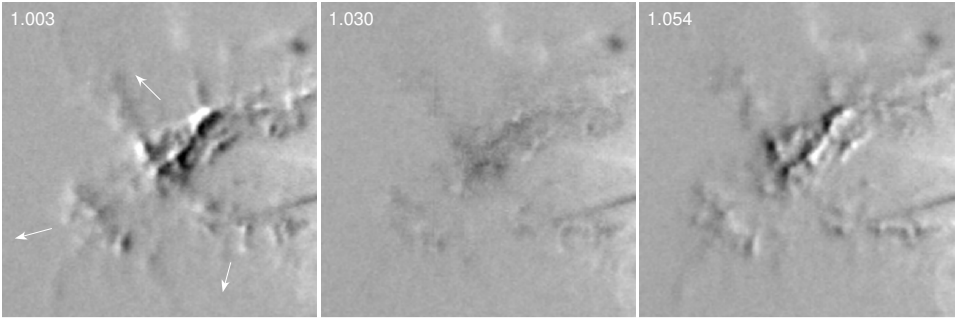

Figure 3.2: An example of a magnification method shown for the residuals with too small magnification factor ($M = 1.003$), appropriate magnification factor ($M = 1.030$), and too large $M = 1.054$. White arrows indicate the general expansion direction of the nebula. Each panel shows a $1' \times 1'$ FOV of the eastern part of the R Aqr bipolar nebula. See text for more details.

structure. However, it permits to find out deviations from this assumption.

In Figure 3.3 an example is presented for the movement of a knot between epochs 1 and 2. The central source from where the knot is ejected during the outburst, is indicated with a star. The observer is on the right, in a point marked with $\oplus$. The line of sight direction is $Z$ and perpendicular to it is the plane of the sky. For very distant objects, the distance between the central star and the observer can be assumed to be identical to the distance between the knot position 1 (or 2) and the observer. We designate this distance as $D$. The knot (black dot) has moved a distance $d$ in the plane of the sky between the initial ejection and the first epoch. Between the epochs 1 and 2 the knot moves a distance $\Delta d$, hence, at the second epoch since the initial ejection, the knot has passed a distance $d + \Delta d$. Considering that the velocity of the knot does not change, corresponding to a ballistic expansion, a simple equation holds $d/t_1 = (d + \Delta d)/t_2$, where $t_1$ is the time from the ejection to epoch 1 and $t_2$, respectively, to epoch 2. With a simple modification of the last formula we get $t_2/t_1 = (d + \Delta d)/d$, which is defined as the magnification factor $M$. Hence,

$$M = \frac{t_2}{t_1} = \frac{(d + \Delta d)}{d},$$
(3.2)

which shows that the magnification factor is dimensionless.

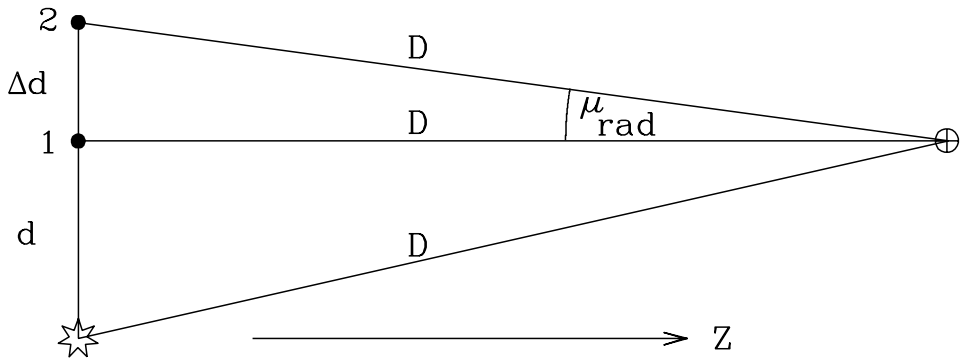

Figure 3.3: An example knot from the observers perspective. See text for more details

In addition, we can define:

$$\Delta t = t_2 - t_1 = (t_2 - t_1) \cdot (\frac{t_1}{t_1}) = t_1(\frac{t_2}{t_1} - 1),$$
(3.3)

where $\Delta t$ is the duration between epoch 1 and 2. Inserting Equation (3.2) into Equation (3.3) delivers

$$t_1 = \frac{\Delta t}{M - 1}.$$
(3.4)

Therefore, with known magnification factor and time difference between the epochs, the age of the knot at the first epoch can be calculated. For convenience we designate this kinematic age as $T$. Thus:

$$T \; [yr] = \frac{\Delta t \; [yr]}{(M - 1)},$$
(3.5)

Similarly, from the Equation (3.2) we know that $\Delta d = d(M-1)$, where $\Delta d$ is the total displacement of a given knot in the plane of the sky between two epochs. When measuring this displacement in arcseconds per unit of time, we obtain the proper motion $\mu = \Delta d/\Delta t$. Hence, knowing the magnification factor $M$, distance $d$ from the central star at the first epoch, and time interval, it is possible to derive $\mu$,

$$\mu \; ['' \; yr^{-1}] = \frac{(M - 1) \cdot d \; ['']}{\Delta t \; [yr]}.$$
(3.6)

This last equation is useful to determine the proper motion of an extended object, when no clear movement of the central coordinate can be detected but the magnification method can be accurately determined.

## 3.4 Calculations in the plane of the sky

From Figure 3.3 it is evident that $\tan(\mu_{rad}) = \Delta d/D$. Considering that the angles are small we obtain $\mu_{rad} = \Delta d/D$, where $\mu_{rad}$ is measured in radians. Considering linear velocity of the knot, the distance $\Delta d$ passed within the time interval $\Delta t$ is $\Delta d = v_{sky} \cdot \Delta t$, where $v_{sky}$ is the tangential velocity defined in Fig 3.1. Hence, $\mu_{rad}$ can be expressed in terms of this tangential velocity as

$$\mu_{rad} \left[ rad \right] = \frac{v_{sky} \left[ km \ s^{-1} \right] \cdot \Delta t \left[ s \right]}{D \left[ km \right]}. \tag{3.7}$$

When measuring $\mu_{rad}$ per unit of time, we receive the proper motion $\mu$:

$$\mu \left[ rad \ s^{-1} \right] = \frac{\mu_{rad} \left[ rad \right]}{\Delta t \left[ s \right]} = \frac{v_{sky} \left[ km \ s^{-1} \right]}{D \left[ km \right]}. \tag{3.8}$$

Thus, from the proper motion, $\mu$, of a given object and the known distance $D$ to the object, it is possible to calculate the component of the expansion velocity in the plane of the sky in $km \ s^{-1}$:

$$v_{sky} \left[ km \ s^{-1} \right] = \mu \left[ rad \ s^{-1} \right] \cdot D \left[ km \right], \tag{3.9}$$

In order to use the formula in convenient units, the following transformation is needed:

$$v_{sky} \left[ km \ s^{-1} \right] = \frac{3.08568025 \times 10^{13}}{206265 \cdot 365.24 \times 24 \times 3600} \cdot \mu \left[ '' \ yr^{-1} \right] \cdot D \left[ pc \right], \tag{3.10}$$

hence,

$$v_{sky} \left[ km \ s^{-1} \right] = 4.74 \cdot \mu \left[ '' \ yr^{-1} \right] \cdot D \left[ pc \right], \tag{3.11}$$

The kinematic age $T$ can be calculated from $\mu$ as well if we know the distance from the central star in the plane of the sky:

$$T \left[ yr \right] = \frac{d \left[ '' \right]}{\mu \left[ '' \ yr^{-1} \right]}. \tag{3.12}$$

Figure 3.4 shows an expanding ring nebula in the plane of the sky. The ring is inclined with respect to the line of sight, resulting in an elliptic shape in the plane of the sky. The solid line indicates the expanding ring at epoch 1, when the example point (black dot 1) of the ring has a distance $d$ from the central star (marked with a star). At the second epoch the nebular ring has expanded until the dashed line. The example point (black dot 2) has moved a distance $\Delta d$ and the ring ellipse has increased by the magnification factor $M$. From Section 3.3 we know that $\Delta d = d(M-1)$. Expansion parallax is given by

$$D\ [km] = \frac{v_{exp}\ [\,\mathrm{km\,s^{-1}}]}{\frac{\Delta d\ [rad]}{\Delta t\ [s]}}, \qquad (3.13)$$

where $D$ is distance to the object, $v_{exp}$ is the expansion velocity of the nebular ring, $\Delta d / \Delta t$ is the angular expansion rate over a period of $\Delta t$. For an extended object like the nebular ring, we cannot directly measure the angular expansion rate but we can measure $d$ and the magnification factor $M$, which are different for different points of the ellipse. Rearranging the equation and transforming it into convenient units similarly to the Equation (3.10) we get

$$D\ [pc] = 0.211\ \frac{v_{exp}\ [\,\mathrm{km\,s^{-1}}]\Delta t\ [yr]}{d\ ['']\cdot(M-1)}. \qquad (3.14)$$

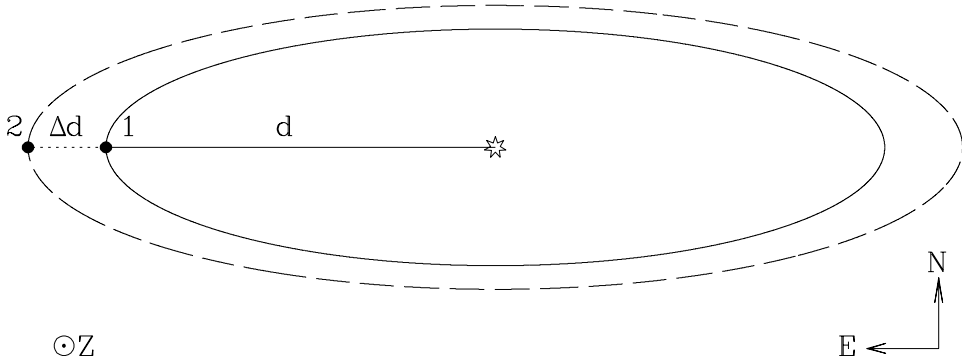

Figure 3.4: Nebular ring expanding in the plane of the sky from epoch 1 (solid line) to epoch 2 (dashed line). The central source is indicated with a star. See text for more details.

## 3.5 Radial velocity

Radial velocities were calculated using the Doppler shift and corrected for the Local Standard of Rest (LSR) as well as for the systemic velocity of the source:

$$v_{rad}\ [km\ s^{-1}] = \frac{\lambda_m\ [\mathring{A}] - \lambda_l\ [\mathring{A}]}{\lambda_l\ [\mathring{A}]}\cdot c\ [km\ s^{-1}] + LSR\ [km\ s^{-1}] - v_{sys}\ [km\ s^{-1}], \qquad (3.15)$$

where $\lambda_m$ and $\lambda_l$ are the measured and the laboratory wavelengths for a given emission line, $c$ is the speed of light, and $v_{sys}$ is the systemic velocity of the source.

The measured Full Width at Half Maximum (FWHM) was firstly instrumentally corrected:

$$FWHM\ [\mathring{A}] = \sqrt{FWHM_m^2\ [\mathring{A}] - FWHM_t^2\ [\mathring{A}]}, \qquad (3.16)$$

where $FWHM_m$ is the FWHM of the measured emission line, while the $FWHM_t$ is theoretical spectral resolution measured as a FWHM of the calibration arc lamp lines. The FWHM in $km\ s^{-1}$ was calculated as follows:

$$FWHM\ [km\ s^{-1}] = \frac{FWHM\ [\mathring{A}] \cdot c\ [km\ s^{-1}]}{\lambda_m\ [\mathring{A}]}, \qquad (3.17)$$

# CHAPTER 4

# A THREE-DIMENSIONAL VIEW OF THE REMNANT OF NOVA PERSEI 1901 (GK PER)

Nova Persei 1901 (GK Per) was discovered by Scottish clergyman Thomas Anderson on 1901 February 21 at a visual magnitude of 2.7. Only one day before, its brightness must have been fainter than 12 mag (Williams 1901). GK Per reached the maximum light on February 22 when it was as bright as Vega, 0.2 mag (Figure 4.1). After the maximum peak it rather quickly, in just 6 days, faded to 2 mag and during the further 2 weeks down to 4 mag. Next, a set of oscillation with an amplitude of a 1.5 mag with a period of about 4 days started. The brightness variations lasted for several months while the star consistently faded. The pre-outburst brightness was reached eleven years later.

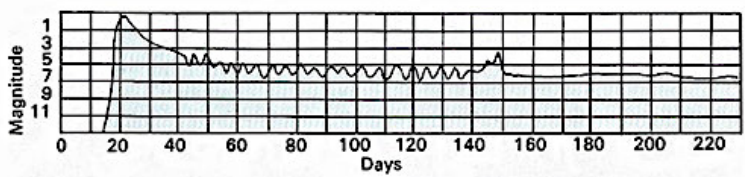

Figure 4.1: Light curve of GK Persei in 1901 from Harvard Annals. Credit: David Darling.

GK Per was the first object around which superluminal light echoes were detected (e.g. Ritchey 1902). The actual ejecta of the nova outburst was discovered in 1916 by Barnard (1916) and is still easily observable now: indeed, it is considered to be the longest lived and most energetic nova remnant ever found.

The earliest optical images of the remnant showed an emission only in the Southwest (SW) side of the remnant (Figure 4.2). Gradually more material became visible around the central source, however, the asymmetric nature of the remnant has remained. Radio observations obtained during 1984–1986 showed that the SW quadrant of the remnant is the source of non-thermal (synchrotron) emission (Seaquist et al. 1989, left panel of Figure 4.3). This would be caused by particle acceleration and enhanced magnetic fields in a fast shock where the expelled matter is meeting some pre-existing material, most probably the result of previous mass ejections from the system, possibly an ancient planetary nebula (Bode et al. 2004). This scenario is supported by the analysis of the X-ray emission from the remnant (Balman 2005, right panel of Figure 4.3). A lumpy, highly asymmetric X-ray nebula is observed, with brightness enhancement in the SW quadrant and wing-like extensions in the Southeast (SE) and Northwest (NW) directions. It is concluded that a density gradient in the circumstellar medium causes the expansion of the X-ray shell to be faster in the NW and SE directions, reaching velocities as high as ∼2600 - 2800 km s$^{-1}$, than toward the SW, where velocities are less than half of that, around ∼1100 km s$^{-1}$. This strong interaction is expected to cause deceleration of the expansion of the SW part of the outflow (Bode et al. 2004). This is one of the main motivation to investigate the nova remnant GK Persei, to see if the deceleration is detectable also in the optical waveband.

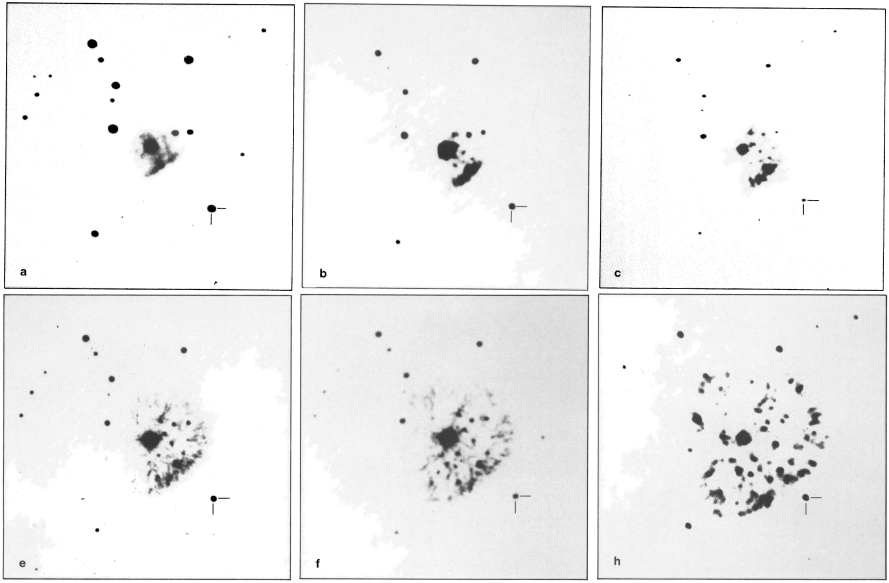

Figure 4.2: Images of GK Per from *(a)* 1917, *(b)* 1937, *(c)* 1942, *(e)* 1949, *(f)* 1959, *(h)* 1977. Figure depicted from Seaquist et al. (1989).

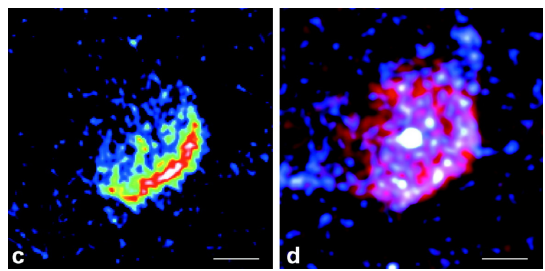

Figure 4.3: GK Per in 5 GHz radio image in 1997 *(c)* and in X-ray emission in 2000 *(d)*. Figure depicted from Bode et al. (2004).

At a distance of half a kilometer (e.g. McLaughlin 1960, this work) and with an expansion velocity of a thousand km s$^{-1}$, the apparent growth of the nebula is about one arcsec per year, which is easily resolvable from ground-based optical imagery even on a timescale of few months. Therefore GK Per offers the opportunity to study the evolution of a stellar outflow in real-time, and to determine not only the expansion velocities but also their derivatives (i.e. acceleration, thus the forces at work), or deviations from pure radial expansion, an important information that is generally not available for dynamical studies of astrophysical outflows. If the expansion in the plane of the sky is coupled with Doppler shift velocities, information about five of the six dimensions of the phase-space is gained, and an almost complete dynamical picture can be drawn. For these reasons, starting on 2004 we embarked in a program of frequent imaging monitoring of the expansion of the GK Per remnant. We present here the results of the work. Doppler shift velocities of 217 knots from long-slit spectroscopy are combined with high-precision measurements of the motion in the plane of the sky for 282 knots. The latter ones are computed using good-quality, ground-based $H\alpha$+[NII] images spanning 25 years, with a particularly intensive time coverage since 2004.

## 4.1 Description of the nebula

The optical nebula of GK Persei (GK Per), displayed in Figure 4.4, has a roundish, knotty morphology. Some deviation from the circular symmetry is however visible because the diameter of the nebula in the NE-SW direction is smaller (105″ as measured in the 2011 image) than in the NW-SE direction (118″). As described in detail by Shara et al. (2012), the nebula is composed of hundreds of knots and filaments of different sizes and brightnesses. Many of the knots have tails pointing either toward or away from the central star. Shara et al. (2012) counted 937 knots in their HST images. Some of them are blended in our lower-resolution ground-based images. On the other hand, due to the limited bandpass of the [NII] F658N HST filter some knots with the largest Doppler-shift displacements, which owing to the projection effects are located near the centre of the remnant, are missed in the HST images.

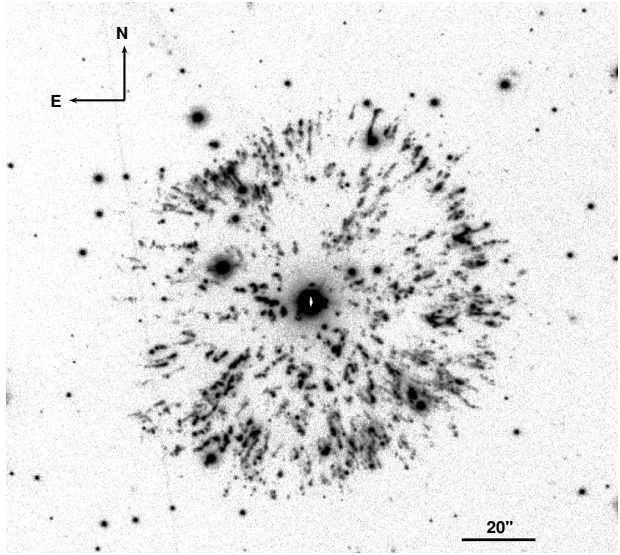

Figure 4.4: Original frame from 2007-09-05 obtained with the NOT. The image shows the filamentary nature of the GK Per remnant. The FOV is $3' \times 3'$. Figure is a reprint from Paper I.

## 4.2 Proper motions

Our multi-epoch imaging allows a precise determination of the apparent expansion of the GK Per nebula in the plane of the sky. This is illustrated in the video which can be seen in Figure 3 in Publication I. The animation includes the early photographic and CCD images, astrometrically matched to the recent ones but in an arbitrary intensity scale and with a generally poor resolution, as well as a selection of the best INT images in the sequence of highly homogeneous data taken starting on 2004, in the same linear intensity scale. In the latter part of the sequence, not only the overall expansion of the outflow is clearly revealed, but also changes in brightness of some of the knots. Latter will be discussed in Section 4.5.

The motions in the plane of the sky for 282 individual knots were determined using the nineteen, carefully registered images obtained between 2004 and 2011. The measured knots are indicated in Figure 4.5. The position of knots at each epoch was computed by fitting independently in the $X$ and $Y$ directions a Gaussian using the IRAF task *imexamine*, with fitting parameters adjusted to the mean characteristic of the knots. Only knots for which a reasonable Gaussian fit can be obtained were considered. Then the proper motion of each knot were determined by means of a least square fit of the positions at the different epochs. Both the distance variation from the central star, as well as the separate motions in the $X$ and $Y$ directions were computed, as illustrated in Figure 4.6. In this way, the proper motion vector is determined, which in addition to the magnitude of the spatial displacement $\mu$ also contains the information on its direction (angle $\alpha = \arctan(\mu_x/\mu_y)$, see Figure 3.1). For all of the knots, a straight line provides an excellent fit to the variation of $d$ during the 7.9-yr lapse of time considered. This is further illustrated in Figure 4.7. Proper motions range from $0''.007$ yr$^{-1}$ to $0''.53$ yr$^{-1}$.

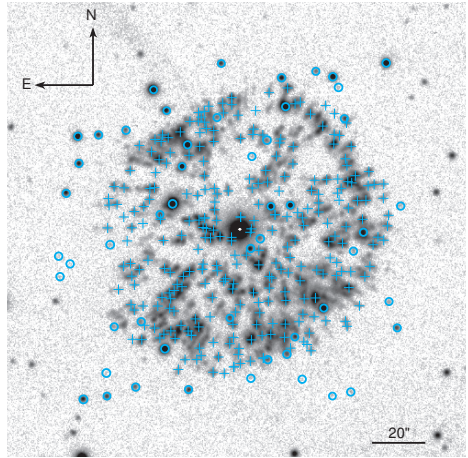

Figure 4.5: Knots with proper motion measurements (crosses) indicated on our reference frame 2004-01-29. Circles show stars inside or nearby the remnant. FOV $3' \times 3'$. Figure is a reprint from Paper I.

The prime focus camera of the INT has well defined and small geometrical distortions[1] in the limited FOV covered by the GK Per nova remnant, and which are robustly removed during the astrometric registering of images. Therefore, errors in the proper motion determination mainly depend on the goodness of the fit of the position of the knots at the different epochs, which in turn depends on the knot's shape and brightness. As the proper motion error, we adopt the formal error of the least square fit: its average value is $0''.010 \pm 0''.006$ yr$^{-1}$.

In the proper motion calculation, we did not use the images earlier than 2004 as they come from a less homogeneous set of observations with generally poorer resolution or astrometric properties, as well as different filters. They are however used to search for signs of acceleration/deceleration of knots in the last decade with respect to the previous 20 years. In this respect, we found no evidence for a systematic acceleration/deceleration, except in a few cases (see the lower three panels in the left column of Figure 4.7). Among these, a modest slow down is the most common case.

---

[1]http://www.ast.cam.ac.uk/∼wfcsur/technical/astrometry/

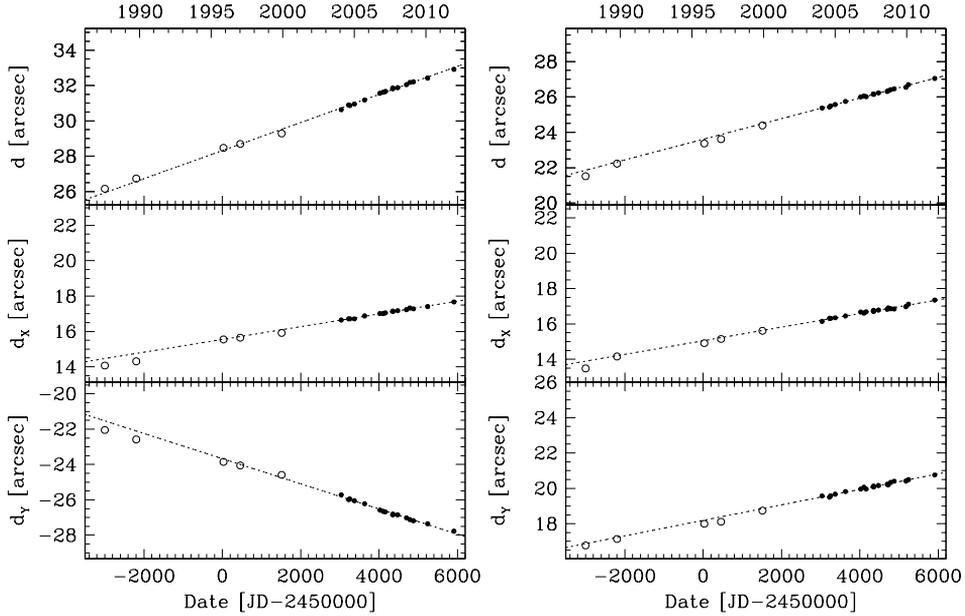

Figure 4.6: Example of proper motion determination of two knots. Upper panels show the variation of distances $d$ with time, and the lower two panels its components along the $X$ and $Y$ axes ($d_x$ and $d_y$) as defined in Figure 3.1. Open circles are data from 1987 to 1999 (not used for the proper motion determination), and filled circles those obtained from 2004 to 2011. Dotted lines are the least square fits of data taken after 2004. Error bars are smaller or equal to sizes of symbols. Figure is depicted from Paper I.

The knots positions, their proper motions $\mu$, $\mu_x$, $\mu_y$, and direction $\alpha$, position angles, radial velocities, and the errors associated to all these quantities, are listed in Table 4.1. An selection is presented in here, while the entire table is available in the Appendix, Table 8.1, and in the electronic version of the Publication I. In the table, individual knots are identified by their X and Y distance in arcsec from the central source ($d_x$, $d_y$) on the plane of the sky at the reference epoch as determined from the adopted least square fits.

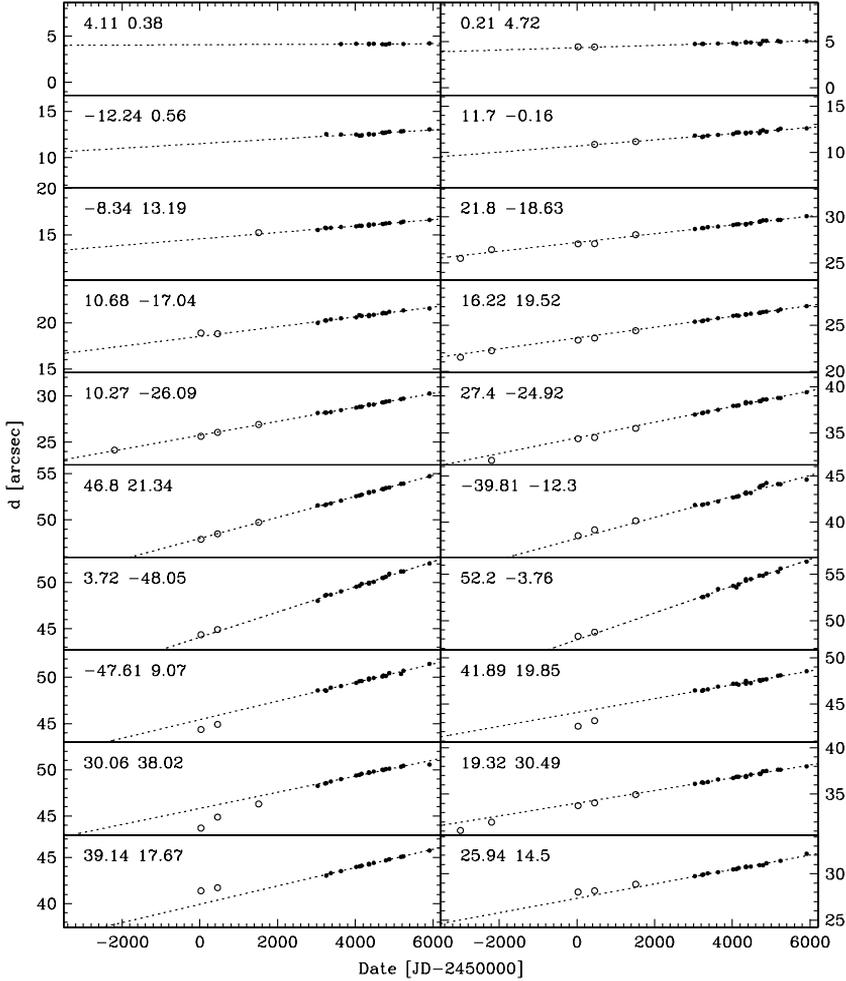

Figure 4.7: More examples of proper motion determinations. In the first seven rows, knots are ordered according to increasing proper motion magnitude (the range spanned by ordinates is the same for all graphs, 10 arcsec). In the three bottom rows, some cases with signs of deceleration/acceleration in the last two decades are shown. At the upper-left corner of each panel, the knot's coordinates ($d_x$ and $d_y$ distances from the centre, in arcsec, extrapolated at the reference epoch, see Table 4.1) are indicated. Error bars are smaller or equal to sizes of symbols. See the caption of Figure 4.6 for the explanation of legends. Figure is a reprint from Paper I.

Table 4.1: Knots positions, proper motions, and radial velocities

| $d_x{}^a$ ('') | $d_y{}^a$ ('') | $\mu$ ('' yr$^{-1}$) | $\sigma_\mu$ ('' yr$^{-1}$) | $\mu_x$ ('' yr$^{-1}$) | $\sigma_{\mu_x}$ ('' yr$^{-1}$) | $\mu_y$ ('' yr$^{-1}$) | $\sigma_{\mu_y}$ ('' yr$^{-1}$) | PA (°) | $\sigma_{PA}$ (°) | $\alpha$ (°) | $\sigma_\alpha$ (°) | $v_{rad}$ (km s$^{-1}$) |
|---|---|---|---|---|---|---|---|---|---|---|---|---|
| 0.06 | -27.43 | 0.242 | 0.005 | 0.005 | 0.004 | -0.242 | 0.005 | 180.1 | 0.3 | 181.2 | 0.9 | – |
| 0.09 | -48.91 | 0.448 | 0.006 | -0.009 | 0.012 | -0.448 | 0.006 | 180.1 | 0.2 | 178.8 | 1.5 | – |
| 0.21 | 4.72 | 0.046 | 0.011 | -0.007 | 0.012 | 0.046 | 0.011 | 357.5 | 2.0 | 8.9 | 14.7 | 808 |
| 0.21 | 4.72 | 0.046 | 0.011 | -0.007 | 0.012 | 0.046 | 0.011 | 357.5 | 2.0 | 8.9 | 14.7 | 919 |
| 0.22 | 33.44 | 0.261 | 0.003 | -0.005 | 0.004 | 0.261 | 0.003 | 359.6 | 0.3 | 1.0 | 0.8 | – |
| 2.25 | -9.25 | 0.081 | 0.004 | 0.010 | 0.010 | -0.080 | 0.005 | 193.7 | 1.0 | 186.8 | 7.1 | -822 |
| 3.41 | 18.67 | 0.153 | 0.013 | 0.003 | 0.019 | 0.155 | 0.014 | 349.7 | 0.5 | 358.9 | 7.1 | 721 |
| 3.69 | -0.88 | 0.048 | 0.017 | 0.041 | 0.015 | -0.033 | 0.022 | 256.6 | 2.5 | 231.4 | 21.0 | -894 |
| 3.72 | -48.07 | 0.497 | 0.012 | 0.057 | 0.007 | -0.494 | 0.012 | 184.4 | 0.2 | 186.5 | 0.9 | – |
| 4.11 | 0.38 | 0.007 | 0.006 | 0.006 | 0.006 | 0.007 | 0.011 | 275.3 | 2.3 | 317.3 | 54.0 | -794 |

The entire table is available in the Appendix (Table 8.1) and in the electronic version of the Publication I.

The first 10 lines are shown here for guidance regarding its form and content.

$^a$Extrapolated to 2004-01-29, or JD2453034.37.

### 4.2.1 Comparison with published data

A study of the proper motions in GK Persei remnant knots was done by Shara et al. (2012) using the 1995 and 1997 HST images listed in Table 2.2. The proper motions of 938 knots is estimated, but no quantitative comparison with our measurements is possible as Shara et al. (2012) did not publish data for individual knots. In spite of the higher resolution of the HST, the seven-times longer baseline of our observations, the large number of measurements (19, which reduce statistical errors), and the accurate knowledge of the astrometric properties of the INT prime-focus camera, provide proper motion measurements which are more accurate by a factor of four (cf. Figure 8 in Shara et al. 2012 with our average error of 0."01 yr$^{-1}$).

Anupama & Prabhu (1993) (AP93) have measured the motion in the plane of the sky of 20 individual bright knots using two ground based images obtained on 1984.63 and 1990.10 at Vainu Bappu Observatory (India), with a similar filter as in our observations. From visual inspection of their Figure 1, it is possible to safely identify 15 knots in common with our measurements. The comparison of their proper motions is presented in Figure 4.8. Individual differences between the two datasets are above the quoted errors, but no systematic effects appear. Given that our measurements come from the analysis of 19 accurately registered high-quality images, we keep our proper motion measurements for the following analysis.

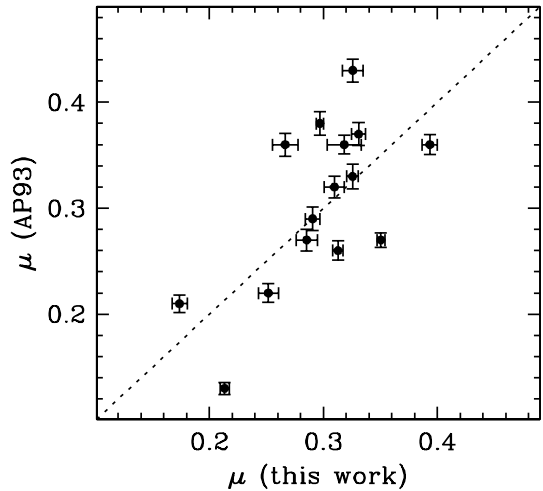

Figure 4.8: Comparison of proper motion of 15 knots in common with Anupama & Prabhu (1993). Figure is a reprint from Paper I.

## 4.3 Radial velocities

Our medium resolution spectra include the Hα line and the [NII] doublet. No other emission lines were detected at the depth of these observations in the relatively small spectral range covered (see Table 2.3). First, the spatial regions which clearly correspond to knots identified in the images obtained at the same epoch were extracted from the long-slit data. In this way, 217 knots could be identified and measured. Then the signal in each region was integrated to create the Hα and [NII] line profiles of each knot. They are generally broad, as also noted by Shara et al. (2012), with an instrument-corrected full width at half maximum up to 200 $km\,s^{-1}$, as illustrated in Figure 4.9. The figure also shows that knots toward the centre of the nebula (right panel) frequently have marked asymmetric shapes, with tails extending to velocities as large as 300 $km\,s^{-1}$. Given the strong projection effects in these central regions, it is not clear if these tails are intrinsic to the knots, associated with related filaments, or caused by apparent superposition of more than one knot.

The radial velocities of each knot were then measured by Gaussian fitting the [NII] 6583 Å line profile. When the line clearly had more than one peak or possessed non-negligible wings, multi-Gaussian fitting was applied, but the secondary components (which takes into account the wings or minor peaks) were not used in the following kinematic analysis. Errors on the radial velocity have been estimated by comparing multiple measurements of the same knots in different nights and with the two different telescopes used. A typical error of 8 $km\,s^{-1}$ is adopted. All velocities were corrected to the Local Standard of Rest (LSR). As the LSR systemic velocity of GK Per was estimated by Bode et al. (2004) to be $45 \pm 4$ $km\,s^{-1}$, radial velocities were also corrected by this value. Table 4.1 lists the radial velocities of knots with respect to the systemic velocity of GK Per. They range from $-989$ to $+967$ $km\,s^{-1}$.

## 4.4 Kinematical analysis

Given its roundish overall morphology, as a first approximation we consider the nebula of GK Per to be a quasi-spherical shell in which knots expand radially from the central star. This hypothesis is broadly consistent with the plot of the observed radial velocities $v_{rad}$ of the knots as a function of their apparent distance from the central star (Figure 4.10, upper panel). The two solid lines in the figure indicate the expected distance-velocity plot for spherical shells of radii of $35''$ and $55''$, and expansion velocities of 400 and 1000 $km\,s^{-1}$, respectively. Most of the knots are comprised within these two limits, suggesting that the knots of the nebula of GK Per form a relatively thick shell expanding with a significant range

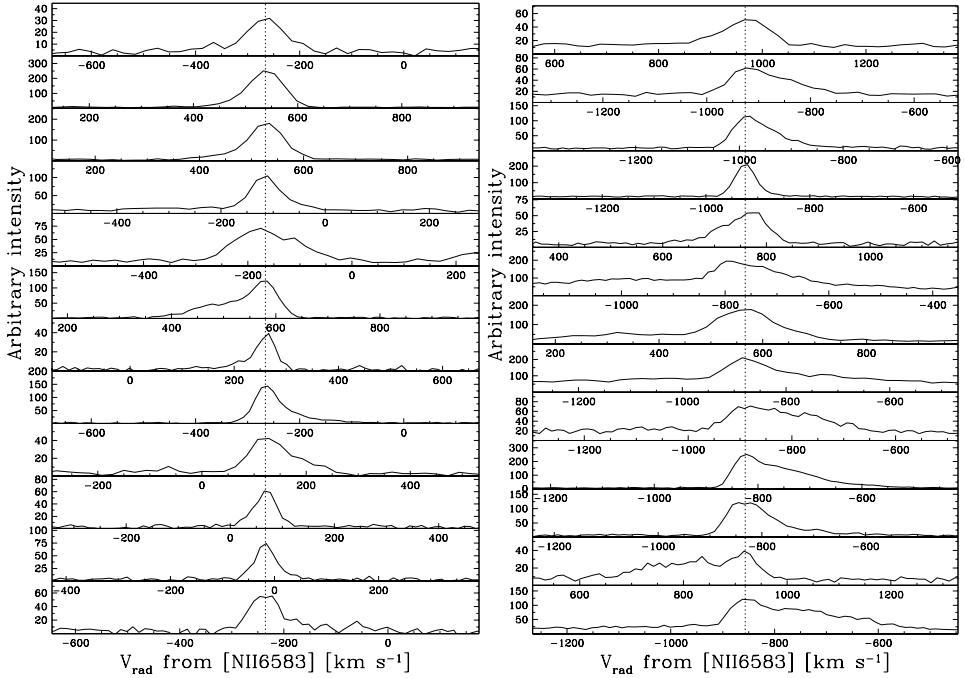

Figure 4.9: Example of intensity profiles of knots on the edges (left panel) and near the centre (right panel) of the remnant. Vertical dotted line indicates the measured radial velocity for the corresponding knot. Figure is a reprint from Paper I.

of velocities.

This conclusion is supported by the behaviour of the proper motions $\mu$. The component of the expansion velocity in the plane of the sky, $v_{sky}$, immediately follows from $\mu$ according to equation 3.10. For a distance of $400 \pm 30$ pc (see next paragraph), $v_{sky}$ of the measured knots varies between 13 and 1005 km s$^{-1}$, and its variation with distance from the centre is also consistent with a spherical thick shell model with parameters as described above (Figure 4.10, lower panel). We note that Harvey et al. (2016), using the kinematical results of this Thesis (Paper I), proposes a different morphology, a cylindrical sphere with polar cones for the nova remnant of GK Per. Indeed, the data does not exclude this possibility (see Figures 9 and 10 in Harvey et al. 2016). However, we emphasise that there are datapoints which do not match with their proposed morphology, while a symmetrical thick shell proposed in this work (see Figure 4.10) is matching just as plausibly, or a bit better, with the datapoints. For that reason we have chosen to favour our own proposed model in this Thesis.

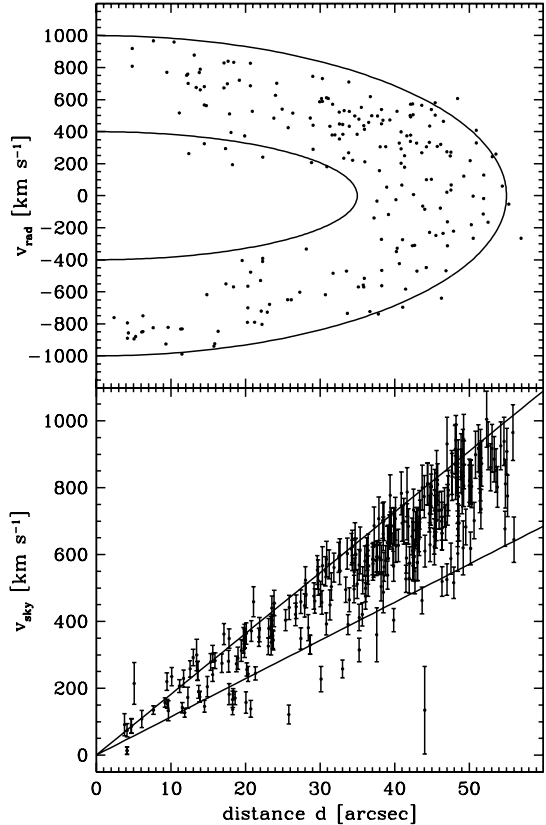

Figure 4.10: Above: Doppler-shift velocities of 217 knots as a function of their distance from the central star projected in the plane of the sky. Solid lines indicate the position-velocity plot for thin spherical shells of radii $35''$ and $55''$. Below: the same, but for the velocity component in the plane of the sky from the measurements of the proper motions of 282 knots. The solid lines indicate velocities of 400 and 1000 $\mathrm{km\,s^{-1}}$. Figure is a reprint from Paper I.

The combination of the velocity components in the plane of the sky and along the line of sight provides the velocity vectors of the knots (see the right panel of the Figure 3.1) once the distance to GK Per is fixed. We could measure both velocity components for 117 knots: their deprojected speed is $v_{exp} = \sqrt{v_{sky}^2 + v_{rad}^2}$, and $\theta_1 = \arctan(v_{sky}/v_{rad})$ is the angle between the line of sight and the velocity vector (as in Figure 3.1). Assuming a purely radial expansion, $\theta_1$ is also the angle that the line joining the central star with a knot forms with the line of sight, $\theta_2$ in Figure 3.1. The deprojected distance from the centre would therefore be $R = d/|\sin\theta_1|$, and its projection along the line of sight is $Z = R\cos\theta_1$. We define the sign of $Z$ so that it is positive when a knot is approaching us (i.e. blue-shifted) and negative when it is receding (red-shifted). We also define $d$ to be positive for all knots in the Northern side of the nebula and negative for those in the Southern half. With these notations, a representation of the "depth" of the nebula along the line of sight (i.e. its extension along the $Z$ direction) is shown in Figure 4.11. In the figure, the observer is on the right side, and the plane $Zd$ represents the folding of all planes perpendicular to the plane of the sky and passing through the central star and the knots considered (right panel of Figure 3.1). 99 knots with a computed error on $R$ smaller than $10''$ are plotted. As stated above, the extension of the nebula along $Z$ depends on the adopted distance $D$. We assume that the distance value that better corresponds to an overall spherical shape of the nebula is our "expansion-parallax" determination of the distance to GK Per. The adopted value, estimated by visual inspection of the variation of the overall shape of the nebula as a function of $D$, is 400±30 pc. This is broadly equivalent to assuming that $v_{sky}$ for knots at the outer apparent edge of the nebula is the same as $v_{rad}$ for knots in the innermost regions.

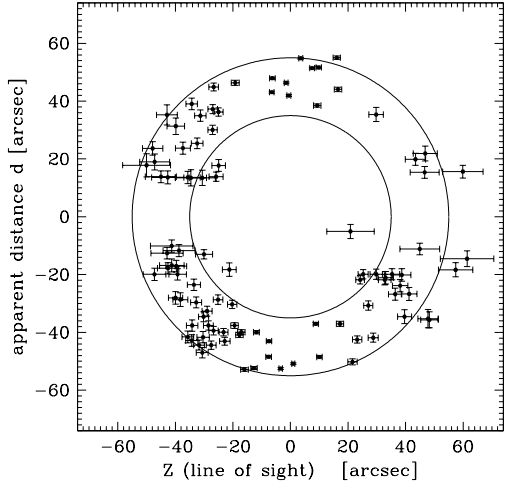

Figure 4.11: Nebular geometry along the line of sight ($Z$ direction, observer on the right side) for the adopted distance to GK Per of 400 pc. Circles indicate inner and outer radii ($35''$ and $55''$ respectively) as in Figure 4.10. See text for details. Figure is a reprint from Paper I.

Figure 4.12 shows the relations between deprojected quantities, namely the speed $v_{exp}$ and the distances $R$ from the central star. As in previous graphs, it confirms that the nova remnant is thick (nearly a half of its outer radius) and that the expansion speed of knots covers the significant range from 250 to 1100 km s$^{-1}$, with the majority of them having a speed between 600 and 1000 km s$^{-1}$. In the figure, we also indicate the expected relationship between velocity and distance for a ballistic ejection (Hubble-like flow) with an age $t_{nova}$=102.9 years, which is the time lapse between the nova maximum of 1901 Feb 22 and our reference epoch of 2004 Jan 29. Points above the solid line have a kinematical age smaller than $t_{nova}$, and those below the line have a larger kinematical age, which indicates that they suffered some deceleration in the course of the expansion. For comparison, in the figure we show as a dotted line also the relation expected for a ballistic expansion with an age of 140 years. The knots age is further investigated in the next section.

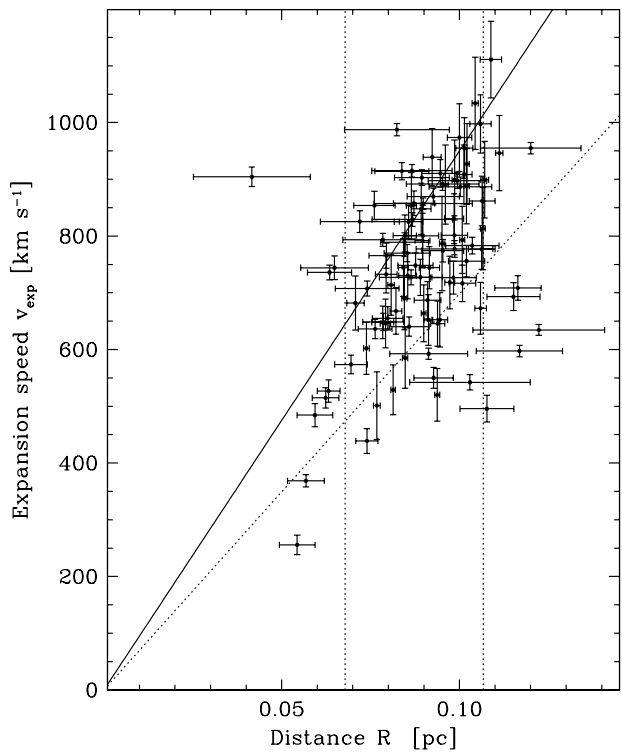

Figure 4.12: Expansion speed vs. linear distance from the central star for $D$=400 pc. The dotted vertical lines indicate the radius limits of $35''$ and $55''$ as in the previous graphs. The solid line indicates the relation expected for ballistic expansion of material ejected in 1901 (102.9 years from our reference epoch). The dotted line shows the same relation for an age of 140 years. Figure is a reprint from Paper I.

### 4.4.1 Kinematical ages

According to equation 3.12 a direct estimate of the kinematical age of each knot can be obtained from the proper motion measurements, where $d$ is measured at our reference epoch (see Figure 3.1).

In Figure 4.13 we present the kinematical age computed in this way as a function of the knots position angle. In the figure, only the 151 knots with an estimated error on the age smaller than 8 years, and with an apparent distance from the centre $d \geq 35''$ (in order to focus on the directions close to the plane of the sky and minimise projection effects) are included. Ages vary between 96 and 170 years, the mean value, weighted by errors, being $118 \pm 12$ years. The horizontal line indicates the nova age $t_{nova}$, and shows that in general knots have suffered only a modest deceleration since their ejection. This conclusion is consistent with the comparison between the knots speeds of up to 1100 km s$^{-1}$ determined above (Figure 4.12), and an initial expansion of the outflow of $\sim$1300 km s$^{-1}$ (Pottasch 1959). Indeed, in the simplistic hypothesis of constant deceleration throughout the knots lifetime (Duerbeck 1987), their observed mean age of 118 years implies that the present-day average speed is 77% of the original ejection velocity. The time $t_{\frac{1}{2}}$ for individual knots to decrease in speed by a factor of two would be 220 years. Even for knots with an apparent age of 140 years (the upper limit for the vast majority of points in Figures 4.12 and 4.13), the velocity would have de-creased by only 42% in the 102.9 years since the nova explosion, and $t_{\frac{1}{2}}$ would be 120 years. The data therefore exclude the short timescale, $t_{\frac{1}{2}}=58$ years estimated by Duerbeck (1987).

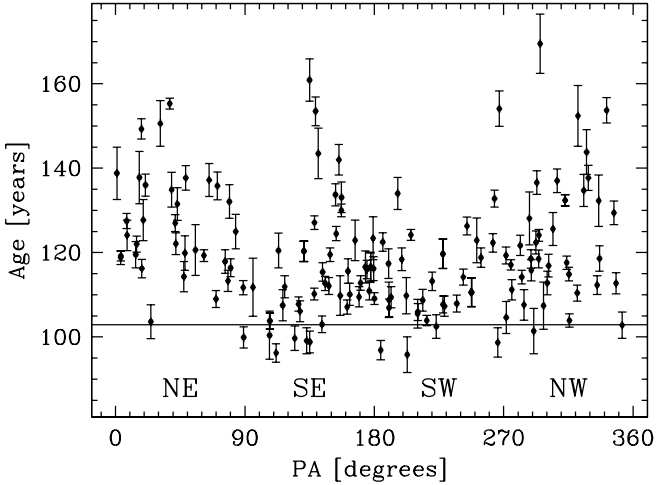

Figure 4.13: Kinematical ages as a function of the knots position angle, for knots with $d \geq 35''$. The horizontal line is the real age of the nova, 102.9 years. Figure is a reprint from Paper I.

### 4.4.2 Deviations from spherical symmetry

The calculations of Duerbeck (1987) were based on the measurement of the radius of the SW side of the nebula from images taken between 1917 and 1984. These measurements were extended by Seaquist et al. (1989), who analysed them using a more physical model which considered the expansion of an isothermal or adiabatic shock created by the interaction of the expanding outflow with a tenuous circumstellar medium. Such an analysis was further extended by Anupama & Kantharia (2005) with images taken till 2003. We refrain from further extending the method, as we found that the measurement of the size of the nebula is highly subjective, owing to the difficulty in defining the nebular edge due to the lack of an abrupt fall of surface brightness, clumpiness (which is enhanced by the better resolution of the recent images), and asymmetries. Indeed, our estimates of the size differ from those of Seaquist et al. (1989) and Anupama & Kantharia (2005) by up to several arcseconds, which prevents any detailed discussion or refinement of the results obtained by these authors.

The proper motion measurements also allow us to measure any deviation from circular symmetry of the expansion pattern in the directions close to the plane of the sky. Anisotropies along the line of sight depends on the adopted distance of the system, and indeed in the previous paragraphs we used the argument inversely, having minimised any anisotropy in order to estimate a distance to the system.

Considering the data in Figure 4.13, the mean weighted age in the different quadrants is: $125 \pm 14$ yr (NE, 34 knots), $115 \pm 11$ yr (SE, 45 knots), $113 \pm 10$ yr (SW, 34 knots), and $121 \pm 11$ yr (NW, 38 knots). We conclude that no strong evidence exists for deviations from the circular symmetry in the expansion pattern. If any, knots in the Northern side of the nebula seem to have suffered a stronger deceleration throughout the remnant lifetime than those in the Southern part. Therefore, our calculations do not support the prediction of Bode et al. (2004) that knots in SW quarter should have been slowed down more than the rest. Smaller proper motions in the NW quadrant are also mentioned in Shara et al. (2012).

In addition, our initial hypothesis of purely radial motions can be tested by comparing the direction of the velocity vectors in the plane of the sky, $\alpha$, with the position angle PA of knots in the plane of the sky, i.e. their radial direction from the central star. As in Figure 4.13, only knots with $d \geq 35''$ are considered in order to limit projection effects. Figure 4.14 shows that velocity vectors are generally aligned along the radial direction but a clear symmetric pattern of non-radial velocities is also observed. The maximum deviations are observed in the NE quadrant, where knots trajectories seems systematically bended toward the East with respect to the radial direction, and the Western part of the nebula, where knots velocity vectors are preferentially bended toward the South.

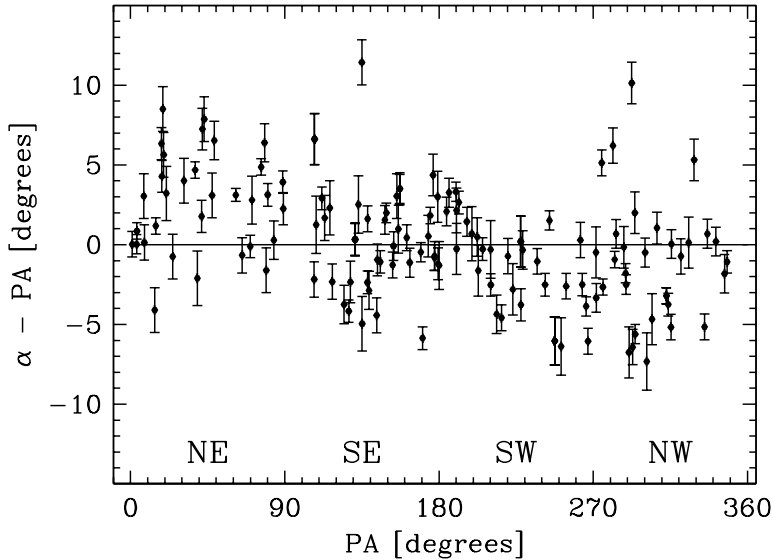

Figure 4.14: Deviations of the proper motion velocity vector $\alpha$ from the position angle PA of knots. Figure is a reprint from Paper I.

## 4.5 Brightness evolution

The GK Per optical nebula suffered clear morphological and brightness changes in the last 50 years. This is illustrated in Figure 4.15, where a $\times 2$ zoom of the 1953 image is compared to a smoothed version of our 2011 image. The main changes are a relative brightening of the whole Eastern side of the nebula compared to the Western side, and in particular the development of a more prominent "bar" in the NE edge (see also Anupama & Kantharia 2005), and a progressive circularization of the SW edge of the nebula, i.e. of the region of the strong interaction observed at radio and X-ray wavelengths. In general, the outflow looks more circular and uniform than in the past.

As most of our images were taken with the same instrument, detector, and filter, the brightness variation of the nebula in the last decade can be quantified. The 15 INT+WFC flux-matched frames marked with an asterisk in Table 2.2 were used. To measure the brightness of the nebula, all stars inside or nearby the remnant were masked. We first measured the brightness variation of the nebula as a whole. In Figure 4.16 we present the evolution of the total flux of the remnant normalised to the value that it had in 1999. It shows that the total $H\alpha$+[NII] flux of the nebula has been linearly decreasing, for a total of 31% over the 12.04 years considered, i.e. at a rate of 2.6% per year, which is similar to what was measured in the radio by Anupama & Kantharia (2005) and attributed to adiabatic expansion of the remnant into the surrounding medium.

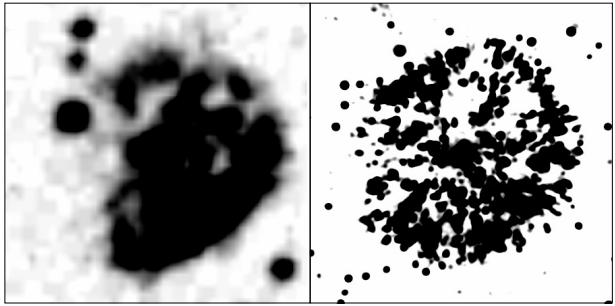

Figure 4.15: Comparison of the 1953 photographic image of GK Per with a smoothed version of our 2011 CCD image. The former is displayed with a ×2 zoom with respect to the latter for easier comparison. North up, East left. Figure is a reprint from Paper I.

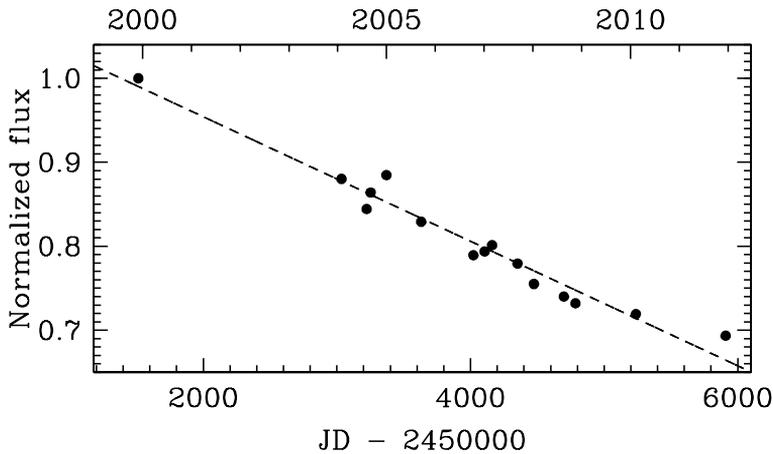

Figure 4.16: Total $H\alpha$+[NII] flux of the remnant from 1999 to 2011. Figure is a reprint from Paper I.

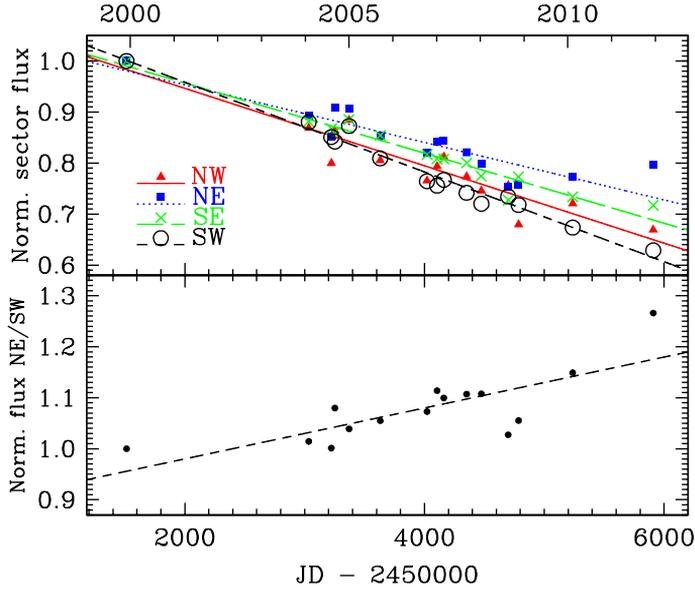

Figure 4.17: Top panel: flux variations for different quadrants. Lower panel: flux ratio variations of NE and SW quadrants.

Second, the flux variation for each of the four quadrants (NE, NW, SE, SW) was computed. As can be seen from Figure 4.17, the NE quadrant shows the shallowest decline, with even some possible re-brightening in the years 2009 to 2011. The quadrant fading at the largest rate is the SW one.

Third, we have also considered the flux variations for a number of individual knots. Their fluxes were measured using circular apertures matched to the knots size, using an external annulus for background subtraction. The 1999 image could not be used for the measurement of individual knots because of its poorer seeing. After excluding knots in crowded areas, we could measure the flux variation of 85 knots. Illustrative examples are presented in Figure 4.18. In the figure, the flux is normalised to the maximum flux of each knot in the period considered. Out of 85 knots, 50% show a similar fading as the nebula as a whole (first two rows in the figure). Some 15% have a constant flux (third row), while some 20% of the knots show instead a monotonic brightening (fourth row in Figure 4.18). Finally, 15% of the knots are rapidly fading or brightening (or both) during the 2004-2011 period (last three rows). The most extreme cases show a sudden brightening of a factor of 5 in one year (bottom-left panel), or a similarly fast brightening and then a dramatic fading on a timescale of less than three years (bottom-right panel). The brightness variations of individual knots is also discussed by Shara et al. (2012).

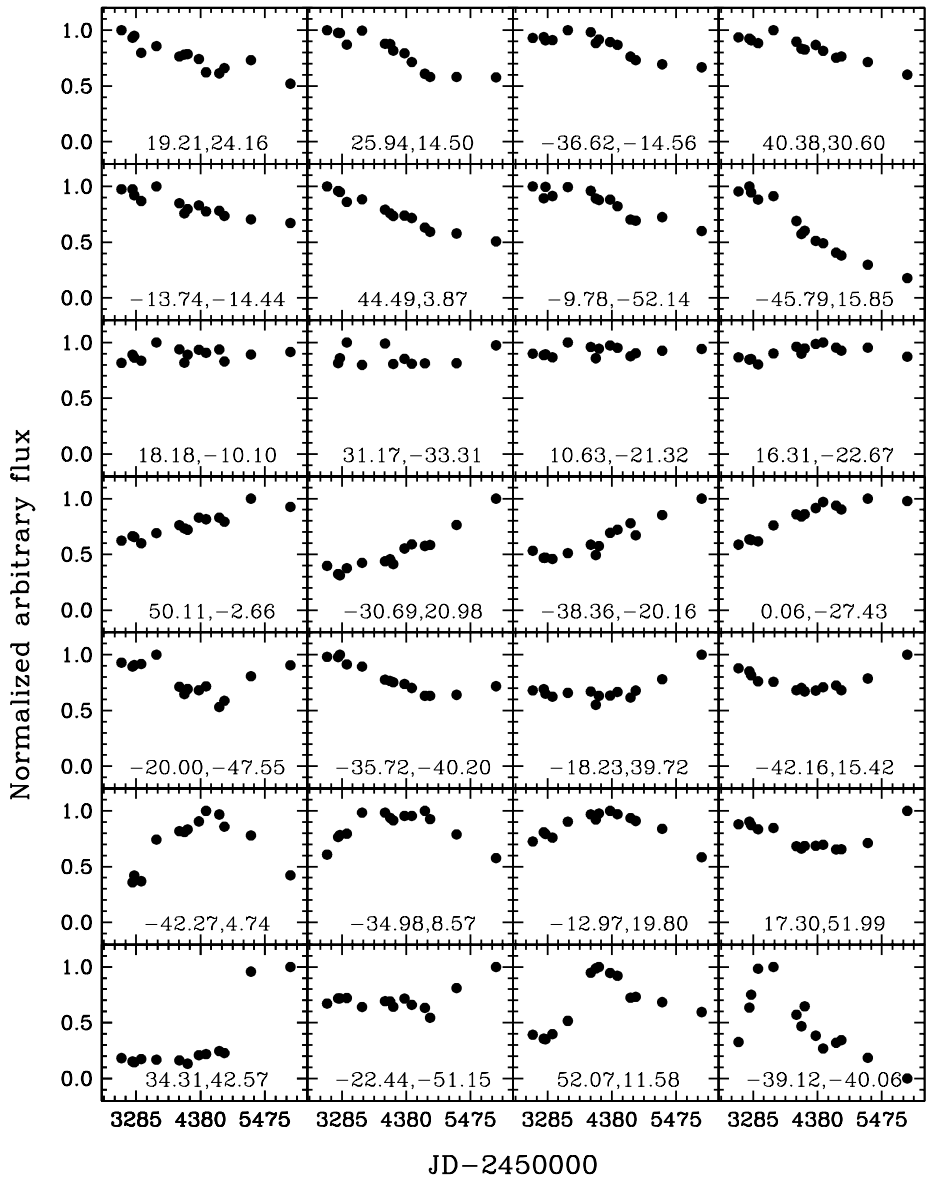

Figure 4.18: Illustrative example of the brightness changes of individual knots. The knots $d_x$ and $d_y$ coordinates as in Table 4.1 are indicated. Figure is a reprint from Paper I.

## 4.6 Discussion

### 4.6.1 Distance

Our kinematical determination of a distance of 400±30 pc for GK Per (Section 4.4) compares well with the value derived using the absolute peak luminosity of novae. We have considered four different relationships between $M_V$ at peak luminosity and the rate of the nova decline (Cohen 1985; Capaccioli et al. 1989; della Valle & Livio 1995; Downes & Duerbeck 2000), depending on $t_2$ (the time in days the nova takes to fade 2 mag from the peak brightness). Assuming the following for GK Per: $t_2 = 7$ d, a peak visual magnitude of $m_v$=0.2 mag, and a reddening of $E(B - V) = 0.3$ (Wu et al. 1989), we have computed the distance to GK Per using the different relationships. The average distance is 440±70 pc, in good agreement with our kinematic estimate.

Our distance also broadly agrees with (and should improve) the estimates from other authors using the expansion parallax of the remnant. The most widely used distance is 470 pc by McLaughlin (1960). Duerbeck (1981) found the distance to be 525 pc. A more recent determination by Slavin et al. (1995) states $455 \pm 30$ pc. Using the parallax of the central star from GAIA DR2 (Gaia Collaboration et al. 2016, 2018) results in a distance of 442±8 pc.

Finally, a rather short distance to GK Per, 337 pc, was determined in Warner (1987) using the Bailey's method (Bailey 1981). This method considers the $K$-band magnitude of the secondary star, which for cataclysmic variables is not easily separated from the disc contribution: therefore the calculated distance should be considered as a lower limit. The largest distance determination is the one by Sherrington & Jameson (1983), 726 pc, using again the $K$ mag and their estimated spectral type for the secondary star (K2 IV).

### 4.6.2 Dynamics of the ejecta

Our kinematical study, combining velocity measurements both in the plane of the sky and along the line of sights, has highlighted several basic properties of the dynamical evolution of the GK Per nova remnant. The outflow consists of a thick shell composed of about a thousand knots and filaments. Knot velocities range from 250 to 1100 km s$^{-1}$, with the majority of them having a velocity between 600 and 1000 km s$^{-1}$. Our analysis support a symmetrical thick shell, however other geometries, such as a cylindrical sphere with polarcones (Harvey et al. 2016) cannot be ruled out. The kinematical ages of knots from the proper motion measurements indicate that most of them have been only modestly slowed down in the 102.9 years from the ejection to the present, with no significant deceleration dur-ing the last 10-20 years. Clear evidence for systematic deviations of the expansion rate with direction is also not found.

On the other hand, some long-term evolution of the overall characteristics of the nebula is highlighted. The most notable changes are a progressive circularization of the SW nebular edge, and a relative brightening of the whole Eastern side compared to the Western part, with the development of a "bar" in the NE edge.

In few words, the nebula seems to expand nearly ballistically, and is progressively becoming rounder and more uniform in terms of surface brightness. These results are somewhat surprising considering the previous interpretation of the evolution of the GK Per remnant. In fact, the asymmetry observed at radio and X-ray wavelengths (Seaquist et al. 1989; Balman 2005) seems to indicate that along the SW direction strong interaction of the remnant with relatively dense circumstellar material has occurred. This is however not reflected in enhanced deceleration of the optical knots along this direction.

The interpretation of the nebular dynamics should take into account these new results. A basic role in the discussion is played by the original circumbinary density distribution into which the nova ejecta expanded. Its most visible expression is the large bipolar nebula first detected from IRAS data by Bode et al. (1987). The strongest argument for the association between this bipolar structure and the nova is geometrical, that is, the emission is symmetrically placed around GK Per. The inherent problems of the classification of the bipolar nebulosity as an old planetary nebula from the primary star of the system were presented by Tweedy (1995). An alternative solution has been proposed by Dougherty et al. (1996): the secondary star could be responsible for the formation of the extended nebula as it evolved off the main sequence triggering a major transfer of material into the primary star causing a born-again asymptotic giant branch event. As revealed by deep optical imaging the extended emission presents a typical bipolar morphology with the polar axis oriented in the NW-SE direction. The nebula shows an asymmetry between the NE and SW edges that Bode et al. (2004) interpreted as the motion of the nebula to the SW assuming high relative velocity with respect to the ISM. Within this picture the non-thermal radio (Anupama & Prabhu 1993) and the X-rays (Balman & Ögelman 1999) emission in the SW direction is interpreted by Bode et al. (2004) to be a consequence of the nova ejecta ploughing the denser medium of the SW side of the bipolar nebula.

We can test this idea in the framework of the existing hydrodynamical models of nebular interaction with the ISM. The central binary is moving with a space velocity of $45 \pm 4$ km s$^{-1}$ (Bode et al. 2004), or $4.6 \times 10^{-5}$ pc yr$^{-1}$. For simplification, we consider the central source as a single star (as successfully done for Mira, see Martin et al. 2007). Moreover, given that the direction of the move-ment is almost perpendicular to the axis of the bipolar lobes we can also assume spherical symmetry (the formation of a bipolar structure is complex and beyond the scope of the present work). With this simplifications at hand we can consider that the interaction process can be similar, at least regarding the discussion that follows, to that of the wind of an AGB star with the ISM (see e.g. Villaver et al. 2003, 2012). In particular, a simulation with a relative speed of $50$ km s$^{-1}$ can be found in Villaver et al. (2012). These simulations show that once a bow-shock structure is formed, the subsequent stellar wind expands unperturbed inside the asymmetric outer cavity created by the interaction. In these conditions, the stellar material ejected inside this cavity can never feel the outer interaction if it is inside the bow-shock structure. The same idea could be applied to GK Per: if both the outer bipolar nebula and the nova ejecta are the result of the evolution of the central binary, and the nova shell is expanding inside the bipolar nebula, then the ejecta cannot see the interaction with the ISM, and should expand unperturbed. This interpretation is consistent with the lack of evidence for deviations of the expansion rate with direction found in this work. In particular, the SW side, where most of the interaction with the circumstellar medium was assumed to have occurred in the past (see radio and X-ray data), does not show any sign of larger deceleration than the rest of the nebula. The asymmetry observed in the radio and X-rays, however, would remain unexplained in this scenario.

An alternative possibility is the existence of a mechanism that allows interaction with the ISM to permeate from the outer structure to the nova shell. An interaction with magnetised ISM could provide this mechanism (Soker & Dgani 1997). In fact, the nature of the radio and X-ray emission could be explained as dense gas clouds moving supersonically through a magnetised low-density medium. Jones et al. (1994) show how the synchrotron emission from relativistic electrons increases rapidly when the cloud begins to fragment under a Rayleigh-Taylor instability. They also show how the interaction of this wind with dense condensations can lead to strong magnetic fields. If relativistic particles are formed as the fast wind is shocked, then the enhanced magnetic field will result in non-thermal radio emission.

At this point we can turn our attention back to the outburst mechanism that leads to the formation of the nova ejecta. Its thick shell geometry can be the result of an ejection with a range of initial velocities, or the dynamical evolution of the interaction of different components. We favour the latter hypothesis, which is supported by the observations of nova shells soon after outburst. It has been argued in the literature that two distinct outflow components of different origin are present at early times: an earlier, slow one from the circumbinary gas, and a later, fast one result of the outburst (see e.g. Cassatella et al. 2004; Williams & Mason 2010). In the case of Nova Cyg 1992, Cassatella et al. (2004) show that the main shell ejected at the moment of the outburst contains the bulk of the ejected matter and is located in the outermost region of the ejecta region, while a less massive impulsive event forms a faster expanding shell that eventually will catch up with the main shell generating a region of shock interaction where the two shells meet. Eventually, the process gives rise to the formation of a new shell formed by the merger of the two. A similar scenario in terms of discrete shell ejection events is proposed by Williams et al. (2008) and Williams & Mason (2010) to explain the absorption line spectra of the early ejecta of 15 nova shells.

In GK Per we are probably witnessing the late evolution of such of a process, and in this respect it is natural that any asymmetry originally present in the dense ejecta (see e.g. the 1917 images in Bode et al. 2004) disappears with time as the combined shell expands outward. This explains the progressive circularization and homogenization of the optical outflow. The original asymmetry can also be understood in terms of the binary interactions in the system (e.g. Sytov et al. 2009).

# CHAPTER 5

# EVER SURPRISING R AQUARII: FROM THE INTRIGUING INNER JET TO A DISCOVERY OF NEW EXTENDED NEBULOSITIES

R Aquarii (R Aqr) is a symbiotic binary system, consisting of a M7III Mira variable and a white dwarf, surrounded by complex nebular structures extending across several arcminutes. At large scales, R Aqr appears as a bipolar, hourglass-like nebula with a prominent toroidal structure at its waist, within which a curved jet-like structure is found. At a distance of about 200 pc R Aqr is the closest known symbiotic binary, and therefore provides a unique opportunity to study in detail the evolution of stellar outflows.

The hourglass nebula of R Aqr was first discovered by Lampland (1922) (see also Figure 5.1) and repeated observations have revealed that it is expanding - at a first approximation - in a ballistic way (Solf & Ulrich 1985). The expansion of the large-scale nebula was used by Baade (1944) to calculate a kinematical age of 600 yr. Solf & Ulrich (1985) refine this value to 640 yrs by applying a kinematical model using a hourglass geometry, with an equatorial expansion velocity of 55 $\mathrm{km\,s^{-1}}$, and assuming that the expansion velocity at each point of the nebula is proportional to the distance from the centre.

The presence of the central jet in R Aqr was first remarked upon by Wallerstein & Greenstein (1980), however Hollis et al. (1999a) showed that the jet was present in observations taken as early as 1934 (see Figure 5.1). Since these earlier observations, the large-scale S-shape of the jet has remained unchanged, while at smaller scales its appearance varies greatly even on short timescales (Paresce et al. 1991; Michalitsianos et al. 1988; Hollis et al. 1990, 1999b; Kellogg et al. 2007, this work). A detailed investigation of the innermost $5''$ of the jet has been carried out using high resolution radio data (e.g Kafatos et al. (1989); Mäkinen et al. (2004)). However, it has been demonstrated that at different wavelengths the appearance and the radial velocity pattern of the jet varies dramatically (Paresce et al. 1991; Hollis & Michalitsianos 1993; Sopka et al. 1982; Solf & Ulrich 1985; Hollis et al. 1990, 1999b).

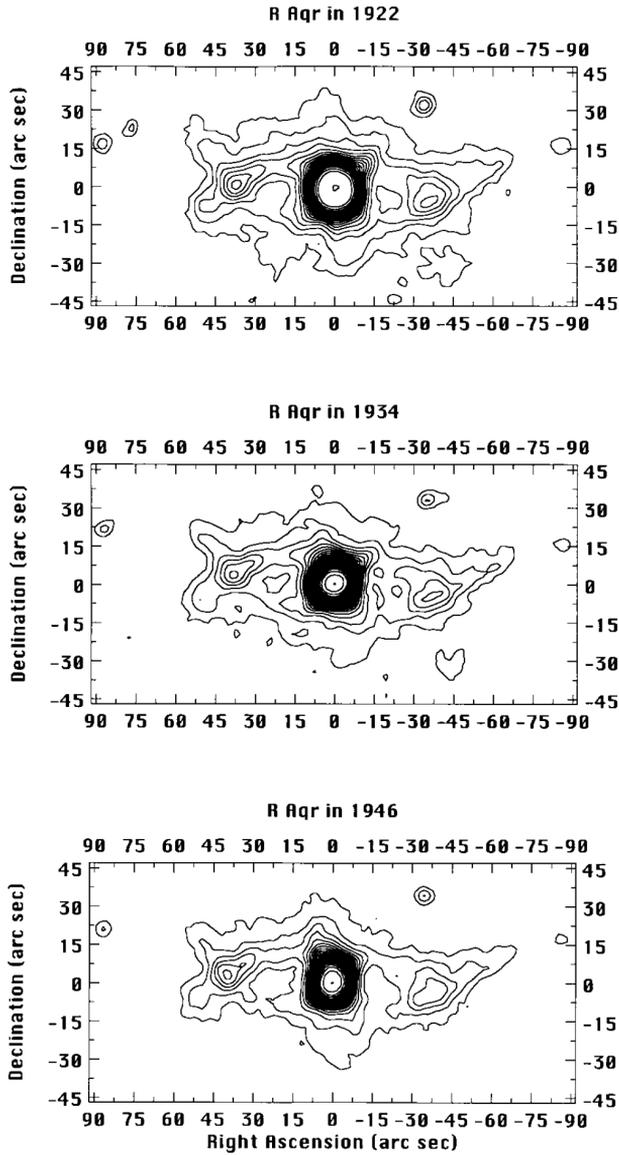

Figure 5.1: Contour plots from Lowell Observatory photogaphic plates of R Aqr nebulosities. The slight elongation in the NE-SW direction around the brigth central area is visible on the 1934 frame, an indication of a forming jet. This elongation is even more apparent on the image from 1946. North up, East to the left. Figure is depicted from Hollis et al. (1999a)

In this Chapter, we present a detailed study of the R Aqr nebula based on deep, narrow-band emission line imaging, acquired over a period of more than two decades, as well as on high spectral resolution integral field spectroscopy of the [O III] 5007 Å emission from the central regions.

## 5.1 Overall properties of the R Aquarii outflows

The R Aquarii (R Aqr) outflow consists of a bright, curved jet structure contained within a larger bipolar nebula (Hollis et al. 1990). The inner regions of the jet extend along the Northeast-Southwest direction, at an apparent angle of 40 degrees from the symmetry axis of the bipolar lobes which instead are very close to the North-South orientation.

Solf & Ulrich (1985) also identified an inner bipolar nebula, oriented along the same symmetry axis as the larger lobes but on a smaller scale. However, this latter structure is not obviously visible in any of our images. Figure 5.2 shows the large scale morphological properties of these outflows and the faint new details revealed by our deep imaging.

### 5.1.1 Structure and expansion of the bipolar nebula

The large system of lobes and their bright equatorial waist are mainly visible in the lower-ionisation light of Hα+[N II], [O II], and [O I] (Figure 5.2). Indeed, Gonçalves et al. (2003) show large [N II]/Hα and [S II]/Hα flux ratios in the ring of the bipolar nebula that they ascribed to shock ionisation.

The appearance of the ring and lobes, at the high resolution provided by our images, is complex. In the light of low-ionisation ions, the ring is broken into knotty and filamentary structures, while it looks much smoother in the light of higher ionisation species such as [O III]. Furthermore, our deep images show for the first time fainter features that we designate as streams in Figure 5.2. Streams appear in the NE and SW direction of the bipolar lobes extending up to $1'.2$ from the central star. On the Southern side, the streams seem to replicate the curved appearance of the outer regions of the jet.

The overall structure of the bipolar nebula does not show notable changes over the 21 years considered (1991 - 2012). We therefore use the magnification method (see Section 3.3) to calculate the expansion in the plane of the sky and hence the age of this structure.

The magnification factor $M$ was found using the 1991 and 2012 Hα+[N II] images. Increasing magnification factors were applied to the 1991 frame, and the differences with the 2012 image computed. A precise determination of the best-fitting magnification factor is limited by the highly inhomogeneous morphology of the nebula. By dividing the nebula into quadrants, we found that the best-fitting $M$ values were as follows: NE 1.030, SE 1.033, SW 1.033. Due to the faintness of the NW region of the nebula no measurement could be obtained from that quadrant, and for this reason we also defined a western (W) region restricted to the prominent part of the waist in that direction where a value of 1.033 to 1.036 was derived. According to the Equation (3.5) and taking into account the time lapse between the two epochs, 21.17 years, the $M$ values above imply an average age, weighted by errors, of the bipolar nebula of $T_{\mathrm{bip}} = 653 \pm 35$ years in 1991. This compares well with previously published results. Similarly to us, using the expansion of the nebula, Baade (1944) found a kinematical age of 600 yrs. This result was further refined by Solf & Ulrich (1985) to 640 yrs by applying a kinematical model using a hourglass geometry, with an equatorial expansion velocity of 55 km s$^{-1}$. Yang et al. (2005) found that R Aqr could have experienced a nova explosion in A.D. 1073 and A.D. 1074. Without knowing the initial conditions of the explosion and of the circumstellar medium (ISM) it is not possible to find further support to the relation between the bipolar nebula and the possible ancient nova outburst, nor to discard it.

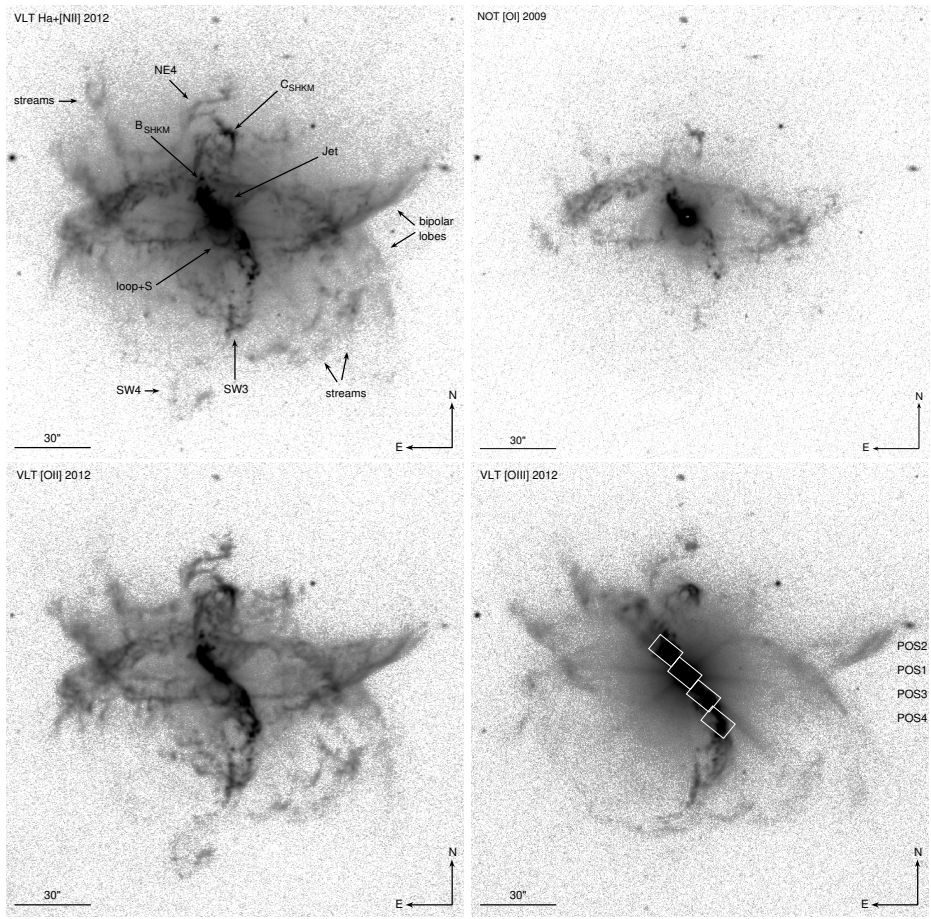

Figure 5.2: VLT 2012 Hα+[N II], [O II], [O III] and NOT 2009 [O I] frames. On the [O III] frame the white boxes represent the spectroscopic observations (see Section 2.2). The FOV of all frames is $3' \times 3'$. North up, East left. Figure is a reprint from Paper II.

The estimated ages of the SW streams indicate that they are likely part of the extended nebula, rather than jet features. The stream in the NE is not detected in our 1991 H$\alpha$+[N II] image therefore no age estimate can be given.

### 5.1.2 New faint outer features

An additional set of images acquired in Terroux Observatory reveal new faint outer features in [O III] and in H$\alpha$. A combined [O III] and H$\alpha$ image is presented in Figure 5.3. It reveals the existence of a thick [O III] arc along the East-West direction with an extent of $6'.4$, and a thinner and fainter H$\alpha$ loop, which extends to the North up to $2'.8$ from the central source. The latter may be related to the streams described in the previous section.

The [O III] arc is confirmed by stacking up all our other [O III] and [O II] long exposure frames, but at a low signal-to-noise ratio. Unfortunately, the H$\alpha$ feature is undetected in our VLT images, as their FOV does not fully cover the region. It is likely that these features are related to the mass loss from the red giant and/or a nova eruption from the white dwarf in the earlier evolutionary stages of the system.

### 5.1.3 Kinematic distance

The combination of our determination of the apparent expansion of the bipolar nebula with the radial velocity measurements of Solf & Ulrich (1985) allows us to derive the expansion parallax of R Aqr nebula. In particular, Solf & Ulrich (1985) found that the equatorial waist can be modelled as an inclined ring expanding at a speed of $v_{\rm exp} = 55$ km s$^{-1}$. As described in the Section 3.4 a distance to the expanding ring can be calculated according to the Equation (3.14). Adopting the average value of the magnification factors determined above, and fitting an ellipse to the nebular ring to measure its major axis, $d$, we obtain a kinematic distance to R Aqr of $178\pm18$ pc.

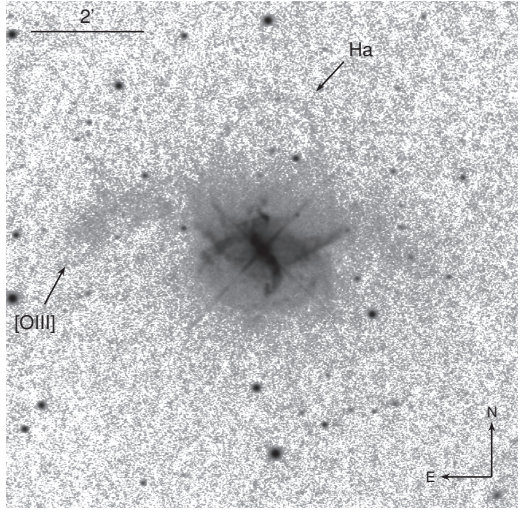

Figure 5.3: Terroux image showing faint outer [O III] and Hα features. FOV=$9' \times 9'$. See text for more details. Figure is a reprint from Paper II.

In general, this kinematic distance is in fair agreement with previous estimates. The nebular kinematics were first used by Baade (1944) to derive a distance of 260pc, later revised down to 180–185 pc (Solf & Ulrich 1985). This value is also in good agreement with the estimate of 181 pc by Lepine et al. (1978) based on the absolute magnitude of R Aqr at 4$\mu$m and an assumed value of $-8.1^m$. Hipparcos parallax measurements by Perryman et al. (1997) result in a slightly larger distance of 197 pc, in strong agreement with the estimate based on the separation of the orbital components measured by the VLA (Hollis et al. 1997b). Parallax measurements of SiO maser spots using VERA have indicated a yet greater distance of 214–218 pc (Kamohara et al. 2010; Min et al. 2014). More recently, the GAIA DR2 (Gaia Collaboration et al. 2016, 2018) parallax would refere to an even larger distance, $320 \pm 29$ pc. However, considering the variability and relatively large brightness of the R Aquarii central star, the latter cannot be considered too reliable.

## 5.2   Structure and expansion of the jet

Unlike the large-scale bipolar nebula, which within our present errors in the determination of the apparent and radial motions is well modelled assuming a mainly ballistic expansion, the jet shows a much more complex and irregular evolution. In our images, features identified by previous authors have brightened or faded, or even broken into multiple components moving along different directions. The bulk motions of various regions of the jet determined using our images are highlighted in Figure 5.4. They demonstrate that, while the overall flow pattern is consistent with a radial expansion of the jet, in some regions there are significant deviations, even perpendicular the radial direction of expansion. Their complex changes of appearance are probably the combination of illumination/ionisation variations as well as shocks. This makes the magnification method inappropriate to describe them. To illustrate this, we will discuss the evolution of individual features across our multi-epoch imagery in the following subsections.

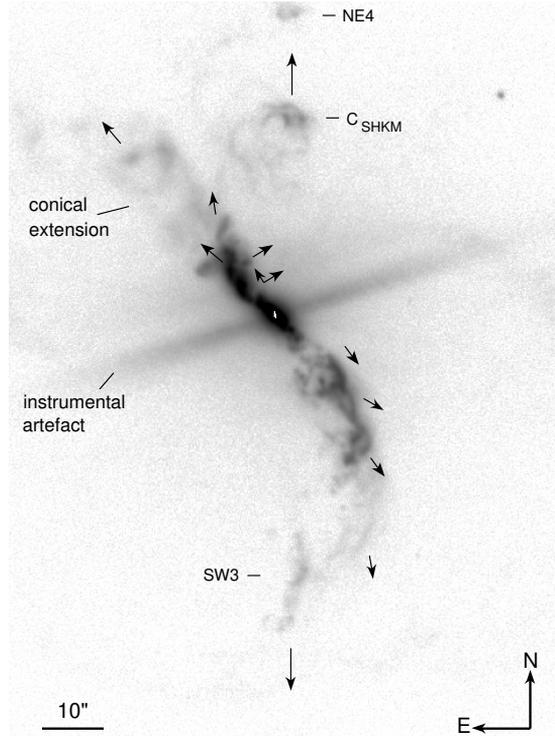

Figure 5.4: NOT 2007 [O III] image, with logarithmic display cuts that highlight the overall structure of the jet. Arrows point the expansion direction. FOV is $1'.5 \times 2'.0$. Figure is a reprint from Paper II.

Figures 5.5 and 5.6 show the evolution over time of the northern and southern region of the jet in the most relevant epochs and filters. Note that the bright central area is often saturated, with strong charge overflow especially in Hα+[N II]. The intensity level of the different frames in both figures was adapted to maximise the visible information. The small blob near the central source in the [O II] 2007 and 2009 frame, pointed to with arrows in Figure 5.5, are a red-leak images of the central star (displaced because of atmospheric differential refraction at the significant airmass of these observations). In Figure 5.6 on the 2009 [O III] frame, instrumental artefacts are visible as faint lighter thick lines emanating from the central star at PAs $135°$ and $225°$.

In order to avoid confusion, some clarification of the nomenclature of the jet features used in the past is in order. We refer to the earliest designated jet features from Sopka et al. (1982, hereafter referred to as SHKM) as $B_{SHKM}$, $C_{SHKM}$, and the "loop". Later publications revealed more details of the jet closer to the central source, which were designated as knots A and B by Kafatos et al. (1983), and knot D and S by Paresce et al. (1988). Knot S is a brightness concentration inside the "loop" feature and is therefore often referred to as loop+S. We keep the designations of the later authors. The aforementioned features, as well as the new ones found in this work, are marked in Figures 5.2, 5.4, 5.5, and 5.6.

Most of our images, as well as previously published data (imaging and spectra), show that the NE jet is brighter than the SW jet at both large scale and in the central area (Paresce & Hack 1994). However, our short exposure [O III] and [O II] VLT images from 2012 reveal that in the central region ($< 2''$), the SW jet is brighter. This is also confirmed by our spectral data from 2012 (Figure 5.8). This seems to have appeared a few years before 2012, as the images from 2002 and 2007 show equal brightnesses for NE and SW in the central area, while by 2009 and 2012 clearly the SW is more prominent. This central area is resolved in the recent high spatial resolution SPHERE images from 2014 by Schmid et al. (2017). They also detect that the SW jet is brighter than its NE counterjet in these central areas.

### 5.2.1 The NE Jet

The evolution of the northern, bright part of the jet is shown in Figure 5.5. It has a complex, knotty, and variable appearance. For instance, the feature that we name as F first appeared in 2007 in the [O III] filter. The 2012 image indicates that, contrary to the general radial expansion, it is seemingly moving towards the West. The same applies to feature G, which appeared in 2011. Their westward lateral movements are indicated by the two arrows in Figure 5.4. Transformed into linear velocities, their motions would imply speeds of between 500 and 900 $\mathrm{km\,s^{-1}}$. This is several times larger than the bulk radial motions from imaging and spectroscopy, indicating that very likely they do not reflect the true physical movement of a clump of material but rather are due to changes in the ionisation conditions of the region.

At a distance of $\sim 20''$ (see Figure 5.4) the northern jet splits into two compo-nents, a brighter one bending toward the North and ending in filamentary features such as C_SHKM and NE4, and a fainter, more diffuse one that seems a straight, conical extension of the innermost jet. The latter is only visible in the [O III] emission line and its cone-like morphology suggest that it may be an illumination effect (Corradi et al. 2011), further indicating that the role of changing illumination/excitation is critical in understanding the structure of the jet.

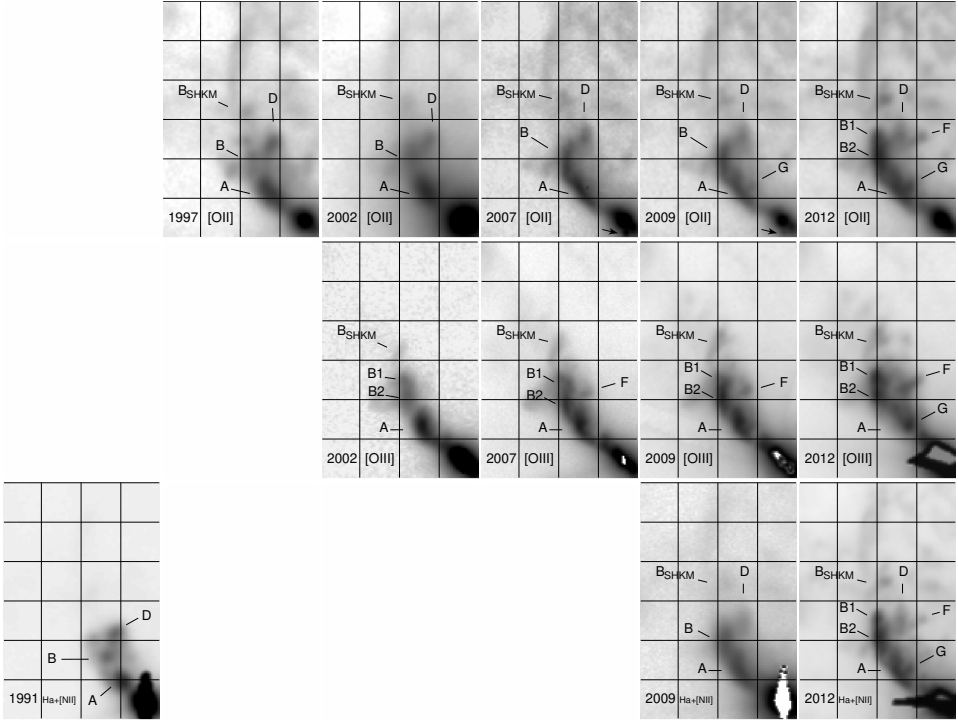

Figure 5.5: NE jet in [O II], [O III], and Hα+[N II] frames from top to bottom respectively. One square is $5'' \times 5''$. The FOV of each frame is $20'' \times 30''$. North is up, East to the left. Figure is a reprint from Paper II.

Burgarella & Paresce (1991) used optical emission line ratios to conclude that the R Aqr jet features are best fitted with local shock-wave models, as was firstly suggested by Solf & Ulrich (1985). Burgarella & Paresce (1991) found that the brightening of knot D (and fading of knot B) came some 15 yrs after the brightening of knot B (in the late 1970s), concluding that the shock wave was propagating outwards at 90–100 km s$^{-1}$. As such, they predicted that the knot D should fade analogously 15 yrs after it was observed to brighten, which would approximately occur in 2003. Inspection of our Hα+[N II] data, Figure 5.5, reveals that in 1991

B and D have similar brightness (just as in observation from 1986 in Burgarella & Paresce (1991)). Our next H$\alpha$+[N II] frame is from 2009 were knot D, indeed, is much fainter than knot B, while at the same time the knot A is the brightest. By 2012, the relative brightness of knot B is around that of knot A, while knot D is the faintest. A similar tendency is detected in our [O II] data where by 1997-2002 knot B has almost disappeared, while knot A and D are clearly brighter. By 2007 knot B starts to brighten again and quite soon (2009) becomes as bright as knot A. At the same time it does appear that knot D is fading. The similar relative brightening of knot B is visible in the [O III] frames. To conclude, the prediction of knot D fading is confirmed (though perhaps at a faster rate than predicted). In addition, the re-brightening of knot B may imply that another shock-wave is passing through the system. If so, it should eventually start again influencing knot D. However, from Figure 5.5 it is evident that, by our latest epochs, knot B has stretched as far out as knot D, but the latter is not brightening together with knot B.

Over the years, features in the central area of the NE jet (Figure 5.5) tend to get elongated along the expansion direction and eventually break into separate components. For instance, feature B breaks into B1 and B2. Similar stretching is happening with features A and D, which was seen to happen at 8 GHz radio band as early as 1992 (Mäkinen et al. 2004). Our data shows that by 2012 in all filters feature A has a very elongated structure, which is clearly indicative of imminent breakup. The delay between optical and radio can be just due to the lower resolution available in optical wavebands.

As indicated with the arrows in Figure 5.4, the outermost NE jet features expand radially during the period considered. Therefore we use the magnification method (features $C_{SHKM}$ and NE4, see Figure 5.2) or measure directly the proper motion (feature $B_{SHKM}$, Figures 5.2 and 5.5) to estimate their kinematical ages. The ages together with approximate distances from the central star are presented in the Table 5.1. Ages are presented for the epochs 1991-07-06 and 2012-09-05. The distances are measured at the 2012-09-05 date because not all features were present on the 1991 image. Due to the extended nature of most of these features, distances from the central source are roughly estimated. Details related to the measuring can be found in Appendix, see Section 8.1.1. Here only the main results are presented.

The measurements of the brightness peak of feature $B_{SHKM}$ indicate an age between 125 and 180 years. The average proper motion, $\mu = 0''.10 \pm 0''.02 \,\mathrm{yr}^{-1}$, is compatible with the calculations in SHKM ($0''.082 \pm 0''.014 \,\mathrm{yr}^{-1}$), indicating a roughly constant expansion velocity over the last 50 years or so. Feature $C_{SHKM}$ was twice as bright in 2012 than in 1991. In SHKM, a slow brightness change for that feature is mentioned, but it is not clear if it was observed to be brightening or dimming. The age found for that feature, 286 yrs, is much younger than the bipolar nebula, though much older than the previously derived jet age of about 100 yrs (Lehto & Johnson 1992; Hollis & Michalitsianos 1993). If the $C_{SHKM}$ feature is part of the jet, it is not surprising that the brightness has changed, as brightness variations have been seen among all features of the jet. The feature also demonstrates the jet's structural differences at different wavelengths, given that $C_{SHKM}$ keeps its structure in time when considering single filter data, but its form varies in different filters. In H$\alpha$+[N II] it has an arched shape, while in [O III] it seems a double arched or circular. In [O I] it has a T-shape structure. For feature NE4, an age of $285\pm61$ years is found. This age is much younger than the bipolar nebula, though older than the jet, just as found for $C_{SHKM}$. Collectively our observations of $B_{SHKM}$, $C_{SHKM}$, and NE4 imply that they are real physical structures, and that their apparent morphological changes are not dominated by the changing ionisation.

Table 5.1: Kinematic ages at epochs 1991-07-06 and 2012-09-05 together with an approximate distances for the epoch 2012-09-05 from the central source for the ballistic features of the jet.

| | Feature | Age 1991 yrs | Age 2012 yrs | Distance $''$ | Method | Comments |
|---|---|---|---|---|---|---|
| North | $B_{SHKM}$ | 125-180 | $145 - 200$ | 17 | direct | Stable expansion velocity over the last 50 yrs. $\mu = 0''.10 \pm 0''.02\,\mathrm{yr}^{-1}$ |
| | $C_{SHKM}$ | $286 \pm 12$ | $307 \pm 12$ | 35 | magnif. | Brightness variations. Structural changes in different wavelengths. |
| | NE4 | $285 \pm 61$ | $306 \pm 61$ | 45 | magnif. | |
| South | loop+S | $160 \pm 40$ | $183 \pm 40$ | 10 | magnif. | "loop" expanding steadily at least since 1960s. Significant brightness change of knot S. |
| | SW3 | $215 \pm 36$ | $236 \pm 36$ | 45 | magnif. | |
| | SW4 | $880 \pm 150$ | $900 \pm 150$ | 75 | magnif. | |

### 5.2.2 The SW Jet

The evolution of the SW jet from 1991 to 2012 in the most relevant filters is presented in Figure 5.6. We refer to the SW jet with the following nomenclature. SW1 consists of several blobs (apart from the loop+S seen in the 1991 H$\alpha$+[N II] frame in Figure 5.6. The rest of the SW jet visible in Figure 5.6 is named SW2. The more extended features SW3 and SW4 are also highlighted in Figure 5.2. The expansion of the SW jet is more uniform than the northern one, showing mostly radial motions (features loop+S, SW4, and SW3 in Figure 5.2 and 5.6). However, based on our observations, a few interesting remarks can be made.

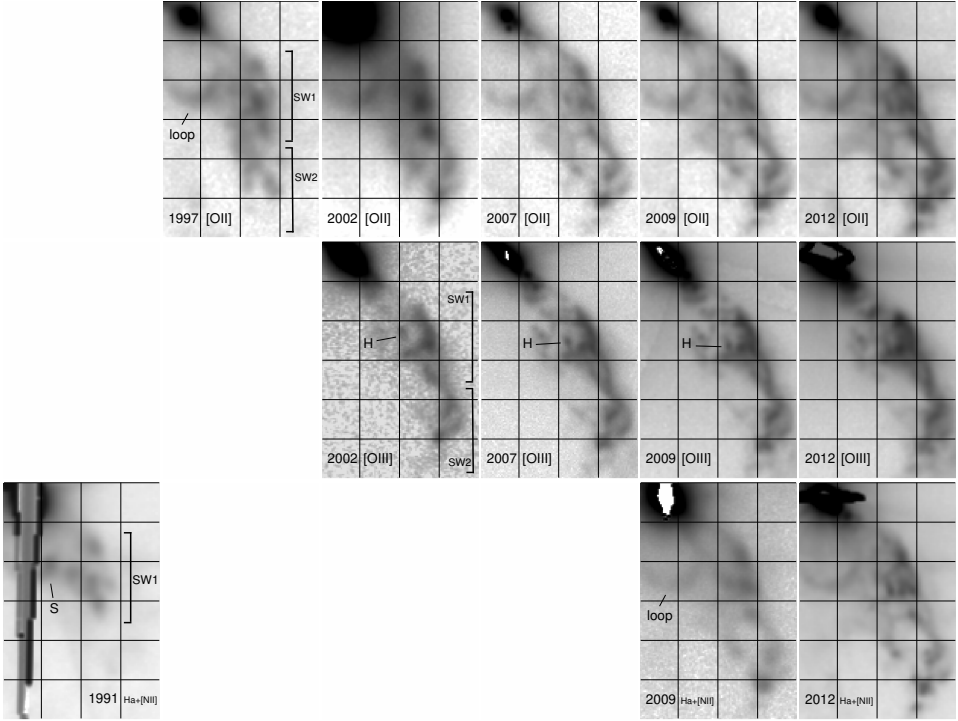

Figure 5.6: SW jet in [O II], [O III], and Hα+[N II] frames. One square is $5'' \times 5''$. The FOV of each frame is $20'' \times 30''$. North up, East left. Figure is a reprint from Paper II.

First of all, the relative brightness of SW2, compared to SW1, has increased over the years. The complex was marginally visible in our 1991 Hα+[N II] frame, and overall is now as bright as SW1. Significant structural changes are also detectable in the SW2 (compare 1997 and 2002 [O II] in Figure 5.6), and more structure has become visible which seems to connect SW2 with SW3 via faint curved filaments (see Figure 5.2).

Feature H, shown in the light of [O III] in Figure 5.6, is found to be moving faster than the surrounding jet at the epochs 2002, 2007, 2009, and then disappears or dissolves into the surrounding jet emission. Its proper motion is on average $0''.30 \pm 0''.01$ yr$^{-1}$. Taking into account the distance to R Aqr found in Section 5.1.3 the average linear velocity would be $\sim 250$ km s$^{-1}$. This is not as fast as the lateral movements detected in the NE jet but it is still faster than the jet in general, again possibly resulting from a shock-wave moving through the system rather than a real matter movement. If we assume constant velocity, it was ejected 38±1 years ago.

In the SW jet, kinematical ages could be calculated using the magnification method for the radially expanding features loop+S, SW3, and SW4 (see Figures 5.2 and 5.6). As for the NE jet, details of the analyses are presented in the Appendix, Section 8.1.2, and in Table 5.1. The feature "loop" preserves its horseshoe shape over our observing period in all filters in which it is detected. The brightness enhancement, knot S, was detected in 1986 by Paresce et al. (1988) is still clearly visible in our 1991 H$\alpha$+[N II] frame (see Figure 5.6). At later epochs, the knot S becomes elongated along the loop until it almost disappears, but that part of the loop stays brighter than the other regions. Figure 5.6 also shows that the loop is slowly expanding towards the south. The age found for the loop+S, 160 yrs, corresponds to an ejection event around the year $1827 \pm 40$. This is comparable with the ejection date $1792 \pm 32$ computed by SHKM. This also shows that no major changes have occurred in the evolution of the loop since their observations in the 1960s. The estimated age of feature SW3, $215 \pm 36$ years is much younger than the bipolar nebula, though older than the jet, just as found for NE4 and $C_{SHKM}$. In the case of the newly-identified hook-shaped feature SW4 we find a rather large age ($880 \pm 150$ years) but as the feature is very faint, we conclude that its age is consistent within the uncertainties with that of the extended bipolar nebula. The age of the ballistic features of the jet increases with the distance from the centre, consistent with continuous/repeated ejection events or that the more extended features have been slowed down by circumstellar material.

## 5.3 Radial velocity measurements

The [O III] 5007 Å emission extends over the entire FOV of the ARGUS IFU pointings, except for very few spaxels. This resulted in ~1200 usable individual spectra in total. Radial velocity measurements were obtained from each lenslet by Gaussian fitting of the [O III] line using the *splot* task in IRAF. Every fit was visually checked and multiple Gaussians were used when needed. The measured radial velocities were corrected first to the Local Standard of rest (LSR), and then to R Aqr systemic velocity of -24.9 km s$^{-1}$ (Gromadzki & Mikołajewska 2009).

The [O III] line profiles are generally complex and sometimes display broad wings, as illustrated in Figure 5.7. We initially limit the discussion to the strongest emission peaks, defined as those spectral components whose integrated flux is larger than 75% of any other component from the multi-Gaussian fit at each spaxel. Results are presented in Figure 5.8, where the radial velocities of the strongest peaks are plotted on top of the reconstructed [O III] image.

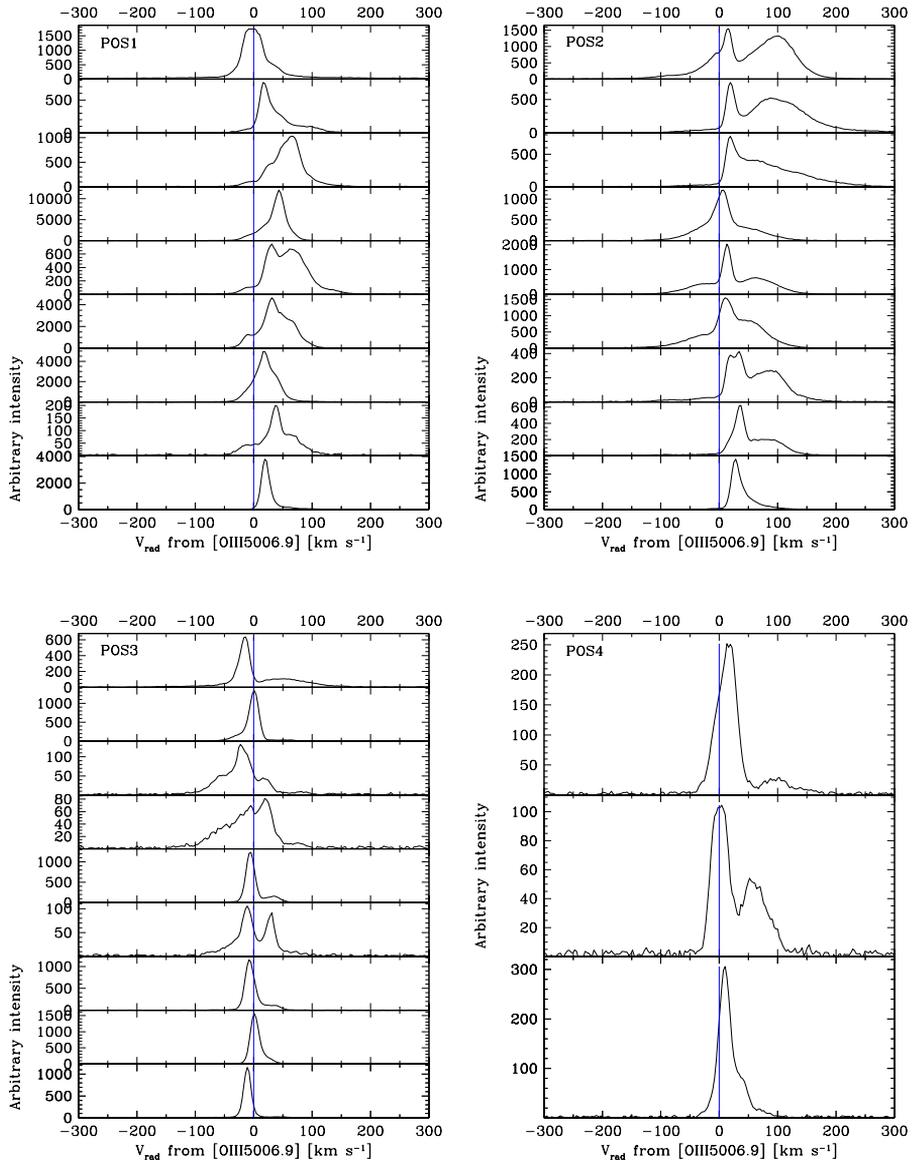

Figure 5.7: Line profile examples from POS1, POS2, POS3, and POS4. Blue vertical line refers to a radial velocity 0 km s⁻¹. All radial velocities are corrected for the systemic velocity. Figure is a reprint from Paper II.

At position 1 (POS1, see Figure 5.7), that is centred on the star, the radial velocity varies from -33 to +72 and the north counterpart mainly red-shifted. This trend continues further away from the centre, at position 2 (POS2, to the North) that is mostly red-shifted (from -5 to +97 km s$^{-1}$) and position 3 (POS3, to the South), which is mostly blue-shifted with a few red-shifted components (from -56 to +59 km s$^{-1}$). At position 4 (POS4), where the southern jet significantly bends, most of the emission becomes red-shifted, ranging from -38 to +137 km s$^{-1}$. This general description is obviously complicated by the very complex line profiles, which include additional emission components as well as extended wings spanning a range of radial velocities as large as $\sim$400 km s$^{-1}$. In general, however, it is clear that on the same side of the central star, both red- and blue-shifted regions are found, which - assuming purely radial motions - is a clear sign of a changing direction of the ejection vector crossing the plane of the sky. This in turn is usually associated to precession of the ejection nozzle as seen at large inclinations.

From Figure 5.8 it is evident that broken up components of feature B, namely B1 and B2 (see Section 5.2.1), have different radial velocities, $V_{\mathrm{rad}_{B1}} = +73 \pm 27\,\mathrm{km\,s^{-1}}$ and $V_{\mathrm{rad}_{B2}} = +22 \pm 22\,\mathrm{km\,s^{-1}}$, respectively. Similarly, we calculate their average FWHM to be $78 \pm 23\,\mathrm{km\,s^{-1}}$ and $32 \pm 24\,\mathrm{km\,s^{-1}}$, respectively. As such, B1 presents almost 3 times the radial velocity of B2 and nearly twice its velocity dispersion. From their motion in the plane of the sky, we see that the entire feature B (including components B1 and B2) has been moving steadily towards the NE (Figure 5.4), implying that their velocities in the plane of the sky should be similar. As such, we can conclude that the true spatial velocity of B2 is significantly lower than that of B1.

Another feature for which we can determine a radial velocity from the data shown in Figure 5.8 is feature F (see Section 5.2.1). It has a uniform radial velocity over the whole elongated feature, $+36 \pm 2\,\mathrm{km\,s^{-1}}$, with a narrow single Gaussian line profiles FWHM $= 16 \pm 1\,\mathrm{km\,s^{-1}}$. This would seem to imply that the feature F is moving almost completely perpendicular to the the line of sight, as its tangential velocity, 500 km s$^{-1}$ (see Section 5.2.1), is much larger than the measured radial velocity.

In an attempt to clarify the overall Doppler-shift kinematics of the jet, and compare with previous observations, we have extracted from the 3 ARGUS data cubes surrounding the central star a position-velocity plot simulating an observation with a long slit, with a width of 1 arcsec, oriented along the general orientation of the jet at PA=40°, and through the central source (Figure 5.9). Hollis et al. (1990) adopted instead PA=29°, which was aligned with the inner jet at that time. Considering the overall PA change of the jet between the Hollis et al. (1990) and us (about 30 years later) it clearly indicates that the jet, on large scales, is rotating counterclockwise (CCW). The CCW evolution of the jet has also been seen in radio observations (Hollis et al. 1997a).

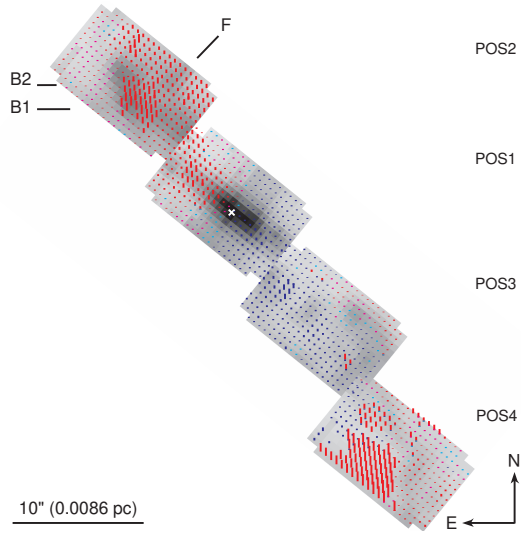

Figure 5.8: Radial velocities of the emission peaks, superimposed onto a greyscale representation of the reconstructed [O III] image from the IFU data. The FOV=$40'' \times 40''$. Radial velocities are indicated by lines, whose colour indicates if emission is blue- or red-shifted, and whose length is proportional to the absolute value of the velocity. Cyan and pink lines represent blue-shifted velocities smaller than -5 km s$^{-1}$ and red-shifted smaller than +5 km s$^{-1}$ respectively. The length of the cyan and pink lines are constant because otherwise they would be too small to be visible. The white X-point marks the position of the central star. The distance used for the linear size is taken from this work, 178 pc. Figure is a reprint from Paper II.

The extracted longslit (Figure 5.9) confirms the overall structure of the jet. However, the figure also highlights the complex variation in velocity profile along the jet. A simple, ballistically-expanding, precessing jet would produce a perfect S-shape, while here we have an overall S-shape but with dramatic variations in the width of the S along the slit (much broader in the NE, perhaps reflective of the multiple components like B1 and B2, and with significant tails out to very high velocities). The SW jet, on the other hand, appears more regular in terms of velocity structure but much more broken spatially with "gaps" in the emission. Furthermore, the artificial long slit spectrum also highlights the acceleration in the inner parts of the jet (POS1). This is consistent with the solution proposed in Mäkinen et al. (2004) that the jet features, after being formed due to increased matter flow at periastron, are accelerated inside the first $1''$ and then ejected as bullets. The outer parts, POS2 and POS3, seem to expand freely. What is unclear throughout the length of the longslit is the contribution of illumination effects - perhaps the matter and velocity structure of the two sides of the jet are symmetrical and the observed differences are the result of differing illumination. The possible illumination beam visible on the [O III] 2007 and 2012 images in NE direction does not seem to move or, at most, moves very little. However, the cone is not seen in the southern direction indicating that this may be an important factor in determining the structure of the observed position-velocity profile.

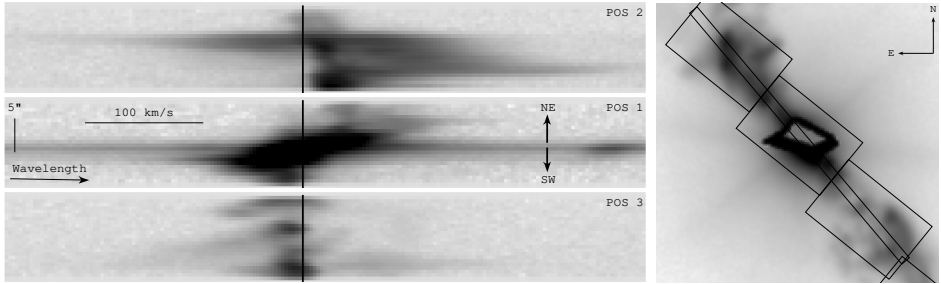

Figure 5.9: *Left.* Extracted long slit spectra with the PA = 40°, extending $18''.7$ from the central star in both direction. The black vertical line is the 0 km s$^{-1}$ radial velocity. In POS1 also the spectrum of the central star is visible. *Right.* The extracted long slit shown over the VLT pointings on top of a matched VLT [O III] frame. The central star is heavily saturated. The FOV is $30'' \times 30''$, North up, East left, slit width $1''$. Figure is a reprint from Paper II.

### 5.3.1 Comparison with previous works

When comparing our RV data with previously published data, at a first sight it seems that there is a notable and intriguing difference. We find that the NE jet is mostly red-shifted and the SW jet mostly blue-shifted, with occasional opposite velocity signs. Only in the farthest part of the SW jet covered by our IFU data, a significant red-shifted component is detected.

Previously published data seem to indicate the contradictory behaviour, that is discussed in the following. A direct comparison with the RV data of SHKM is difficult because they do not mention if their radial velocities are corrected for the systemic velocity. Contrary to our finding, Solf & Ulrich (1985) found that the southern jet is mostly red-shifted and the northern one blue-shifted. Also, they measured RVs for features A and B of around -55 km s$^{-1}$ (based on an [N II] 6583 Å emission line spectrum), which have an opposite sign compared to our measurement for feature B +20 to +70 km/s with respect to the systemic velocity. Only near the central star is there some agreement with our measurements, as Solf & Ulrich (1985) found a negative radial velocity, -20 $km\,s^{-1}$, in the southern jet. Overall, Solf & Ulrich (1985) say that the southern jet is red-shifted and the northern blue-shifted but if one looks more carefully at their Figure 5 it is evident that Solf & Ulrich (1985) have red- and blue-shifted components present at almost all measured positions.

Hollis et al. (1990) present long slit observations of the same emission line as we ([O III] 5007 Å) but also find that overall the NE jet peak emission is approaching and the SW receding, contrary to our measurements. Though, it is clear from their Figure 1 that in the NE up to $7''$ from the centre both blue- and red-shifted components are equally bright. Reconstructing their slit on our earlier data we see that most of their SW faint structure is somewhere at our POS4, which is also red-shifted. Hollis et al. (1999b) employed a Fabry-Perot imaging spectrometer to obtain velocity maps of the [N II] 6583 Å emission line, again finding that the northern jet is blue-shifted and the south red-shifted. They too remark that the FWHM is extremely large (up to several hundred $km\,s^{-1}$) and that the velocity structure at each position is complex, often presenting multiple components, just as we find in our data.

Considering all the above mentioned observations and the nature of our measurements, the apparent discrepancies may be explained by brightness variations of the different line profile components, which would cause that e.g. a previously faint blue component has become significantly brighter than its red component. Furthermore, the higher spectral resolution of our data would allow us to better define the characteristics of the outflow for different components.

It is unlikely that the differences in radial velocity measurements with previous epochs are mainly due to precession effects, given the long ~2000 yr precessing period (Michalitsianos et al. 1988).

We conclude that, at the intermediate scales of the R Aqr jet investigated in this article, most of the observed changes in the morphology and velocity are driven by changes in the illumination and ionising conditions of circumstellar gas at the different epochs of observations, as well as of local dynamical effects, rather than to a structural change of the entire jet of R Aqr.

Only in the innermost regions, significant structural changes are observed, as shown by Schmid et al. (2017). In these regions, some of the discrepancy with the RV data between different authors could indeed be related to the actual motion of the ionised material. The discrepancies that Schmid et al. (2017) find gives further indication of the complexity of the R Aqr system.

## 5.4 Conclusions on R Aquarii

We have presented a multi-epoch morphological and kinematical study of the nebula of R Aqr.

New morphological features (referred to as the arc and loop) outside the known hourglass nebula add richness and complexity to the circumbinary gas distribution and the mass loss history from the system.

The large bipolar nebula of R Aqr is expanding ballistically and does not show major structural changes during the observed period. An average age of it was calculated to be $T_{\mathrm{bip}} = 653 \pm 35$ years in 1991. Determination of its apparent expansion allows us to determine a distance to R Aqr of $178\pm18$ pc, consistent, within errors, with Solf & Ulrich (1985) and Lepine et al. (1978).

The jet is experiencing a more complex evolution. At large scales the jet is mainly expanding radially from the central star. However, apparently closer to the central source prominent high velocity lateral movements are detected. In addition, structural and brightness changes of several features are detected. In the northern jet, closer to the centre, knotty features tend to become elongated until they break into separate blobs. Some features are fading, some brightening, and in some cases the position and shape depends on the observed wavelength. Even so, the overall S-shape of the jet did not change significantly during the last 30 years.

Our high resolution radial velocity measurements of the jet present a somewhat controversial behaviour with respect to previously published data. We find that the northern jet is mostly red-shifted, while the southern counterpart is blue-shifted. Formerly published results show the opposite. We are inclined to believe that this discrepancy is due to the general complexity of the line profiles (multiple components, wide wings) and combination of higher spectral and spatial resolution of our data, which allows a more detailed view than previously published long-slit spectra.

The overall conclusion of our study is that the evolution of the jet cannot be described by purely radial expansion, and the combined action of changing ionisation, illumination, shocks and precession should be added, although it is difficult to disentangle the importance of each effect over the others.

# The large-scale nebula of B[e] supergiant MWC 137

The enigmatic object MWC 137 (V1308 Ori) belongs to the group of Galactic B[e] stars. It is surrounded by the optical nebula Sh 2-266 and located in the center of a cluster. The evolutionary state of MWC 137 has long been debated. Suggestions ranged from pre-main sequence (Hillenbrand et al. 1992; Berrilli et al. 1992; The et al. 1994; Hein Bertelsen et al. 2016) to post- main-sequence, spreading over a large luminosity range (Herbig & Kameswara Rao 1972; Finkenzeller & Mundt 1984; Esteban & Fernandez 1998; Oksala et al. 2013).

Significant progress in the star's classification was achieved by Muratore et al. (2015), who modeled the emission from hot $^{13}CO$ gas in the vicinity of MWC 137 that was first detected by Oksala et al. (2013). The presence of measurable amounts of $^{13}CO$ implies a significant enrichment of the circumstellar material in $^{13}C$ (Kraus 2009). As stellar evolution models show (e.g. Ekström et al. 2012), this isotope is processed inside the star and via mixing processes transported to the surface, from which it is liberated into the environment by mass loss events. With the discovery of hot, circumstellar $^{13}CO$ emission, a pre-main sequence evolutionary phase of MWC 137 could finally be excluded.

The evolved nature of MWC 137 is further consolidated by the studies of Mehner et al. (2016). These authors investigated the whole cluster and determined a cluster age of $> 3$ Myr. With a mass of 10-15 $M_\odot$ for MWC 137, this object has clearly evolved off the main sequence.

Moreover, Mehner et al. (2016) discovered a jet with several individual knots emanating from MWC 137 with high velocities. Estimates of the age of these knots revealed that the jet must be much younger than the large-scale optical nebula. The position angle of the jet is aligned with the polarization angle and hence perpendicular to the circumstellar disk traced on small scales by intrinsic polarization in H$\alpha$ (Oudmaijer & Drew 1999). Further confirmation for a rotat-ing circumstellar disk was provided by the rotationally broadened CO band head emission reported by Muratore et al. (2015).

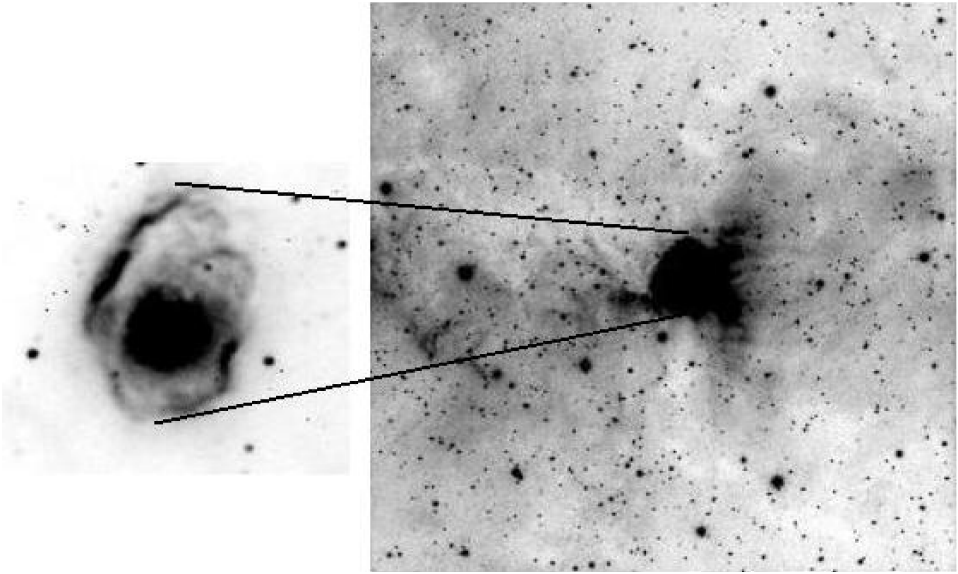

Figure 6.1: Hα image of the enviroments of MWC 137. The bipolar nebula is shown as inset. The FOV of the image on the right is $12'.5 \times 12'.5$. North up, east left. Figure is depicted from Marston & McCollum (2008).

The large-scale structure seen in Hα images (e.g., Marston & McCollum 2008 and Figure 6.1) led Esteban & Fernandez (1998) to suggest that Sh 2-266 could be a ring nebula produced by the interaction of the stellar wind with the interstellar medium. On the other hand, as the morphology of Sh 2-266 is reminiscent of bipolar ring nebulae detected around two early B-type supergiants, Sher 25 (Brandner et al. 1997) and SBW1 (Smith et al. 2007), Muratore et al. (2015) proposed that the nebula material might have been ejected during the blue supergiant phase so that MWC 137 might be transiting from a B[e] supergiant into a blue supergiant with a bipolar ring nebula.

We carried out an observational campaign combining data from various wavelength regimes (optical, infrared, and radio) and on different spatial scales. These data are aimed to investigate in detail the structure and kinematics of the environment of MWC 137 on both the large and small scale: Publication III (Kraus et al. 2017) of these Thesis. In this Chapter only the measurements and analyses which were carried out by the defender of these Thesis, together with the discussion, are presented: the large scale optical nebula and the jet of MWC 137.

## 6.1 Kinematics of the large-scale optical nebula

The structure of the optical nebula around MWC 137 is resolved in the long expo-sure ALFOSC narrow band image in the H$\alpha$+[N II] filter (left panel of Figure 6.2). The nebula has an oval shape with arc-like structures along its outer boundary and a size of $80'' \times 60''$, similar to what was resolved in earlier observations (Marston & McCollum 2008; Mehner et al. 2016). The ionized nebula displays an asymme-try with respect to the stellar position. It is more extended to the North. We also note diffuse emission at larger distances, in particular South-East and South-West of the main nebula.

To investigate the nebula kinematics, we make use of the long-slit spectra. The orientation and nebula coverage of the three available slit positions are presented in the left panel of Figure 6.2.

For the radial velocity (RV) measurements we used the two nebula lines [N II] 6583 Å and [S II] 6716 Å because they are the more intense ones from the dou-

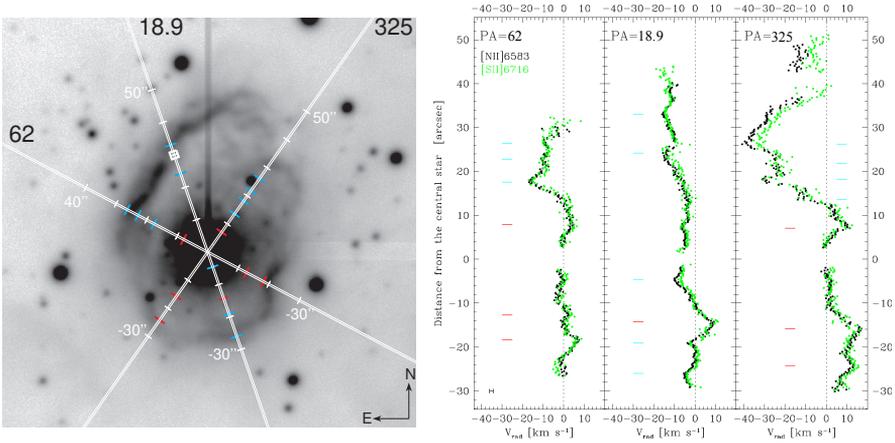

Figure 6.2: *Left.* Positions of the NOT slits (white) and their PA values (black numbers) overlaid on the zoomed (FOV of $2' \times 2'$) ALFOSC H$\alpha$ image. A distance scale in steps of $10''$ is marked with white ticks along the slits in both directions from the central star. Red and cyan ticks mark positions of red- and blue-shifted RV extrema, respectively. PA = 18.9 is along the jet and the white square indicates knot c. *Right.* Radial velocities of the [N II] 6583 Å (black) and [S II] 6716 Å (green) emission lines. Positive distance from the central star on the image is towards the North, negative towards the South. Red and cyan ticks mark red- and blue-shifted velocity extrema determined for the [N II] 6583 Å line. The average error bar for all the measurements is in the lower left corner.

Table 6.1: Radial velocity measurements of the nebula of MWC 137

| PA | $\lambda_{\text{lab}}$ | $v_{rad}$ | $\sigma_{v_{rad}}$ | Dist |
|---|---|---|---|---|
| $^{\circ}$ | Å | $\text{km s}^{-1}$ | $\text{km s}^{-1}$ | $''$ |
| 325 | 6716.44 | 3.85 | 0.90 | -29.84 |
| 325 | 6716.44 | 5.23 | 0.90 | -29.63 |
| 325 | 6716.44 | 11.03 | 0.90 | -29.42 |
| 325 | 6716.44 | 8.27 | 0.90 | -29.21 |
| 325 | 6716.44 | 9.83 | 0.90 | -29.00 |
| ... | | | | |

The entire table is published in the Appendix (Table 8.2) and in the electronic version of the Publication III. The first 5 lines are shown here for guidance regarding its form and content.

blets. These lines can be traced from the stellar position out to the edges of the optical nebula. As the images are strongly saturated in H$\alpha$, in particular in the vicinity of the central star's position, this line was excluded from the analysis. The measurements were done from the 2D spectra by fitting the line profiles, line by line with a single Gaussian. Occasionally, a cosmic ray fell on top of the line which could not be removed. Also, in some parts of the nebula the intensity was too weak for reliable measurements. These regions were thus excluded. The measurements were corrected for heliocentric velocity and systemic velocity of +42.0 $\text{km s}^{-1}$ (Kraus et al. 2017), and the resulting RV values along the three individual positionings of the slit are shown in the right panel of Figure 6.2. The full information on the final RVs is available in the Appendix (Table 8.2) and in the eletronic version of the Publication III. In Table 6.1 an example of the first five lines is presented.

Both nebula lines display the same kinematic behavior, with RV values ranging from -41 to +18 $\text{km s}^{-1}$. The errors of the individual measurements are relatively small, on the order of 0.9 $\text{km s}^{-1}$ on average as shown by the error bar in the lower left corner of the radial velocity plot. They were estimated based on the RMS of the wavelength calibration, whereas the contribution from the fitting is negligible. Precise error values are included in Table 6.1.

The [S II] 6716 Å line is considerably weaker than [N II] 6583 Å resulting sometimes in a larger scatter of the measured values, in particular in regions of lower intensity as is the case at the edges of the nebula structure and in the northern region of PA = 325$^{\circ}$. Along this position angle the intensity in both lines basically disappears at distances 38$''$ - 42$''$ to the North so that the velocities could not be measured resulting in a gap in Figure 6.2.

Table 6.2: Extrema of the RV values of [N II] 6583 Å marked in Figure 6.2.

| PA | $v_{rad}$ | Dist | PA | $v_{rad}$ | Dist | PA | $v_{rad}$ | Dist |
|---|---|---|---|---|---|---|---|---|
| ° | km s$^{-1}$ | ″ | ° | km s$^{-1}$ | ″ | ° | km s$^{-1}$ | ″ |
| 62 | 7.32 | -18.36 | 18.9 | -5.41 | -25.99 | 325 | 13.35 | -24.20 |
| | 2.27 | -12.69 | | -3.68 | -19.06 | | 17.04 | -15.80 |
| | 4.82 | 7.89 | | 9.16 | -14.23 | | 11.39 | 7.09 |
| | -16.81 | 17.55 | | -10.01 | -4.57 | | -13.84 | 13.60 |
| | -10.98 | 22.80 | | -15.56 | 24.20 | | -31.00 | 18.22 |
| | -9.71 | 26.37 | | -15.93 | 33.02 | | -29.14 | 21.79 |
| | | | | | | | -40.79 | 26.20 |

As a general trend we note that the emission from the northern nebula parts is predominantly blue-shifted, whereas the southern regions have mostly red-shifted emission. However, deviations from this general trend exist. For instance, the RV measurements for PA = 325° and PA = 62° display a red-shifted emission in the northern nebula part in the vicinity of the star (5″ - 15″). In addition, PA = 18.9° displays three southern regions with blue-shifted emission. In fact, we observe a series of extrema in the RV measurements along each slit position. They are marked by cyan and red ticks in both panels of Figure 6.2 and represent maxima in blue- and red-shifted velocities, respectively. These values are also listed in Table 6.2.

Most remarkable is the RV variation along PA = 325°. It displays the highest amplitudes in both the blue- and red-shifted emission while the slit passes through nebula regions which appear less intense. The blue-shifted emission seen towards the North contains a series of pronounced maxima. The image of the nebula also shows that this region seems to have generally more structure. However, there is another star (identified as B3-5 by Mehner et al. 2016) in the close vicinity of PA = 325° at a distance of $\sim 25''$, which might influence the kinematics of the nebula.

Close inspection of the positions of the major velocity extrema along all three slits in comparison with the intensity structure of the nebula reveals that they typically precede intensity accumulations.

## 6.2  Jet

The PA=18.9° was positioned such that it covered parts of the jet discovered by Mehner et al. (2016). As this jet appears slightly tilted (see their Figure 8) and the slit is rather narrow, only one clear signal from the jet is seen in our data: the emission from knot c. This feature is located at a distance of 30″ from the central star and is marked by a white box in the left panel of Figure 6.2. On the 2D image, knot c has a slightly elongated shape spreading over about $10 \times 10$ pixels. This size corresponds to $2''.1$. We observe a clear inclination of the feature on the 2D image, meaning that the knot displays a velocity gradient with a smaller blue-shift for larger distance from the star.

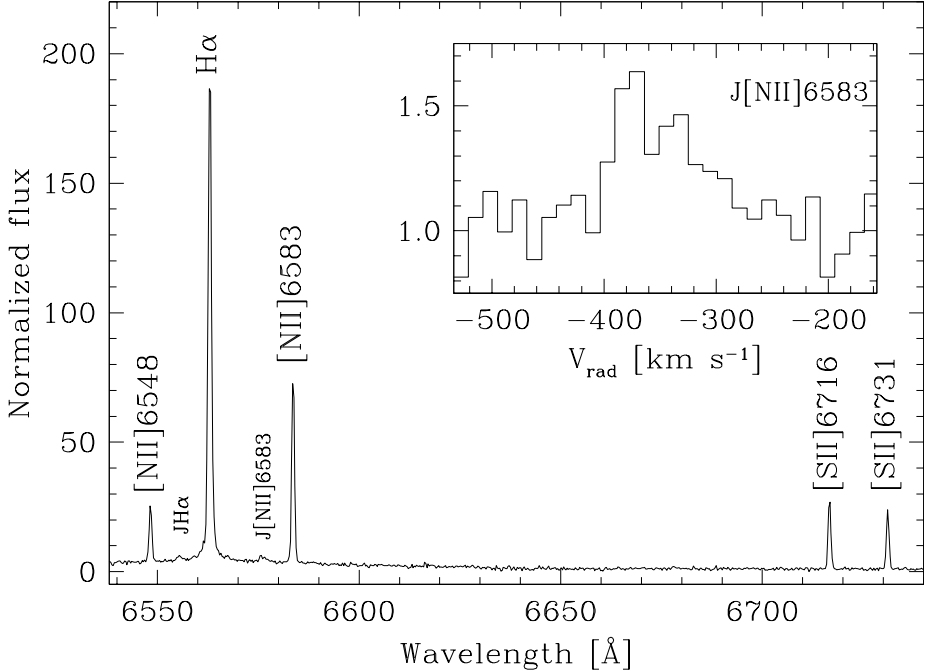

Figure 6.3: ALFOSC spectrum of the nebula emission and the emission of the jet component c (denoted as J Hα and J [N II] 6583Å). The J [N II] 6583Å line of knot c in the insert indicates sub-structure over a large velocity range that was previously not reported. Figure is a reprint from Paper III.

The intensity of the knot is too low to perform a line-by-line measurement of the velocities over the feature. Therefore, we added up the ten lines in the spatial direction. The resulting nebula plus knot c emission spectrum is shown in Figure 6.3. The jet emission is very faint and seen only in Hα and [N II] 6583 Å (labeled as J Hα and J [N II] 6583 Å in the plot), while Mehner et al. (2016) discovered it also in [N II] 6548 Å and the two [S II] lines in their long-exposure image. The strongest jet emission line is J [N II] 6583 Å from which we derive a mean knot radial velocity of -356.5 km s$^{-1}$(corrected with systemic velocity) in agreement with the estimates of Mehner et al. (2016) from their low-resolution spectra[1].

A blow-up of the jet emission feature J [N II] is shown in the inset of Figure 6.3. It demonstrates that the line profile of knot c is rather broad and double-peaked compared to the narrow, single-peaked regular nebula line. Considering the double-peaked profile shape in combination with the elongated inclined structure identified on the 2D image, we tend to believe that knot c might consist of at least two individual sub-structures. Fitting both with individual Gaussians, we obtain a radial velocity of $-378.9$ km s$^{-1}$and FWHM of 25.4 km s$^{-1}$for the narrow blue component, and a radial velocity of $-334.0$ km s$^{-1}$and FWHM of 55.9 km s$^{-1}$for the broad red component.

## 6.3 Discussion

The modelling in this section was performed by the co-author of Paper III, M. Kraus, in collaboration with the defender of the current thesis.

The optical nebula of MWC 137 displays some structure that might be approximated by a double-ring. This is visualized by the two identical, parallel ellipses overlaid on the ALFOSC image in Figure 6.4. We measured the semimajor ($a = 31''.9 \pm 1''.0$) and semiminor ($b = 12''.7 \pm 0''.5$) axis, the alignment of the ellipses (here, of the semiminor axis PA($b$) $= 31°.10 \pm 0°.50$), the distance ($c = 20''.2 \pm 0''.2$) between the two ellipses, and the alignment of the line connecting the two ellipses (PA($c$) $= 58°.11 \pm 0°.50$).

The equal dimensions of the ellipses and their parallel arrangement suggest that they might represent the outer rims of a double-cone. However, since the alignment of the semiminor axis is not parallel to the alignment of the line connecting the two ellipses, this double-cone must be sheared in one direction.

To test whether such a structure might be a reasonable scenario to explain the observed nebula kinematics, we use the observed parameters as constraints and compute the geometric shape of a sheared double-cone. We start by defining a double-cone in North-South direction, which we align with the z-axis. The top and bottom circle have radius $R$, which is given by the length of the semimajor axis $a$ of the ellipse. The height of each cone is $h$ so that the object extends from $z = -h$ to $z = +h$, and the angle $\alpha$ represents half the opening angle of the cone. Then, we apply a shear $\Delta y$ along the $y$-axis. This shear is characterized by the angle $\xi$ defined via $z = \Delta y \tan \xi$. Finally, we rotate the double-cone first around the $y$-axis with angle $\theta$ and then around the $z$-axis with angle $\phi$, where $\theta$ and $\phi$ represent the usual spherical coordinates. This rotated double-cone structure is then projected into the $y - z$ plane, which we identify with the plane of the sky.

---

[1]Note that Mehner et al. (2016) did not correct their measurements for the systemic velocity.

To compute the kinematics, we assume that the gas is streaming radially with constant velocity and purely along the cone surfaces. This allows us to determine the regions of blue- and red-shifted emission. To reproduce the observed quantities and position angles we found the values of $h = 17''.81$, $\alpha = 60°.86$, $\xi = 137°$ for the double-cone parameters, and $\theta = 119°$ and $\phi = 63°$ for the rotation angles. The projection of this geometrical shape to the plane of the sky is shown in the top left panel of Figure 6.5. In blue and red we mark regions with blue- and red-shifted gas kinematics, respectively.

Next, we overlaid the position angles of the slits and determined the intersection points with the projected double-cone. During the observations the three PAs were positioned such that they intersect at the location of the star. However, with respect to the chosen geometrical model it is important to note that the center of the double-cone was found to be located $5''.5$ North of the stellar position. The extracted velocity values at the intersection points, normalized to the constant out-flow velocity, are shown in the top right panel of Figure 6.5. For comparison we included the observed velocities, which we normalized arbitrarily.

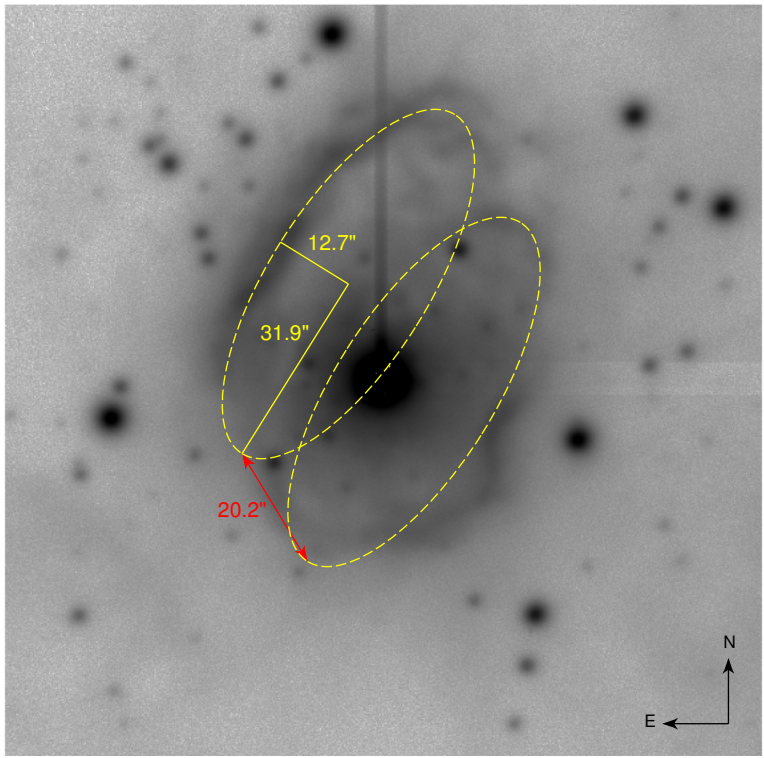

Figure 6.4: Measured quantities of the putative double-ring (or cone) structure of the optical nebula. The image size is $2' \times 2'$. Figure is a reprint from Paper III.

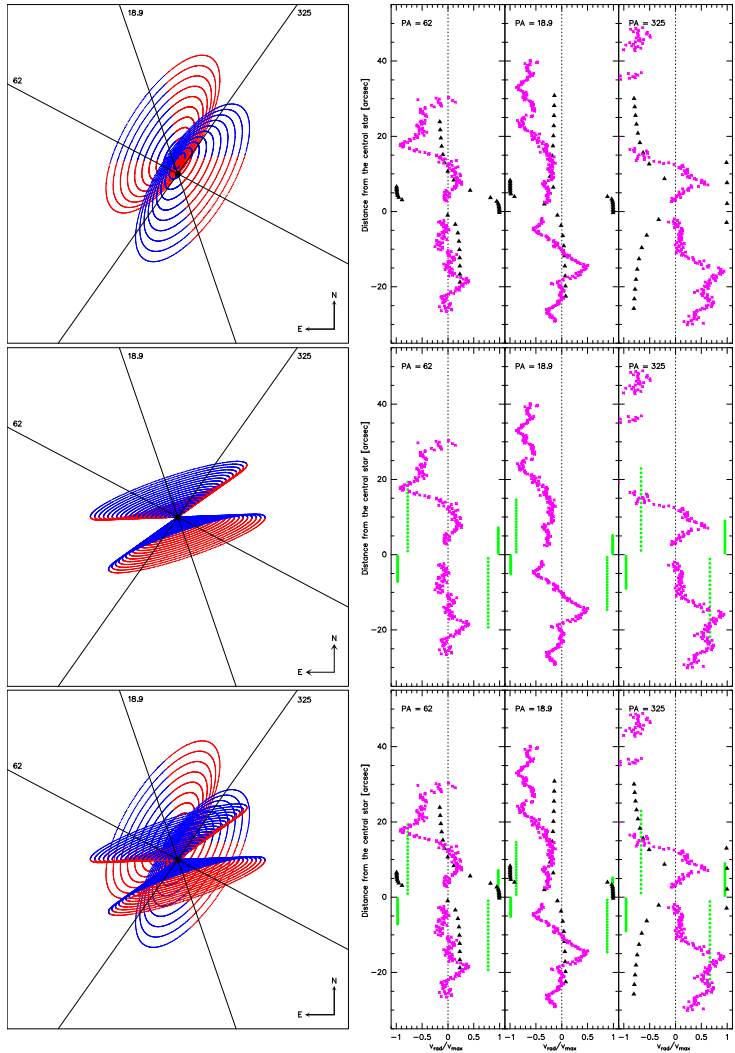

Figure 6.5: *Left:* Nebula kinematics for the scenario of a sheared double-cone structure as estimated from the optical nebula (top), for a regular double-cone centered on the stellar position and with the jet PA as symmetry axis (middle), and a combination of both (bottom). In both cases, the north-eastern cone points away and the south-western one towards the observer. Colors indicate regions with blue- and red-shifted velocities. Image size corresponds to $2' \times 2'$. *Right:* Model kinematics along the PAs (black triangles and green dots) compared to observations (purple crosses).

In the northern nebula regions our model produces blue-shifted emission in agreement with the observations. Also the red-shifted northern emission in the vicinity of the star along PA=62° agrees with the model predictions. Note that the densely populated regions at distances $0'' - 10''$ North of the star along PA=62° and PA=18°.9 with both high blue- and red-shifted velocities originates from the overlap regions of the two cones. What cannot be reproduced is the variability in the velocities and the high amplitudes observed in the blue-shifted gas. In the southern nebula our model produces red-shifted emission. There is some agreement with the observations along PA= 62°and PA=18°.9, however, for PA=325° the observations display strong red-shifted emission in contrast to the theoretical predictions.

Although some of the kinematics can be reproduced, we are aware that the chosen geometrical scenario is too simple to account for all observed features. Still, this exercise provides important insight. The fact that the star MWC 137 is not located in the center of the geometrical double-ring structure suggests that the large-scale optical nebula was most probably not formed exclusively by the wind of a single star but rather by the combined winds of the early-type stars identified by Mehner et al. (2016) in the center of the cluster and in the vicinity of MWC 137. In addition, the jet which was found by Mehner et al. (2016) to originate from MWC 137, is not aligned with the axis of the double-cone. Instead, the jet is perpendicular to the circumstellar disk of MWC 137 which is traced by the intrinsic polarization in H$\alpha$ (Oudmaijer & Drew 1999) and the rotationally broadened CO bands (Muratore et al. 2015). Under the presence of a circumstellar disk, the current wind of MWC 137 is most likely also channelled into a double-cone, but with a different orientation. One might speculate whether this inner disc and the jet might point towards MWC 137 having a compact companion as, e.g., in the case of CI Cam (see, e.g., Clark 2006). The detection of an X-ray source in the vicinity of MWC 137 by Evans et al. (2014) might support such a suggestion. However, the uncertainty in position of the X-ray source renders it difficult to unambiguously identify MWC 137 as the optical counterpart.

Unfortunately, the optical image provides no clear indication of the postulated second double-cone, which might imply a (much) lower wind density. Besides the position angle of the jet and the request that the velocity in the southern part of the slit PA=325° should be red-shifted, we have no additional observational constraints for a possible model. Therefore, and simply for demonstration purposes, we compute a circular double-cone centered on the star's position. To align it with the PA of the jet we only altered the rotation angle $\theta$, for which we find a value of $159°$ but kept the same value for $\phi$. For the height and the cone opening angle we adopted values of $10''$ and $71°$. They were chosen such that the double-cone remains small but wide enough to guarantee predominantly red-shifted emission in the southern intersection region with PA=$325°$. The velocities were again computed under the assumption of a purely radial outflow along the cone surfaces.

The projection to the plane of the sky and the velocities along the slit positions are shown in the middle panels of Figure 6.5. As before, the northern cone points away and the southern one towards the observer. The symmetric intersections with the slit positions and the chosen constant outflow velocity result in symmetric patterns of constant values in both red- and blue-shifted velocities. Also with this model some of the observed features might be reproduced.

Finally, we combine these two double-cones and present their projection and velocities in the bottom panels of Figure 6.5. Obviously, various observed features might be better reproduced with these combined double-cones, while others still cannot. This example shows that the real physical scenario is much more complex than what can be explained by our simplistic model. However, without further constraints from observations it will not be possible to reconstruct a reliable scenario.

# CHAPTER 7
# CONCLUSIONS

In this Thesis three ionised gaseous stellar outflows were studied: the nebula of the symbiotic binary R Aquarii, the classical nova remnant of GK Persei, the nebula and a jet of the massive B[e] supergiant MWC 137.

The following main conclusions are drawn:

- New extended nebulosities at greater distances from the central star than the previously known bipolar nebula were discovered. Modelling of the hourglass nebula provides a kinematic distance of $178 \pm 18$ pc. The age of the hourglass nebula is calculated to be $653 \pm 35$ years, while that of the jet features range from 125 to 290 years. Closer to the central star, the jet shows apparent, lateral fast moving components, with apparent displacement of up to 900 $km\,s^{-1}$. Considering the generally slow velocities otherwise present, $\leq 140$ $km\,s^{-1}$, and the difficulty to be explained by real matter motions, we conclude that these lateral movements must be attributed to changing ionisation and/or illumination effects. We find that the northern jet is mostly red-shifted and southern blue-shifted, contrary to previously published data. We believe this apparent discrepancy is a result of the complex behaviour of the jet, the abovementioned changes in ionisation and/or illumination, as well as superior quality of the present data as compared to previous works.

- GK Persei is modelled as a thick spherical shell consisting of filaments expanding with a significant range of velocities, mostly between 600 and 1000 $km\,s^{-1}$. No clear asymmetric expansion is detected, and the overall deceleration of the expansion is found to be modest, contrary to previous predictions. A kinematic distance of $400 \pm 30$ pc is derived.

- As a first approximation, the nebula of MWC 137 can be described using a double-cone model, but additional complex velocity patterns are identified, and its formation scenario remains an open issue.

# CHAPTER 8
# FUTURE WORK

Several conclusions were drawn in this Thesis. However, a number of issues should be further investigated in order to fully understand the formation and evolution of these outflows and the predominant physical mechanisms at work in the production and shaping of this kind of nebulae.

It was deduced that GK Persei is a spherical thick shell expanding relatively homogeneously. However, the data presented does not completely rule out different models such as that proposed in Harvey et al. (2016) which considers a cylindrical sphere with polar cones. In order to fully understand the nature of the remnant continuous monitoring at extended wavelengths is needed.

In the case of R Aquarii, it is likely that several phenomena that we have observed are not related to the physics of the ejection in the central engine, but rather on changes in the evolution and corresponding radiative flux from the central sources and the interactions with the ambient medium. It is important to continue obtaining new images to see their subsequent evolution and fully understand the puzzling behaviour observed. A complete coverage of all jet features, both in imaging and in spectroscopy to obtain their radial velocities and excitation state, is needed to fully understand the evolution of this active jet.

The model of the double-cone nebula of the massive supergiant MWC 137 needs further refinement. For that purpose we have already obtained additional spectral data, which will be reduced and analysed in the near future. In the field of massive stars we have continued to investigate the large scale shells of B[e] stars in the northern sky, which was started by Marston & McCollum (2008). MWC 137 was the first object on the list. Several other targets will be studied. Additionally, we plan to expand our study to the southern sky. Massive stars are the most beneficial to enrich the Universe with heavy elements. Their shells are a direct component in that chain. Disentangling their evolution helps to shed light into the mass loss of massive stars, which is still not fully understood.

Also for symbiotic stars and classical novae, it will be necessary to extend this kind of spatiokinematical studies to other objects which, while not spatially resolved in the same details as for the targets of this Thesis, will provide an overall view of the phenomena and highlight the physical mechanisms that are common to these classes of evolved binary stars. Concluding, we will continue following the three stellar outflows that we have presented in this Thesis, and at the same time we will enlarge the number of targets to be able to draw more general conclusions about the importance and physics of mass loss in the late stages of stellar evolution. To this aim, we will use the best optical and infrared facilities available worldwide, that nowadays offer powerful instrumentation able to do deep imaging at high spatial resolution as well as 2D spectroscopy.

# SUMMARY IN ESTONIAN

## Hilises evolutsiooni faasis olevate tähtede väljapaisatud udukogud ja joad.

Tähed sünnivad gaasi ja tolmu pilvest, mis oma raskusjõu mõjul kokku tõmbub. Kui proto-tähe mass on suurem kui 0.08 $M_\odot$, on tal küllalt energiat, et saavutada tsentri piirkonnas piisav temperatuur, $T = 15 \times 10^6$ K, vesiniku tuumasünteesiks. Tähed veedavad enamuse oma elueast sünteesides tuumas vesinikku heeliumiks. Seda etappi nimetatakse peajadaperioodiks. Väiksema massiga tähed veedavad peaaegu kogu oma pika eluea peajadal, piirjuhtudel kuni triljoneid aastaid. Suure massiga tähed, $M > 8$ $M_\odot$, elavad tormilist ja kiiret elu, vaid mõni miljon aastat.

Pärast peajada sõltub tähe edasine areng tema algmassist. Kuni 0.8 $M_\odot$ tähel ei jätku piisavalt energiat, et alustada tuumas heeliumi sünteesi süsinikuks. Täht langeb enda raskusjõu mõjul kokku, moodustades niinimetatud valge kääbuse, mis aeglaselt jahtub, kuni ta ei ole enam vaadeldav. Kui tähel on rohkem massi, siis algab tuumas heeliumi ja järkjärgult ka raskemate elementide süntees, kuni ränini välja (tähtede puhul, mille algmass on suurem kui 8 $M_\odot$). Samal ajal kui tuumas sünteesitakse aina raskemaid elemente, jätkub kergemate elementide süntees tähe väliskihtides.

Tähtede elutsüklites, peale peajada, esineb erinevat tüüpi massikaoga etappe, kuni tähe evolutsiooni lõppfaasini, kui toimub väiksem või suurem plahvatus vastavalt tähe algmassile. Väiksema massiga tähtede elu lõppu iseloomustab planetaarudu, mis tekib tähest eemale paisatud vesiniku ja heeliumi kihtidest. Tähest endast jääb järele peamiselt süsinikust koosnev valge kääbus. Suure massiga tähed võivad plahvatada supernoovana, mille käigus sünteesitakse ka rauast raskemaid elemente. Sellisel juhul võib tsentraalsest tähest saada ülimassiivne ja suure tihedusega neutrontäht, must auk või üldse mitte midagi.

Asjaolu, et enamik tähti on kaksik- või mitmiktähed, teeb tähtede evolutsiooni keeruliseks. Ühe tähe arengus toimuv massikadu mõjutab otseselt tähesüsteemi teist komponenti. Niimoodi koos evolutsioneerudes läbivad mitmiktähed kordamööda evolutsiooni etappe, olenevalt teise komponendi mõjust jätavad osa etappe vahele ja/või kordavad neid. Ükskõik, milline on tähtede algmass ja millise plahvatusliku mehhanismiga nende elu lõppeb, jäävad nendest maha mitmesuguse kujuga jäänukid. Tähtede jäänukid helendavad kas kuuma tsentraalse tähe kiirguse tõttu ja/või väljapaisatud aine enda liikumise tõttu, mis tekitab lööklained oma teekonnal hõredasse tähtedevahelisse ruumi. Kuigi tähed on kogu oma elu sfäärilised gaasikerad, on nende jäänukid enamasti asfäärilised: uni- ja bipolaarsed, filamentaarsed, mitmes suunas kontsentrilised ringid, kollimeeritud joad. Miks jäänukid on ebasümmeetrilised, ei ole hetkel täpselt teada. Tõenäoliselt on oluliseks faktoriks tähtede mitmiklus ning tähtede magnetväljad, kuid täpsetes põhjustes ei ole üksmeelele jõutud. Jälgides jäänukite evolutsiooni ajas, on võimalik teha järeldusi nende kuju tekkimise kohta. Saamaks täielikku ettekujutust nende tekkemehhanismidest, on vaja nende arengut jälgida kolmemõõtmelises ruumis. Vaatekiire suunalist kiiruse komponenti, radiaalkiirust, on võrdlemisi lihtne arvutada Doppleri nihke kaudu spetrijoontes, aga vaatekiirega risti olevaid kiiruse komponente, liikumisi taevasfääril, on võimalik mõõta ainult väheste tähtede väljapursete korral. Nüüdisaegsete teleskoopide ja instrumentidega, mis võimaldavad piisavalt head nurklahutust, on see saavutatav. Lisaks, tuleb valida sobivad objektid: piisavalt lähedal ja heledad, et nende areng taevasfääril oleks jälgitav inimese elueaga võrreldava aja jooksul. Sel juhul saame unikaalse võimaluse mõõta kõiki kolme kiiruse komponenti ning teha detailset dünaamilist ja morfoloogilist analüüsi, mõistmaks väljapursete geomeetriat. See oligi käesoleva doktoritöö eesmärk: uurida hilises evolutsiooni faasis olevate tähtede väljapaisatud ainehulkade struktuuri ja kinemaatikat. Selleks valisime kolm objekti: klassikaline noova jäänuk GK Persei, udukogude ja jugadega sümbiootiline kaksiktäht R Aquarii ja massiivne B[e] ülihiid MWC 137. Uurimiseks valitud objektid evolutsioneeruvad nii kiiresti, et paari aastaga, isegi paari kuuga on võimalik näha struktuurilisi ja heleduslikke muutusi taevasfääril. See on astronoomias harv nähe!

Valitud objektide monitooring meie töörühma poolt algas 1990ndate alguses. Aastate jooksul on kogutud mitmetel epohhidel fotomeetrilise andmeid, milledele on lisatud fikseeritud epohhi spektraalandmed. Saadud andmekogu võimaldas meil jõuda järgmistele tulemustele:

- Noova GK Persei jäänuki paisumine on võrdlemisi homogeenne, kuigi varem oli ennustatud, et jäänuki edelaosa peaks olema märkimisväärselt aeglustunud võrreldes ülejäänud osadega. Jäänuk on paks sfääriline kiht, milles filamendid liiguvad tsentraalsest tähest väljapoole peaaegu ballistiliselt ja võrdlemisi laias kiiruste vahemikus, peamiselt 600 kuni 1000 $\mathrm{km\,s^{-1}}$. Üldine aeglustumine on väike, jällegi vastuolus eelnevate uuringutega.

- R Aquarii näitab eriskummalist ja huvitava palet. Tema liivakella kujuline udukogu paisub ballistiliselt, nagu ka kollimeeritud joa kaugemad osad. Joa tsentraalsele tähele lähemal asuvad osad näitavad lateraalseid väga kiireid liikumisi, mis arvestades valdavalt väiksemaid kiirusi joa kõigis teistes osades, peab olema põhjustatud muutuvast ionisatsiooniolekust ja/või valgustatuse efektist. Lisaks leidsime, et põhjasuunaline juga on punanihkes ja lõunasuunaline sininihkes, kuigi eelmised uuringud näitasid vastupidist tulemust. Oleme veendunud, et see erisus on põhjustatud joa keerulisest evolutsioonist ning asjaolust, et meie spektraalandmetel on oluliselt suurem spektraallahutus kui varasemalt avaldatud andmetel.

- MWC 137 kohta kogutud andmete põhjal saame väita, et tema jäänuki morfoloogia on kõige paremini lähendatav bipolaarse udukogu mudeliga, mille orbitaaltasandist paiskub välja filamentidest koosnev juga.

Kuigi antud töös saadi detailseid tulemusi iga objekti kohta, jäid mitmed küsimused ka vastamata, võimaldades meil tulevikus jätkata nende ja teiste huvitavate objektide uurimist.

# ACKNOWLEDGEMENTS

Firstly I would like to thank my parents, Aime and Ants Liimets, thanks to your support I was always able to dedicate fully on my studies.

During my long period of PhD studies, there have been many helpful people to whom I am very grateful. It would be impossible to mention you all individually but I thank you all for your support and guidance. Bringing out few names, in no particular order. Peeter Tenjes, I thank you for giving me honest advice, on a difficult moment of hesitation, which lead me to decide to continue and to choose astronomy field after all. Taavi Tuvikene, who was my first mentor, introducing me to Linux and tolerating my first baby steps in astronomy. My supervisor, since the bachelor degree, Indrek Kolka, whose door has always been open to me with whatever problem I had, scientific or personal. My PhD supervisor Romano Corradi, for guiding me thoroughly through various steps, from data reduction to writing papers, but also giving me space to make my own decisions. My eternal office mate, Tõnis Eenmäe, my personal IT help as well as the teacher on telescopes in Tõravere. My student Kristiina Verro, whose constant cheerful mood and a repeating question "When is the article ready?", kept me always hard working. Dave, thank you for stepping in and saving the last paper and with that probably the whole Thesis! Miguel, thank you for hosting me in Madrid and teaching me everything about the magnification method. Michaela, thank you for proof reading several chapters of the Thesis. Chicas, meeting you has been a match made in heaven!

Last but definitely not least I would like to thank my husband, Boris. Thank you for being who you are. I hope one day all your dreams will come true. As you said, our two little boys did not really help but they are wonderful.

Several institutes have been hosting me during this PhD: Tartu Observatory, University of Tartu, Isaac Newton Group of Telescopes, Nordic Optical Telescope, Centro de Astrophísica en La Palma, Astronomical Institute of the Czech Academy of Sciences. I thank you all for the fruitful and inspiring enviroment.

Based on observations made with the Isaac Newton Telescope, operated by the Isaac Newton Group, and with the Nordic Optical Telescope, operated by the Nordic Optical Telescope Scientific Association, on the island of La Palma in the Spanish Observatorio del Roque de los Muchachos of the Instituto de Astrofísica de Canarias. The data presented here were obtained in part with AL-FOSC, which is provided by the Instituto de Astrofísica de Andalucia (IAA) under a joint agreement with the University of Copenhagen and NOTSA. This work has made use of data from the European Space Agency (ESA) mission *Gaia* (`https:`

//www.cosmos.esa.int/gaia), processed by the *Gaia* Data Processing and Analysis Consortium (DPAC, https://www.cosmos.esa.int/web/gaia/dpac/consortium). Funding for the DPAC has been provided by national institutions, in particular the institutions participating in the *Gaia* Multilateral Agreement.

The research presented in this Thesis has been financially supported by the Estonian Ministry for Education and Science (grant IUT40-1 and IUT 26-2), European Regional Development Fund (TK133), and GAČR (grant 17-02337S). This research was also partially supported by European Social Fund's Doctoral Studies and Internationalisation Programme DoRa and Kristjan Jaak Scholarship, which are carried out by Foundation Archimedes in collaboration with the Estonian Ministry of Education and Research.

# ORIGINAL PUBLICATIONS

## APPENDIX

### 8.1 Ballistically expanding jet features of R Aquarii

In this appendix we present more details related to the age analyses of the ballistic jet features.

#### 8.1.1 NE jet

Due to the structural change in later epochs of feature $B_{SHKM}$ it is difficult to obtain precise measurements. For the proper motion measurements the brightness enhancement was used in filters H$\alpha$+[N II], [O II], and [O III] .

For the rest of the NE features the magnification method was applied. For feature $C_{SHKM}$, the H$\alpha$+[N II] 1991 and 2012 frames needed a further flux correction due to the fact that the feature was about twice as bright in 2012 than in 1991, after using the nebula for flux matching. For rematching, the $C_{SHKM}$ feature itself was used. Following the same methodology as described in Section 5.1.1, the smallest residuals were found in the frame with magnification factor of $M = 1.074 \pm 0.003$.

Due to the change of shape and/or varying exposure time, and hence different level of details detectable in H$\alpha$+[N II] from feature NE4 between 1991 and 2012, we chose a different filter for the magnification method. After careful visual examination of our data we decided to use the [O II] 2009 and 2012 frames. Again, the same methodology was followed as for the bipolar nebula, first convolving to the worst seeing and then flux matching using the nebula. The smallest residuals for NE4 are for $M = 1.010 \pm 0.002$.

#### 8.1.2 SW jet

The longer baseline from 1991 to 2012 in the H$\alpha$+[N II] frames was unsuitable for the magnification method to be used for the feature loop+S due to the dramatic change in brightness of knot S as well as the extreme saturation streaks exactly on top of the feature on 1991 frame. Given that the nebula is, in general, very similar in the [O II] filter, we used its 1997 and 2012 observations (15.14 yrs time interval) to estimate a tentative value for the magnification of $M = 1.09 \pm 0.02$. L Just as for the NE4 feature, the [O II] 2009 and 2012 frames were used to derive the age of the SW3 feature via the magnification method. The best fitting $M$ was found to be $1.013 \pm 0.002$. In a case of the feature SW4 the same H$\alpha$+[N II] 1991 and 2012 frames, as for the nebular age in Section 5.1.1, were usable.

## 8.2 Tables

Table 8.1: Positions, proper motions, and radial velocities of the knots of the nova remnant GK Persei. The published online table is available at https://iopscience.iop.org/article/10.1088/0004-637X/761/1/34/meta

| $d_x{}^a$ ($''$) | $d_y{}^a$ ($''$) | $\mu$ ($''\,yr^{-1}$) | $\sigma_\mu$ ($''\,yr^{-1}$) | $\mu_x$ ($''\,yr^{-1}$) | $\sigma_{\mu_x}$ ($''\,yr^{-1}$) | $\mu_y$ ($''\,yr^{-1}$) | $\sigma_{\mu_y}$ ($''\,yr^{-1}$) | PA ($°$) | $\sigma_{PA}$ ($°$) | $\alpha$ ($°$) | $\sigma_\alpha$ ($°$) | $v_{rad}$ (km s$^{-1}$) |
|---|---|---|---|---|---|---|---|---|---|---|---|---|
| 0.06 | -27.43 | 0.242 | 0.005 | 0.005 | 0.004 | -0.242 | 0.005 | 180.1 | 0.3 | 181.2 | 0.9 | – |
| 0.09 | -48.91 | 0.448 | 0.006 | -0.009 | 0.012 | -0.448 | 0.006 | 180.1 | 0.2 | 178.8 | 1.5 | – |
| 0.21 | 4.72 | 0.046 | 0.011 | -0.007 | 0.012 | 0.046 | 0.011 | 357.5 | 2.0 | 8.9 | 14.7 | 808 |
| 0.21 | 4.72 | 0.046 | 0.011 | -0.007 | 0.012 | 0.046 | 0.011 | 357.5 | 2.0 | 8.9 | 14.7 | 919 |
| 0.22 | 33.44 | 0.261 | 0.003 | -0.005 | 0.004 | 0.261 | 0.003 | 359.6 | 0.3 | 1.0 | 0.8 | – |
| 2.25 | -9.25 | 0.081 | 0.004 | 0.010 | 0.010 | -0.080 | 0.005 | 193.7 | 1.0 | 186.8 | 7.1 | -822 |
| 3.41 | 18.67 | 0.153 | 0.013 | 0.003 | 0.019 | 0.155 | 0.014 | 349.7 | 0.5 | 358.9 | 7.1 | 721 |
| 3.69 | -0.88 | 0.048 | 0.017 | 0.041 | 0.015 | -0.033 | 0.022 | 256.6 | 2.5 | 231.4 | 21.0 | -894 |
| 3.72 | -48.07 | 0.497 | 0.012 | 0.057 | 0.007 | -0.494 | 0.012 | 184.4 | 0.2 | 186.5 | 0.9 | – |
| 4.11 | 0.38 | 0.007 | 0.006 | 0.006 | 0.006 | 0.007 | 0.011 | 275.3 | 2.3 | 317.3 | 54.0 | -794 |
| 4.11 | 0.38 | 0.007 | 0.006 | 0.006 | 0.006 | 0.007 | 0.011 | 275.3 | 2.3 | 317.3 | 54.0 | -890 |
| 4.12 | -40.28 | 0.331 | 0.006 | 0.052 | 0.005 | -0.327 | 0.006 | 185.8 | 0.2 | 189.1 | 0.9 | – |
| 4.41 | 15.30 | 0.149 | 0.005 | 0.023 | 0.005 | 0.148 | 0.005 | 343.9 | 0.6 | 351.2 | 2.1 | – |
| 4.64 | 34.95 | 0.166 | 0.014 | -0.054 | 0.015 | 0.174 | 0.015 | 352.4 | 0.3 | 17.4 | 4.6 | 383 |
| 4.87 | -23.07 | 0.205 | 0.006 | 0.049 | 0.004 | -0.199 | 0.006 | 191.9 | 0.4 | 193.9 | 1.2 | – |
| 4.92 | 36.92 | 0.362 | 0.011 | 0.140 | 0.016 | 0.347 | 0.010 | 352.4 | 0.3 | 338.1 | 2.3 | 498 |
| 5.07 | 3.39 | 0.057 | 0.013 | 0.053 | 0.010 | 0.022 | 0.011 | 303.8 | 1.6 | 292.9 | 10.5 | – |
| 5.66 | 43.67 | 0.071 | 0.069 | 0.065 | 0.047 | 0.063 | 0.073 | 352.6 | 0.2 | 314.1 | 39.2 | 435 |
| 6.05 | -34.73 | 0.300 | 0.009 | 0.069 | 0.003 | -0.293 | 0.009 | 189.9 | 0.3 | 193.2 | 0.6 | – |
| 6.59 | 17.05 | 0.072 | 0.006 | 0.028 | 0.007 | 0.066 | 0.005 | 338.9 | 0.5 | 337.3 | 5.4 | – |
| 6.72 | 9.37 | 0.075 | 0.006 | 0.006 | 0.008 | 0.088 | 0.005 | 324.4 | 0.8 | 356.1 | 5.1 | 701 |
| 6.72 | -26.59 | 0.184 | 0.010 | 0.048 | 0.008 | -0.178 | 0.009 | 194.2 | 0.3 | 195.0 | 2.6 | – |

Table 8.1: continued

| $d_x$ [a] (") | $d_y$ [a] (") | $\mu$ (" yr$^{-1}$) | $\sigma_\mu$ (" yr$^{-1}$) | $\mu_x$ (" yr$^{-1}$) | $\sigma_{\mu_x}$ (" yr$^{-1}$) | $\mu_y$ (" yr$^{-1}$) | $\sigma_{\mu_y}$ (" yr$^{-1}$) | PA (°) | $\sigma_{PA}$ (°) | $\alpha$ (°) | $\sigma_\alpha$ (°) | $v_{rad}$ (km s$^{-1}$) |
|---|---|---|---|---|---|---|---|---|---|---|---|---|
| 6.80 | -7.49 | 0.124 | 0.011 | 0.095 | 0.015 | -0.081 | 0.004 | 222.2 | 0.9 | 229.4 | 4.6 | 959 |
| 6.80 | -12.82 | 0.077 | 0.007 | -0.011 | 0.015 | -0.092 | 0.003 | 208.0 | 0.7 | 173.0 | 9.2 | -617 |
| 7.16 | -30.37 | 0.301 | 0.004 | 0.068 | 0.008 | -0.293 | 0.003 | 193.3 | 0.3 | 193.1 | 1.4 | — |
| 7.99 | -38.74 | 0.362 | 0.008 | 0.090 | 0.004 | -0.351 | 0.008 | 191.7 | 0.2 | 194.3 | 0.7 | — |
| 8.27 | 34.77 | 0.276 | 0.006 | 0.072 | 0.006 | 0.267 | 0.006 | 346.6 | 0.3 | 344.8 | 1.2 | — |
| 8.40 | -46.96 | 0.439 | 0.017 | 0.094 | 0.010 | -0.429 | 0.017 | 190.1 | 0.2 | 192.3 | 1.4 | — |
| 8.49 | -46.91 | 0.446 | 0.009 | 0.078 | 0.012 | -0.439 | 0.010 | 190.3 | 0.2 | 190.0 | 1.6 | — |
| 9.33 | 21.80 | 0.181 | 0.006 | 0.100 | 0.013 | 0.154 | 0.004 | 336.8 | 0.4 | 327.0 | 3.6 | 541 |
| 10.00 | 17.53 | 0.135 | 0.003 | 0.080 | 0.003 | 0.110 | 0.003 | 330.3 | 0.5 | 324.0 | 1.3 | — |
| 10.27 | -26.10 | 0.276 | 0.005 | 0.090 | 0.007 | -0.261 | 0.007 | 201.5 | 0.3 | 199.0 | 1.5 | 745 |
| 10.53 | 50.27 | 0.456 | 0.010 | 0.101 | 0.005 | 0.444 | 0.010 | 348.2 | 0.2 | 347.1 | 0.7 | — |
| 10.63 | -21.32 | 0.221 | 0.008 | 0.074 | 0.004 | -0.209 | 0.008 | 206.5 | 0.4 | 199.6 | 1.2 | — |
| 10.68 | -17.05 | 0.192 | 0.009 | 0.107 | 0.012 | -0.160 | 0.006 | 212.1 | 0.5 | 213.8 | 3.2 | -700 |
| 11.16 | -38.26 | 0.298 | 0.008 | 0.090 | 0.004 | -0.284 | 0.009 | 196.3 | 0.2 | 197.7 | 0.9 | — |
| 11.53 | -17.66 | 0.242 | 0.015 | 0.064 | 0.008 | -0.246 | 0.020 | 213.1 | 0.5 | 194.7 | 2.1 | -720 |
| 11.70 | -0.17 | 0.121 | 0.010 | 0.119 | 0.010 | -0.057 | 0.007 | 269.2 | 0.8 | 244.3 | 3.3 | 759 |
| 11.89 | -34.31 | 0.307 | 0.007 | 0.104 | 0.009 | -0.289 | 0.007 | 199.1 | 0.3 | 199.8 | 1.7 | — |
| 12.70 | 5.55 | 0.092 | 0.005 | 0.093 | 0.005 | 0.017 | 0.002 | 293.6 | 0.7 | 280.7 | 1.3 | 325 |
| 12.70 | 5.55 | 0.092 | 0.005 | 0.093 | 0.005 | 0.017 | 0.002 | 293.6 | 0.7 | 280.7 | 1.3 | 566 |
| 13.06 | -12.12 | 0.184 | 0.006 | 0.137 | 0.005 | -0.123 | 0.006 | 227.1 | 0.5 | 228.0 | 1.8 | 833 |
| 13.18 | 27.01 | 0.254 | 0.006 | 0.109 | 0.005 | 0.229 | 0.007 | 334.0 | 0.3 | 334.5 | 1.2 | 433 |
| 14.18 | 14.21 | 0.083 | 0.016 | 0.069 | 0.013 | 0.048 | 0.018 | 315.1 | 0.5 | 304.9 | 11.2 | — |
| 14.40 | 31.88 | 0.270 | 0.011 | 0.158 | 0.028 | 0.225 | 0.010 | 335.7 | 0.3 | 325.0 | 4.9 | 458 |
| 14.78 | 33.29 | 0.275 | 0.013 | 0.134 | 0.009 | 0.242 | 0.011 | 336.1 | 0.3 | 331.0 | 2.0 | — |
| 14.88 | -14.41 | 0.073 | 0.012 | 0.051 | 0.008 | -0.052 | 0.010 | 225.9 | 0.5 | 224.1 | 7.1 | -790 |

Table 8.1: continued

| $d_x{}^a$ (") | $d_y{}^a$ (") | $\mu$ (" yr$^{-1}$) | $\sigma_\mu$ (" yr$^{-1}$) | $\mu_x$ (" yr$^{-1}$) | $\sigma_{\mu_x}$ (" yr$^{-1}$) | $\mu_y$ (" yr$^{-1}$) | $\sigma_{\mu_y}$ (" yr$^{-1}$) | PA (°) | $\sigma_{PA}$ (°) | $\alpha$ (°) | $\sigma_\alpha$ (°) | $v_{rad}$ (km s$^{-1}$) |
|---|---|---|---|---|---|---|---|---|---|---|---|---|
| 16.15 | -39.35 | 0.387 | 0.015 | 0.150 | 0.006 | -0.357 | 0.016 | 202.3 | 0.2 | 202.8 | 1.2 | – |
| 16.22 | 19.52 | 0.213 | 0.004 | 0.141 | 0.005 | 0.160 | 0.006 | 320.3 | 0.4 | 318.7 | 1.4 | 516 |
| 16.31 | -22.67 | 0.236 | 0.005 | 0.141 | 0.003 | -0.190 | 0.005 | 215.7 | 0.3 | 216.5 | 0.9 | – |
| 16.43 | -34.49 | 0.308 | 0.003 | 0.131 | 0.004 | -0.278 | 0.003 | 205.5 | 0.2 | 205.2 | 0.7 | – |
| 16.64 | -25.83 | 0.290 | 0.007 | 0.132 | 0.003 | -0.260 | 0.007 | 212.8 | 0.3 | 206.9 | 0.8 | -484 |
| 17.06 | 4.85 | 0.148 | 0.013 | 0.144 | 0.012 | 0.035 | 0.017 | 285.9 | 0.5 | 283.5 | 6.5 | 395 |
| 17.19 | -6.26 | 0.088 | 0.011 | 0.064 | 0.012 | -0.080 | 0.012 | 250.0 | 0.5 | 218.5 | 6.6 | 194 |
| 17.30 | 51.99 | 0.357 | 0.007 | 0.111 | 0.005 | 0.339 | 0.008 | 341.6 | 0.2 | 341.8 | 0.9 | – |
| 17.52 | -15.72 | 0.232 | 0.008 | 0.147 | 0.005 | -0.184 | 0.011 | 228.1 | 0.4 | 218.7 | 1.8 | 627 |
| 18.18 | -10.10 | 0.196 | 0.007 | 0.158 | 0.007 | -0.119 | 0.005 | 240.9 | 0.5 | 233.1 | 1.8 | – |
| 18.32 | -43.31 | 0.491 | 0.021 | 0.178 | 0.012 | -0.457 | 0.021 | 202.9 | 0.2 | 201.3 | 1.6 | 607 |
| 19.21 | 24.16 | 0.309 | 0.016 | 0.214 | 0.009 | 0.224 | 0.015 | 321.5 | 0.3 | 316.3 | 2.3 | – |
| 19.32 | 30.49 | 0.251 | 0.009 | 0.120 | 0.010 | 0.221 | 0.006 | 327.6 | 0.3 | 331.5 | 2.1 | – |
| 19.98 | 46.20 | 0.424 | 0.011 | 0.164 | 0.006 | 0.392 | 0.011 | 336.6 | 0.2 | 337.3 | 0.9 | – |
| 20.00 | -3.77 | 0.128 | 0.007 | 0.120 | 0.007 | -0.053 | 0.006 | 259.3 | 0.5 | 246.1 | 2.9 | – |
| 20.95 | 6.31 | 0.198 | 0.012 | 0.171 | 0.020 | 0.116 | 0.036 | 286.8 | 0.4 | 304.1 | 8.7 | -803 |
| 21.41 | -33.64 | 0.213 | 0.009 | 0.111 | 0.012 | -0.181 | 0.007 | 212.5 | 0.2 | 211.6 | 2.9 | – |
| 21.79 | -18.63 | 0.174 | 0.007 | 0.136 | 0.006 | -0.108 | 0.007 | 229.5 | 0.3 | 231.5 | 2.1 | 289 |
| 21.79 | -18.63 | 0.174 | 0.007 | 0.136 | 0.006 | -0.108 | 0.007 | 229.5 | 0.3 | 231.5 | 2.1 | 440 |
| 21.87 | 46.81 | 0.460 | 0.009 | 0.233 | 0.006 | 0.399 | 0.008 | 335.0 | 0.2 | 329.8 | 0.8 | -166 |
| 22.17 | -13.33 | 0.233 | 0.008 | 0.189 | 0.008 | -0.138 | 0.006 | 239.0 | 0.4 | 233.8 | 1.7 | – |
| 22.64 | 4.42 | 0.172 | 0.003 | 0.168 | 0.003 | 0.037 | 0.006 | 281.1 | 0.4 | 282.3 | 2.0 | – |
| 22.83 | -30.88 | 0.370 | 0.004 | 0.196 | 0.006 | -0.315 | 0.003 | 216.5 | 0.2 | 211.9 | 0.8 | – |
| 23.08 | 1.57 | 0.200 | 0.003 | 0.200 | 0.003 | 0.006 | 0.005 | 273.9 | 0.4 | 271.6 | 1.4 | – |
| 23.30 | -25.23 | 0.329 | 0.007 | 0.219 | 0.006 | -0.246 | 0.007 | 222.7 | 0.3 | 221.6 | 1.1 | – |

Table 8.1: continued

| $d_x$ $^a$ (") | $d_y$ $^a$ (") | $\mu$ (" yr$^{-1}$) | $\sigma_\mu$ (" yr$^{-1}$) | $\mu_x$ (" yr$^{-1}$) | $\sigma_{\mu_x}$ (" yr$^{-1}$) | $\mu_y$ (" yr$^{-1}$) | $\sigma_{\mu_y}$ (" yr$^{-1}$) | PA (°) | $\sigma_{PA}$ (°) | $\alpha$ (°) | $\sigma_\alpha$ (°) | $v_{rad}$ (km s$^{-1}$) |
|---|---|---|---|---|---|---|---|---|---|---|---|---|
| 24.32 | -41.96 | 0.460 | 0.015 | 0.229 | 0.016 | -0.399 | 0.011 | 210.1 | 0.2 | 209.8 | 1.8 | -181 |
| 24.61 | 7.79 | 0.064 | 0.014 | 0.071 | 0.017 | -0.011 | 0.037 | 287.6 | 0.4 | 261.3 | 29.1 | 425 |
| 25.16 | -37.79 | 0.418 | 0.010 | 0.205 | 0.008 | -0.365 | 0.011 | 213.7 | 0.2 | 209.3 | 1.2 | – |
| 25.60 | 42.41 | 0.360 | 0.008 | 0.157 | 0.008 | 0.325 | 0.005 | 328.9 | 0.2 | 334.2 | 1.3 | – |
| 25.79 | -44.28 | 0.484 | 0.009 | 0.225 | 0.003 | -0.429 | 0.010 | 210.2 | 0.2 | 207.7 | 0.7 | – |
| 25.94 | 14.50 | 0.285 | 0.009 | 0.266 | 0.010 | 0.108 | 0.005 | 299.2 | 0.3 | 292.1 | 1.2 | – |
| 26.90 | -31.81 | 0.368 | 0.007 | 0.234 | 0.009 | -0.284 | 0.004 | 220.2 | 0.2 | 219.5 | 1.1 | – |
| 27.09 | 18.20 | 0.292 | 0.006 | 0.253 | 0.007 | 0.146 | 0.005 | 303.9 | 0.3 | 299.9 | 1.1 | – |
| 27.40 | -24.93 | 0.309 | 0.009 | 0.230 | 0.004 | -0.208 | 0.011 | 227.7 | 0.3 | 227.9 | 1.6 | -139 |
| 27.40 | -24.93 | 0.309 | 0.009 | 0.230 | 0.004 | -0.208 | 0.011 | 227.7 | 0.3 | 227.9 | 1.6 | -272 |
| 27.87 | -14.71 | 0.266 | 0.009 | 0.223 | 0.010 | -0.146 | 0.008 | 242.2 | 0.3 | 236.8 | 1.8 | – |
| 28.50 | 13.05 | 0.237 | 0.016 | 0.218 | 0.017 | 0.093 | 0.016 | 294.6 | 0.3 | 293.1 | 3.8 | 571 |
| 28.54 | 6.56 | 0.243 | 0.006 | 0.233 | 0.006 | 0.069 | 0.004 | 282.9 | 0.3 | 286.6 | 0.9 | – |
| 28.57 | 9.89 | 0.241 | 0.004 | 0.237 | 0.004 | 0.054 | 0.006 | 289.1 | 0.3 | 282.8 | 1.5 | – |
| 30.06 | 38.02 | 0.318 | 0.015 | 0.214 | 0.007 | 0.236 | 0.015 | 321.7 | 0.2 | 317.7 | 2.1 | – |
| 30.16 | -16.15 | 0.325 | 0.005 | 0.294 | 0.005 | -0.141 | 0.005 | 241.8 | 0.3 | 244.3 | 0.9 | – |
| 31.10 | 45.52 | 0.409 | 0.012 | 0.230 | 0.012 | 0.338 | 0.010 | 325.7 | 0.2 | 325.8 | 1.6 | – |
| 31.17 | -33.31 | 0.445 | 0.012 | 0.288 | 0.015 | -0.339 | 0.008 | 223.1 | 0.2 | 220.3 | 1.6 | – |
| 31.47 | 18.38 | 0.323 | 0.008 | 0.280 | 0.008 | 0.160 | 0.004 | 300.3 | 0.3 | 299.8 | 0.9 | – |
| 31.77 | 2.43 | 0.296 | 0.006 | 0.296 | 0.006 | 0.004 | 0.006 | 274.4 | 0.3 | 270.8 | 1.1 | – |
| 32.22 | -4.35 | 0.312 | 0.004 | 0.305 | 0.005 | -0.077 | 0.003 | 262.3 | 0.3 | 255.7 | 0.5 | – |
| 32.74 | -10.26 | 0.315 | 0.005 | 0.292 | 0.005 | -0.123 | 0.007 | 252.6 | 0.3 | 247.2 | 1.2 | – |
| 34.06 | 31.44 | 0.350 | 0.003 | 0.271 | 0.004 | 0.223 | 0.002 | 312.7 | 0.2 | 309.5 | 0.5 | 21 |
| 34.06 | 31.44 | 0.350 | 0.003 | 0.271 | 0.004 | 0.223 | 0.002 | 312.7 | 0.2 | 309.5 | 0.5 | 275 |
| 34.19 | 34.98 | 0.471 | 0.007 | 0.329 | 0.007 | 0.337 | 0.007 | 315.7 | 0.2 | 315.7 | 0.9 | – |

Table 8.1: continued

| $d_x{}^a$ ('') | $d_y{}^a$ ('') | $\mu$ ('' yr$^{-1}$) | $\sigma_\mu$ ('' yr$^{-1}$) | $\mu_x$ ('' yr$^{-1}$) | $\sigma_{\mu_x}$ ('' yr$^{-1}$) | $\mu_y$ ('' yr$^{-1}$) | $\sigma_{\mu_y}$ ('' yr$^{-1}$) | PA (°) | $\sigma_{PA}$ (°) | $\alpha$ (°) | $\sigma_\alpha$ (°) | $v_{rad}$ (km s$^{-1}$) |
|---|---|---|---|---|---|---|---|---|---|---|---|---|
| 34.31 | 42.57 | 0.495 | 0.009 | 0.316 | 0.010 | 0.382 | 0.009 | 321.1 | 0.2 | 320.4 | 1.1 | – |
| 34.49 | -2.28 | 0.262 | 0.009 | 0.263 | 0.010 | 0.002 | 0.007 | 266.2 | 0.3 | 270.5 | 1.5 | -570 |
| 35.01 | -16.72 | 0.307 | 0.005 | 0.281 | 0.005 | -0.125 | 0.003 | 244.5 | 0.2 | 246.0 | 0.6 | – |
| 35.13 | -11.90 | 0.302 | 0.013 | 0.275 | 0.015 | -0.129 | 0.008 | 251.3 | 0.3 | 244.9 | 1.8 | – |
| 35.39 | 13.48 | 0.373 | 0.020 | 0.365 | 0.021 | 0.092 | 0.009 | 290.9 | 0.3 | 284.1 | 1.6 | – |
| 35.50 | 18.63 | 0.373 | 0.019 | 0.327 | 0.017 | 0.180 | 0.013 | 297.7 | 0.2 | 298.9 | 2.1 | – |
| 35.99 | 10.85 | 0.190 | 0.040 | 0.175 | 0.041 | 0.078 | 0.011 | 286.8 | 0.3 | 294.0 | 5.9 | -739 |
| 36.12 | 34.59 | 0.425 | 0.006 | 0.326 | 0.007 | 0.274 | 0.004 | 313.8 | 0.2 | 310.0 | 0.7 | – |
| 36.74 | 15.19 | 0.325 | 0.007 | 0.278 | 0.007 | 0.178 | 0.008 | 292.5 | 0.2 | 302.6 | 1.3 | – |
| 36.90 | -15.25 | 0.361 | 0.011 | 0.319 | 0.010 | -0.173 | 0.009 | 247.5 | 0.2 | 241.5 | 1.5 | 203 |
| 36.90 | -15.25 | 0.361 | 0.011 | 0.319 | 0.010 | -0.173 | 0.009 | 247.5 | 0.2 | 241.5 | 1.5 | 293 |
| 36.90 | -15.25 | 0.361 | 0.011 | 0.319 | 0.010 | -0.173 | 0.009 | 247.5 | 0.2 | 241.5 | 1.5 | 399 |
| 37.21 | -4.90 | 0.307 | 0.005 | 0.305 | 0.006 | -0.038 | 0.006 | 262.5 | 0.3 | 262.8 | 1.1 | – |
| 38.26 | 26.07 | 0.369 | 0.011 | 0.321 | 0.009 | 0.183 | 0.011 | 304.3 | 0.2 | 299.6 | 1.6 | – |
| 38.27 | -33.47 | 0.474 | 0.011 | 0.355 | 0.012 | -0.314 | 0.007 | 228.8 | 0.2 | 228.5 | 1.2 | -17 |
| 38.53 | 1.17 | 0.369 | 0.013 | 0.369 | 0.013 | -0.010 | 0.006 | 271.7 | 0.2 | 268.4 | 0.9 | – |
| 38.67 | 39.17 | 0.479 | 0.006 | 0.367 | 0.008 | 0.311 | 0.006 | 315.4 | 0.2 | 310.2 | 0.8 | -266 |
| 38.82 | 23.53 | 0.388 | 0.009 | 0.358 | 0.008 | 0.159 | 0.013 | 301.2 | 0.2 | 293.9 | 1.8 | – |
| 38.89 | -35.30 | 0.488 | 0.009 | 0.339 | 0.006 | -0.352 | 0.010 | 227.8 | 0.2 | 224.0 | 1.0 | 60 |
| 39.14 | 17.67 | 0.363 | 0.007 | 0.325 | 0.008 | 0.161 | 0.009 | 294.3 | 0.2 | 296.3 | 1.3 | – |
| 39.35 | 8.00 | 0.330 | 0.007 | 0.316 | 0.007 | 0.101 | 0.007 | 281.5 | 0.2 | 287.7 | 1.1 | – |
| 39.58 | -21.12 | 0.393 | 0.007 | 0.338 | 0.007 | -0.200 | 0.004 | 241.9 | 0.2 | 239.4 | 0.7 | – |
| 40.22 | -15.32 | 0.304 | 0.017 | 0.267 | 0.015 | -0.152 | 0.012 | 249.1 | 0.2 | 240.4 | 2.4 | 100 |
| 40.22 | -15.32 | 0.304 | 0.017 | 0.267 | 0.015 | -0.152 | 0.012 | 249.1 | 0.2 | 240.4 | 2.4 | 306 |
| 40.38 | 30.60 | 0.370 | 0.007 | 0.291 | 0.009 | 0.229 | 0.003 | 307.2 | 0.2 | 308.2 | 1.0 | – |

Table 8.1: continued

| $d_x{}^a$ ('') | $d_y{}^a$ ('') | $\mu$ ('' yr$^{-1}$) | $\sigma_\mu$ ('' yr$^{-1}$) | $\mu_x$ ('' yr$^{-1}$) | $\sigma_{\mu_x}$ ('' yr$^{-1}$) | $\mu_y$ ('' yr$^{-1}$) | $\sigma_{\mu_y}$ ('' yr$^{-1}$) | PA (°) | $\sigma_{PA}$ (°) | $\alpha$ (°) | $\sigma_\alpha$ (°) | $v_{rad}$ (km s$^{-1}$) |
|---|---|---|---|---|---|---|---|---|---|---|---|---|
| 40.55 | 17.23 | 0.322 | 0.007 | 0.311 | 0.007 | 0.093 | 0.006 | 293.0 | 0.2 | 286.6 | 1.1 | -229 |
| 41.89 | 19.85 | 0.274 | 0.011 | 0.259 | 0.009 | 0.092 | 0.009 | 295.4 | 0.2 | 289.7 | 2.0 | – |
| 42.64 | 10.70 | 0.408 | 0.013 | 0.404 | 0.014 | 0.069 | 0.024 | 284.1 | 0.2 | 279.7 | 3.4 | – |
| 43.38 | -27.80 | 0.477 | 0.009 | 0.397 | 0.008 | -0.265 | 0.006 | 237.3 | 0.2 | 236.3 | 0.8 | – |
| 44.49 | 3.87 | 0.381 | 0.004 | 0.377 | 0.004 | 0.067 | 0.005 | 275.0 | 0.2 | 280.1 | 0.8 | – |
| 45.34 | -17.09 | 0.410 | 0.032 | 0.381 | 0.030 | -0.153 | 0.012 | 249.3 | 0.2 | 248.2 | 2.2 | 122 |
| 45.73 | 10.83 | 0.277 | 0.014 | 0.269 | 0.014 | 0.067 | 0.003 | 283.3 | 0.2 | 284.0 | 0.9 | – |
| 46.52 | 16.20 | 0.425 | 0.009 | 0.408 | 0.009 | 0.123 | 0.003 | 289.2 | 0.2 | 286.7 | 0.6 | – |
| 46.80 | 21.34 | 0.414 | 0.005 | 0.394 | 0.004 | 0.135 | 0.004 | 294.5 | 0.2 | 288.9 | 0.6 | -115 |
| 47.18 | -13.35 | 0.413 | 0.008 | 0.392 | 0.007 | -0.130 | 0.006 | 254.2 | 0.2 | 251.6 | 0.8 | – |
| 48.33 | -5.42 | 0.366 | 0.006 | 0.362 | 0.005 | -0.056 | 0.004 | 263.6 | 0.2 | 261.1 | 0.7 | – |
| 48.73 | 16.67 | 0.435 | 0.008 | 0.416 | 0.008 | 0.128 | 0.004 | 288.9 | 0.2 | 287.1 | 0.6 | – |
| 50.11 | -2.66 | 0.326 | 0.009 | 0.323 | 0.009 | -0.052 | 0.004 | 267.0 | 0.2 | 260.9 | 0.8 | -266 |
| 52.07 | 11.58 | 0.467 | 0.007 | 0.457 | 0.006 | 0.094 | 0.004 | 282.5 | 0.2 | 281.6 | 0.5 | – |
| 52.20 | -3.77 | 0.530 | 0.019 | 0.526 | 0.018 | -0.074 | 0.005 | 265.9 | 0.2 | 262.0 | 0.6 | 243 |
| 52.21 | 16.82 | 0.428 | 0.021 | 0.408 | 0.021 | 0.130 | 0.008 | 287.9 | 0.2 | 287.7 | 1.3 | -53 |
| 53.57 | 1.49 | 0.449 | 0.008 | 0.449 | 0.008 | 0.008 | 0.013 | 271.6 | 0.2 | 271.1 | 1.6 | – |
| 53.97 | 5.35 | 0.488 | 0.010 | 0.487 | 0.010 | 0.025 | 0.005 | 275.7 | 0.2 | 273.0 | 0.5 | – |
| -48.11 | 0.72 | 0.482 | 0.012 | -0.482 | 0.012 | -0.012 | 0.009 | 89.1 | 0.2 | 91.4 | 1.0 | – |
| -47.83 | 11.48 | 0.327 | 0.035 | -0.323 | 0.035 | 0.053 | 0.015 | 76.5 | 0.2 | 80.7 | 2.7 | – |
| -47.62 | 9.07 | 0.367 | 0.011 | -0.358 | 0.011 | 0.079 | 0.009 | 79.2 | 0.2 | 77.6 | 1.4 | – |
| -45.79 | 15.85 | 0.357 | 0.009 | -0.343 | 0.009 | 0.101 | 0.009 | 70.9 | 0.2 | 73.7 | 1.5 | – |
| -44.70 | 9.33 | 0.403 | 0.009 | -0.404 | 0.008 | 0.038 | 0.009 | 78.2 | 0.2 | 84.6 | 1.2 | – |
| -44.18 | -21.82 | 0.458 | 0.016 | -0.403 | 0.014 | -0.219 | 0.014 | 116.3 | 0.2 | 118.6 | 1.7 | – |
| -42.39 | -13.27 | 0.428 | 0.008 | -0.393 | 0.006 | -0.176 | 0.013 | 107.4 | 0.2 | 114.0 | 1.6 | 503 |

Table 8.1: continued

| $d_x$ $^a$ (") | $d_y$ $^a$ (") | $\mu$ (" yr$^{-1}$) | $\sigma_\mu$ (" yr$^{-1}$) | $\mu_x$ (" yr$^{-1}$) | $\sigma_{\mu_x}$ (" yr$^{-1}$) | $\mu_y$ (" yr$^{-1}$) | $\sigma_{\mu_y}$ (" yr$^{-1}$) | PA (°) | $\sigma_{PA}$ (°) | $\alpha$ (°) | $\sigma_\alpha$ (°) | $v_{rad}$ (km s$^{-1}$) |
|---|---|---|---|---|---|---|---|---|---|---|---|---|
| -42.39 | -13.27 | 0.428 | 0.008 | -0.393 | 0.006 | -0.176 | 0.013 | 107.4 | 0.2 | 114.0 | 1.6 | 580 |
| -42.27 | 4.74 | 0.340 | 0.011 | -0.338 | 0.011 | 0.036 | 0.007 | 83.6 | 0.2 | 83.9 | 1.2 | – |
| -42.16 | 15.42 | 0.412 | 0.007 | -0.387 | 0.007 | 0.142 | 0.005 | 69.9 | 0.2 | 69.8 | 0.7 | 464 |
| -41.05 | -4.03 | 0.369 | 0.023 | -0.367 | 0.022 | -0.041 | 0.018 | 95.6 | 0.2 | 96.4 | 2.9 | – |
| -39.94 | 21.70 | 0.381 | 0.004 | -0.345 | 0.004 | 0.164 | 0.002 | 61.5 | 0.2 | 64.6 | 0.4 | – |
| -39.82 | -12.31 | 0.415 | 0.023 | -0.401 | 0.023 | -0.108 | 0.006 | 107.2 | 0.2 | 105.0 | 1.1 | 573 |
| -39.63 | -27.39 | 0.483 | 0.014 | -0.416 | 0.016 | -0.249 | 0.007 | 124.6 | 0.2 | 120.9 | 1.2 | – |
| -39.33 | -16.96 | 0.355 | 0.012 | -0.322 | 0.013 | -0.150 | 0.007 | 113.3 | 0.2 | 115.0 | 1.4 | 538 |
| -39.29 | 0.77 | 0.352 | 0.005 | -0.352 | 0.005 | -0.017 | 0.004 | 88.9 | 0.2 | 92.8 | 0.7 | – |
| -39.12 | -40.06 | 0.340 | 0.025 | -0.241 | 0.024 | -0.240 | 0.019 | 135.7 | 0.2 | 134.9 | 3.6 | – |
| -38.44 | 6.73 | 0.335 | 0.006 | -0.334 | 0.006 | 0.040 | 0.004 | 80.1 | 0.2 | 83.2 | 0.7 | – |
| -38.44 | -29.35 | 0.448 | 0.007 | -0.376 | 0.005 | -0.246 | 0.006 | 127.4 | 0.2 | 123.2 | 0.7 | – |
| -38.36 | -20.16 | 0.387 | 0.009 | -0.350 | 0.008 | -0.166 | 0.007 | 117.7 | 0.2 | 115.4 | 1.1 | – |
| -36.63 | 30.90 | 0.272 | 0.014 | -0.215 | 0.012 | 0.167 | 0.011 | 49.9 | 0.2 | 52.2 | 2.3 | 68 |
| -36.62 | -14.56 | 0.410 | 0.009 | -0.373 | 0.010 | -0.170 | 0.003 | 111.7 | 0.2 | 114.6 | 0.7 | 527 |
| -35.72 | -40.20 | 0.423 | 0.005 | -0.272 | 0.006 | -0.324 | 0.007 | 138.4 | 0.2 | 140.0 | 0.8 | – |
| -35.29 | 2.36 | 0.323 | 0.028 | -0.324 | 0.028 | -0.014 | 0.012 | 86.2 | 0.3 | 92.5 | 2.1 | -516 |
| -35.24 | -27.87 | 0.424 | 0.010 | -0.343 | 0.012 | -0.249 | 0.009 | 128.3 | 0.2 | 126.0 | 1.3 | – |
| -34.98 | 8.57 | 0.306 | 0.007 | -0.303 | 0.007 | 0.047 | 0.002 | 76.2 | 0.3 | 81.1 | 0.5 | – |
| -34.74 | -34.93 | 0.498 | 0.013 | -0.382 | 0.009 | -0.323 | 0.018 | 135.2 | 0.2 | 130.2 | 1.7 | – |
| -33.46 | -11.10 | 0.217 | 0.013 | -0.205 | 0.013 | -0.073 | 0.006 | 108.3 | 0.3 | 109.6 | 1.8 | -558 |
| -32.90 | 11.20 | 0.317 | 0.008 | -0.301 | 0.007 | 0.101 | 0.010 | 71.2 | 0.3 | 71.4 | 1.7 | – |
| -32.76 | 15.16 | 0.263 | 0.007 | -0.237 | 0.009 | 0.113 | 0.004 | 65.2 | 0.3 | 64.5 | 1.1 | – |
| -32.17 | 28.67 | 0.360 | 0.012 | -0.278 | 0.013 | 0.228 | 0.015 | 48.3 | 0.2 | 50.7 | 2.3 | 102 |
| -31.12 | 15.50 | 0.317 | 0.008 | -0.280 | 0.009 | 0.149 | 0.007 | 63.5 | 0.3 | 62.0 | 1.4 | – |

Table 8.1: continued

| $d_x{}^a$ (") | $d_y{}^a$ (") | $\mu$ (" $yr^{-1}$) | $\sigma_\mu$ (" $yr^{-1}$) | $\mu_x$ (" $yr^{-1}$) | $\sigma_{\mu_x}$ (" $yr^{-1}$) | $\mu_y$ (" $yr^{-1}$) | $\sigma_{\mu_y}$ (" $yr^{-1}$) | PA (°) | $\sigma_{PA}$ (°) | $\alpha$ (°) | $\sigma_\alpha$ (°) | $v_{rad}$ (km s$^{-1}$) |
|---|---|---|---|---|---|---|---|---|---|---|---|---|
| -31.09 | -35.97 | 0.310 | 0.007 | -0.214 | 0.008 | -0.224 | 0.004 | 139.2 | 0.2 | 136.3 | 1.2 | – |
| -30.92 | 33.13 | 0.345 | 0.010 | -0.270 | 0.011 | 0.219 | 0.007 | 43.0 | 0.2 | 50.9 | 1.4 | – |
| -30.69 | 20.98 | 0.308 | 0.015 | -0.265 | 0.011 | 0.159 | 0.017 | 55.6 | 0.3 | 59.0 | 2.9 | 339 |
| -29.93 | -27.89 | 0.413 | 0.013 | -0.289 | 0.017 | -0.295 | 0.005 | 133.0 | 0.2 | 135.5 | 1.8 | 339 |
| -29.42 | 25.69 | 0.284 | 0.006 | -0.235 | 0.005 | 0.162 | 0.006 | 48.9 | 0.2 | 55.4 | 1.2 | 473 |
| -29.14 | -39.58 | 0.477 | 0.009 | -0.313 | 0.010 | -0.362 | 0.008 | 143.6 | 0.2 | 139.2 | 1.1 | – |
| -28.91 | 3.87 | 0.251 | 0.009 | -0.251 | 0.009 | 0.018 | 0.004 | 82.4 | 0.3 | 85.9 | 1.0 | – |
| -28.69 | 32.40 | 0.341 | 0.005 | -0.234 | 0.007 | 0.248 | 0.005 | 41.5 | 0.2 | 43.3 | 1.0 | – |
| -28.37 | -24.73 | 0.313 | 0.006 | -0.235 | 0.005 | -0.207 | 0.006 | 131.1 | 0.3 | 131.4 | 1.0 | 306 |
| -28.37 | -24.73 | 0.313 | 0.006 | -0.235 | 0.005 | -0.207 | 0.006 | 131.1 | 0.3 | 131.4 | 1.0 | 449 |
| -28.37 | -24.73 | 0.313 | 0.006 | -0.235 | 0.005 | -0.207 | 0.006 | 131.1 | 0.3 | 131.4 | 1.0 | 539 |
| -28.17 | 34.77 | 0.332 | 0.010 | -0.199 | 0.008 | 0.265 | 0.012 | 39.0 | 0.2 | 36.9 | 1.7 | – |
| -27.85 | -3.90 | 0.252 | 0.004 | -0.249 | 0.004 | -0.038 | 0.005 | 98.0 | 0.3 | 98.7 | 1.1 | – |
| -27.69 | -38.19 | 0.409 | 0.008 | -0.246 | 0.008 | -0.327 | 0.005 | 144.1 | 0.2 | 143.1 | 1.0 | – |
| -27.43 | -14.13 | 0.207 | 0.008 | -0.184 | 0.009 | -0.095 | 0.003 | 117.3 | 0.3 | 117.2 | 1.3 | – |
| -26.78 | 24.45 | 0.317 | 0.010 | -0.246 | 0.010 | 0.201 | 0.006 | 47.6 | 0.3 | 50.7 | 1.4 | 414 |
| -26.12 | 33.78 | 0.275 | 0.002 | -0.186 | 0.003 | 0.204 | 0.002 | 37.7 | 0.2 | 42.4 | 0.5 | – |
| -25.74 | -23.09 | 0.289 | 0.004 | -0.232 | 0.003 | -0.174 | 0.005 | 131.9 | 0.3 | 126.8 | 0.9 | 478 |
| -25.60 | -28.70 | 0.349 | 0.004 | -0.243 | 0.004 | -0.251 | 0.004 | 138.3 | 0.2 | 135.9 | 0.7 | – |
| -25.41 | 28.27 | 0.311 | 0.007 | -0.238 | 0.009 | 0.205 | 0.005 | 42.0 | 0.3 | 49.2 | 1.3 | – |
| -25.40 | -15.54 | 0.275 | 0.009 | -0.230 | 0.008 | -0.151 | 0.004 | 121.5 | 0.3 | 123.3 | 1.2 | – |
| -25.23 | -25.20 | 0.222 | 0.007 | -0.125 | 0.004 | -0.188 | 0.008 | 135.0 | 0.3 | 146.4 | 1.4 | -570 |
| -25.16 | -36.99 | 0.397 | 0.006 | -0.229 | 0.005 | -0.324 | 0.005 | 145.8 | 0.2 | 144.7 | 0.7 | – |
| -24.00 | -29.71 | 0.266 | 0.011 | -0.104 | 0.011 | -0.258 | 0.010 | 141.1 | 0.2 | 158.1 | 2.2 | – |
| -23.96 | -22.18 | 0.316 | 0.009 | -0.244 | 0.010 | -0.201 | 0.005 | 132.8 | 0.3 | 129.5 | 1.3 | 531 |

Table 8.1: continued

| $d_x$[a] (") | $d_y$[a] (") | $\mu$ (" yr⁻¹) | $\sigma_\mu$ (" yr⁻¹) | $\mu_x$ (" yr⁻¹) | $\sigma_{\mu_x}$ (" yr⁻¹) | $\mu_y$ (" yr⁻¹) | $\sigma_{\mu_y}$ (" yr⁻¹) | PA (°) | $\sigma_{PA}$ (°) | $\alpha$ (°) | $\sigma_\alpha$ (°) | $v_{rad}$ (km s⁻¹) |
|---|---|---|---|---|---|---|---|---|---|---|---|---|
| -23.62 | -39.79 | 0.387 | 0.005 | -0.186 | 0.004 | -0.340 | 0.005 | 149.3 | 0.2 | 151.3 | 0.6 | – |
| -22.87 | -19.92 | 0.299 | 0.020 | -0.219 | 0.013 | -0.203 | 0.020 | 131.1 | 0.3 | 132.8 | 3.2 | 379 |
| -22.44 | -51.15 | 0.509 | 0.022 | -0.197 | 0.010 | -0.469 | 0.024 | 156.3 | 0.2 | 157.3 | 1.5 | – |
| -21.72 | 35.88 | 0.279 | 0.010 | -0.161 | 0.004 | 0.228 | 0.010 | 31.2 | 0.2 | 35.2 | 1.4 | 7 |
| -21.47 | -34.95 | 0.366 | 0.007 | -0.183 | 0.006 | -0.317 | 0.007 | 148.4 | 0.2 | 150.0 | 0.9 | – |
| -21.25 | -1.99 | 0.129 | 0.004 | -0.130 | 0.003 | 0.008 | 0.011 | 95.4 | 0.4 | 86.5 | 4.8 | – |
| -21.11 | -20.80 | 0.292 | 0.006 | -0.194 | 0.003 | -0.219 | 0.009 | 134.6 | 0.3 | 138.5 | 1.2 | 612 |
| -20.60 | -17.02 | 0.222 | 0.006 | -0.161 | 0.005 | -0.154 | 0.005 | 129.6 | 0.4 | 133.6 | 1.3 | -650 |
| -20.00 | -47.55 | 0.388 | 0.011 | -0.129 | 0.006 | -0.366 | 0.010 | 157.2 | 0.2 | 160.7 | 1.0 | – |
| -19.36 | 13.20 | 0.178 | 0.015 | -0.145 | 0.011 | 0.104 | 0.018 | 55.7 | 0.4 | 54.3 | 5.2 | – |
| -18.89 | -37.04 | 0.311 | 0.006 | -0.147 | 0.003 | -0.274 | 0.007 | 153.0 | 0.4 | 151.7 | 0.8 | 506 |
| -18.54 | 27.33 | 0.136 | 0.010 | -0.068 | 0.008 | 0.118 | 0.008 | 34.2 | 0.3 | 30.0 | 3.5 | – |
| -18.49 | 3.29 | 0.144 | 0.006 | -0.145 | 0.006 | 0.007 | 0.003 | 79.9 | 0.5 | 87.1 | 1.2 | – |
| -18.34 | -15.25 | 0.240 | 0.009 | -0.202 | 0.008 | -0.131 | 0.006 | 129.8 | 0.4 | 123.0 | 1.7 | -727 |
| -18.23 | 39.72 | 0.422 | 0.016 | -0.171 | 0.009 | 0.386 | 0.015 | 24.7 | 0.2 | 23.9 | 1.4 | – |
| -17.71 | -38.60 | 0.299 | 0.008 | -0.110 | 0.007 | -0.279 | 0.009 | 155.4 | 0.2 | 158.4 | 1.4 | -309 |
| -16.26 | -32.59 | 0.292 | 0.004 | -0.131 | 0.005 | -0.261 | 0.003 | 153.5 | 0.3 | 153.4 | 0.9 | – |
| -15.73 | -47.49 | 0.433 | 0.011 | -0.171 | 0.018 | -0.399 | 0.007 | 161.7 | 0.2 | 156.8 | 2.2 | – |
| -15.40 | 12.58 | 0.191 | 0.006 | -0.138 | 0.006 | 0.133 | 0.006 | 50.8 | 0.5 | 46.1 | 1.9 | -791 |
| -15.26 | -7.12 | 0.145 | 0.010 | -0.120 | 0.010 | -0.086 | 0.006 | 115.0 | 0.6 | 125.5 | 2.9 | 673 |
| -15.19 | 39.83 | 0.314 | 0.006 | -0.128 | 0.010 | 0.287 | 0.005 | 20.9 | 0.2 | 24.1 | 1.7 | – |
| -15.09 | 42.96 | 0.357 | 0.014 | -0.151 | 0.008 | 0.325 | 0.013 | 19.4 | 0.2 | 25.0 | 1.4 | – |
| -15.01 | -35.54 | 0.297 | 0.004 | -0.099 | 0.005 | -0.280 | 0.004 | 157.1 | 0.2 | 160.6 | 0.9 | – |
| -14.93 | 3.58 | 0.147 | 0.007 | -0.140 | 0.004 | 0.049 | 0.020 | 76.5 | 0.6 | 70.7 | 7.3 | -847 |
| -14.78 | -43.31 | 0.427 | 0.007 | -0.135 | 0.006 | -0.405 | 0.007 | 161.2 | 0.2 | 161.6 | 0.8 | – |

Table 8.1: continued

| $d_x{}^a$ ($''$) | $d_y{}^a$ ($''$) | $\mu$ ($'' yr^{-1}$) | $\sigma_\mu$ ($'' yr^{-1}$) | $\mu_x$ ($'' yr^{-1}$) | $\sigma_{\mu_x}$ ($'' yr^{-1}$) | $\mu_y$ ($'' yr^{-1}$) | $\sigma_{\mu_y}$ ($'' yr^{-1}$) | PA ($°$) | $\sigma_{PA}$ ($°$) | $\alpha$ ($°$) | $\sigma_\alpha$ ($°$) | $v_{rad}$ (km s$^{-1}$) |
|---|---|---|---|---|---|---|---|---|---|---|---|---|
| -14.65 | 30.60 | 0.245 | 0.005 | -0.108 | 0.003 | 0.220 | 0.006 | 25.6 | 0.3 | 26.2 | 0.9 | – |
| -14.59 | 35.62 | 0.257 | 0.026 | -0.100 | 0.013 | 0.236 | 0.030 | 22.3 | 0.2 | 22.9 | 3.7 | -117 |
| -14.15 | -46.07 | 0.438 | 0.010 | -0.136 | 0.007 | -0.416 | 0.009 | 162.9 | 0.2 | 161.8 | 0.9 | – |
| -14.14 | 41.32 | 0.244 | 0.011 | -0.113 | 0.003 | 0.219 | 0.012 | 18.9 | 0.2 | 27.4 | 1.4 | – |
| -13.74 | -14.44 | 0.184 | 0.006 | -0.100 | 0.007 | -0.158 | 0.006 | 136.4 | 0.5 | 147.7 | 2.0 | 686 |
| -13.74 | -14.44 | 0.184 | 0.006 | -0.100 | 0.007 | -0.158 | 0.006 | 136.4 | 0.5 | 147.7 | 2.0 | 827 |
| -13.44 | 0.88 | 0.158 | 0.022 | -0.159 | 0.022 | -0.012 | 0.005 | 86.3 | 0.7 | 94.4 | 1.8 | 681 |
| -13.44 | 0.88 | 0.158 | 0.022 | -0.159 | 0.022 | -0.012 | 0.005 | 86.3 | 0.7 | 94.4 | 1.8 | 769 |
| -13.43 | -12.52 | 0.096 | 0.005 | -0.073 | 0.004 | -0.062 | 0.008 | 133.0 | 0.5 | 130.2 | 3.8 | -569 |
| -13.27 | 25.14 | 0.201 | 0.014 | -0.096 | 0.016 | 0.177 | 0.016 | 27.8 | 0.3 | 28.3 | 4.6 | – |
| -12.97 | 19.80 | 0.183 | 0.005 | -0.100 | 0.005 | 0.153 | 0.004 | 33.2 | 0.4 | 33.3 | 1.6 | 702 |
| -12.97 | 43.42 | 0.329 | 0.015 | -0.104 | 0.016 | 0.312 | 0.013 | 16.6 | 0.2 | 18.5 | 2.8 | – |
| -12.89 | 4.36 | 0.133 | 0.011 | -0.136 | 0.013 | 0.012 | 0.004 | 71.3 | 0.7 | 84.9 | 1.8 | 661 |
| -12.89 | 4.36 | 0.133 | 0.011 | -0.136 | 0.013 | 0.012 | 0.004 | 71.3 | 0.7 | 84.9 | 1.8 | 790 |
| -12.87 | 39.46 | 0.278 | 0.005 | -0.115 | 0.005 | 0.255 | 0.004 | 18.1 | 0.7 | 24.4 | 1.0 | – |
| -12.73 | 14.72 | 0.165 | 0.012 | -0.109 | 0.010 | 0.124 | 0.011 | 40.9 | 0.5 | 41.2 | 3.5 | – |
| -12.69 | -19.68 | 0.203 | 0.005 | -0.077 | 0.010 | -0.191 | 0.006 | 147.2 | 0.4 | 158.1 | 2.6 | – |
| -12.68 | -2.82 | 0.154 | 0.006 | -0.141 | 0.006 | -0.069 | 0.003 | 102.6 | 0.7 | 116.1 | 1.5 | 675 |
| -12.29 | 37.13 | 0.337 | 0.006 | -0.130 | 0.006 | 0.311 | 0.007 | 18.3 | 0.2 | 22.6 | 1.0 | – |
| -12.25 | 0.56 | 0.091 | 0.015 | -0.091 | 0.015 | -0.002 | 0.012 | 87.4 | 0.8 | 91.4 | 7.4 | 752 |
| -11.38 | 45.30 | 0.391 | 0.010 | -0.068 | 0.009 | 0.386 | 0.009 | 14.1 | 0.2 | 10.0 | 1.4 | – |
| -10.91 | -31.85 | 0.318 | 0.005 | -0.101 | 0.008 | -0.301 | 0.004 | 161.1 | 0.3 | 161.5 | 1.3 | – |
| -10.77 | -2.93 | 0.109 | 0.006 | -0.095 | 0.005 | -0.066 | 0.004 | 105.2 | 0.9 | 124.6 | 2.2 | -832 |
| -10.10 | -24.82 | 0.234 | 0.006 | -0.080 | 0.006 | -0.220 | 0.008 | 157.9 | 0.4 | 160.0 | 1.5 | -602 |
| -9.78 | -52.14 | 0.484 | 0.011 | -0.093 | 0.004 | -0.475 | 0.011 | 169.4 | 0.2 | 168.9 | 0.6 | – |

Table 8.1: continued

| $d_x{}^a$ (") | $d_y{}^a$ (") | $\mu$ (" yr$^{-1}$) | $\sigma_\mu$ (" yr$^{-1}$) | $\mu_x$ (" yr$^{-1}$) | $\sigma_{\mu_x}$ (" yr$^{-1}$) | $\mu_y$ (" yr$^{-1}$) | $\sigma_{\mu_y}$ (" yr$^{-1}$) | PA (°) | $\sigma_{PA}$ (°) | $\alpha$ (°) | $\sigma_\alpha$ (°) | $v_{rad}$ (km s$^{-1}$) |
|---|---|---|---|---|---|---|---|---|---|---|---|---|
| -9.75 | -41.20 | 0.345 | 0.013 | -0.050 | 0.012 | -0.342 | 0.011 | 166.7 | 0.2 | 171.6 | 2.0 | – |
| -9.67 | -0.07 | 0.070 | 0.015 | -0.070 | 0.015 | -0.018 | 0.001 | 90.4 | 1.0 | 104.1 | 3.1 | 771 |
| -9.60 | 32.37 | 0.206 | 0.004 | -0.059 | 0.004 | 0.197 | 0.004 | 16.5 | 0.3 | 16.8 | 1.2 | – |
| -9.56 | 36.39 | 0.308 | 0.005 | -0.084 | 0.002 | 0.297 | 0.005 | 14.7 | 0.3 | 15.9 | 0.5 | – |
| -9.32 | 15.18 | 0.096 | 0.015 | -0.026 | 0.007 | 0.096 | 0.015 | 31.5 | 0.5 | 15.2 | 4.7 | 511 |
| -9.30 | 7.36 | 0.067 | 0.005 | -0.054 | 0.005 | 0.040 | 0.004 | 51.6 | 0.8 | 53.3 | 3.5 | -989 |
| -9.17 | 10.21 | 0.100 | 0.007 | -0.054 | 0.007 | 0.086 | 0.008 | 41.9 | 0.7 | 32.1 | 4.0 | – |
| -8.53 | -13.06 | 0.160 | 0.005 | -0.077 | 0.004 | -0.141 | 0.004 | 146.8 | 0.6 | 151.3 | 1.4 | – |
| -8.51 | 3.51 | 0.083 | 0.004 | -0.080 | 0.004 | 0.023 | 0.005 | 67.6 | 1.0 | 73.7 | 3.2 | -926 |
| -8.34 | 13.19 | 0.127 | 0.005 | -0.044 | 0.006 | 0.122 | 0.005 | 32.3 | 0.6 | 19.8 | 2.4 | -924 |
| -8.28 | -14.98 | 0.191 | 0.009 | -0.063 | 0.011 | -0.183 | 0.008 | 151.1 | 0.6 | 160.9 | 3.2 | 840 |
| -7.82 | -46.55 | 0.418 | 0.006 | -0.111 | 0.005 | -0.405 | 0.007 | 170.5 | 0.2 | 164.6 | 0.7 | – |
| -7.28 | -2.41 | 0.071 | 0.004 | -0.054 | 0.004 | -0.061 | 0.003 | 108.3 | 1.2 | 138.6 | 2.6 | -825 |
| -6.32 | 46.40 | 0.367 | 0.005 | -0.069 | 0.009 | 0.361 | 0.006 | 7.8 | 0.2 | 10.8 | 1.4 | – |
| -6.07 | 42.84 | 0.349 | 0.010 | -0.050 | 0.006 | 0.345 | 0.010 | 8.1 | 0.2 | 8.2 | 1.1 | – |
| -5.91 | -18.82 | 0.181 | 0.015 | -0.045 | 0.009 | -0.176 | 0.015 | 162.6 | 0.5 | 165.7 | 3.1 | – |
| -5.74 | -52.55 | 0.454 | 0.015 | -0.045 | 0.010 | -0.451 | 0.014 | 173.8 | 0.2 | 174.3 | 1.3 | 260 |
| -5.30 | -17.85 | 0.090 | 0.008 | -0.034 | 0.012 | -0.083 | 0.006 | 163.4 | 0.5 | 157.8 | 7.4 | – |
| -4.30 | 14.26 | 0.108 | 0.014 | -0.025 | 0.004 | 0.105 | 0.015 | 16.8 | 0.6 | 13.6 | 2.7 | – |
| -3.74 | -41.69 | 0.361 | 0.007 | -0.021 | 0.003 | -0.360 | 0.007 | 174.9 | 0.2 | 176.7 | 0.5 | -470 |
| -3.12 | -4.02 | 0.113 | 0.032 | -0.065 | 0.015 | -0.093 | 0.033 | 142.2 | 1.9 | 145.1 | 11.3 | -879 |
| -3.11 | -2.73 | 0.039 | 0.010 | -0.053 | 0.007 | 0.002 | 0.009 | 131.3 | 2.3 | 87.6 | 10.0 | -856 |
| -3.02 | 48.44 | 0.408 | 0.006 | -0.026 | 0.004 | 0.408 | 0.005 | 3.6 | 0.2 | 3.6 | 0.6 | – |
| -3.00 | 45.88 | 0.386 | 0.004 | -0.031 | 0.003 | 0.384 | 0.004 | 3.7 | 0.2 | 4.6 | 0.5 | – |
| -2.68 | -21.64 | 0.186 | 0.004 | -0.001 | 0.004 | -0.187 | 0.005 | 172.9 | 0.4 | 179.7 | 1.4 | -392 |

Table 8.1: continued

| $d_x{}^a$ (") | $d_y{}^a$ (") | $\mu$ (" yr$^{-1}$) | $\sigma_\mu$ (" yr$^{-1}$) | $\mu_x$ (" yr$^{-1}$) | $\sigma_{\mu_x}$ (" yr$^{-1}$) | $\mu_y$ (" yr$^{-1}$) | $\sigma_{\mu_y}$ (" yr$^{-1}$) | PA (°) | $\sigma_{PA}$ (°) | $\alpha$ (°) | $\sigma_\alpha$ (°) | $v_{rad}$ (km s$^{-1}$) |
|---|---|---|---|---|---|---|---|---|---|---|---|---|
| -2.68 | -21.64 | 0.186 | 0.004 | -0.001 | 0.004 | -0.187 | 0.005 | 122.9 | 0.4 | 179.7 | 1.4 | -530 |
| -2.59 | -34.60 | 0.341 | 0.004 | -0.018 | 0.004 | -0.340 | 0.004 | 175.7 | 0.3 | 176.9 | 0.7 | – |
| -2.39 | -39.38 | 0.356 | 0.007 | 0.006 | 0.008 | -0.357 | 0.007 | 176.5 | 0.2 | 180.9 | 1.3 | – |
| -2.33 | -19.76 | 0.168 | 0.006 | 0.045 | 0.006 | -0.174 | 0.006 | 173.3 | 0.5 | 194.5 | 1.8 | -405 |
| -2.33 | -19.76 | 0.168 | 0.006 | 0.045 | 0.006 | -0.174 | 0.006 | 173.3 | 0.5 | 194.5 | 1.8 | -477 |
| -2.33 | -19.76 | 0.168 | 0.006 | 0.045 | 0.006 | -0.174 | 0.006 | 173.3 | 0.5 | 194.5 | 1.8 | -565 |
| -2.27 | -50.15 | 0.431 | 0.013 | -0.026 | 0.006 | -0.430 | 0.013 | 177.4 | 0.2 | 176.6 | 0.8 | – |
| -2.27 | -50.16 | 0.431 | 0.012 | -0.025 | 0.006 | -0.430 | 0.012 | 177.4 | 0.2 | 176.7 | 0.9 | – |
| -1.87 | -9.27 | 0.116 | 0.011 | -0.028 | 0.016 | -0.113 | 0.010 | 168.6 | 1.0 | 166.0 | 7.6 | – |
| -1.70 | -30.09 | 0.120 | 0.018 | -0.078 | 0.033 | -0.115 | 0.019 | 176.8 | 0.3 | 145.7 | 12.0 | 586 |
| -1.18 | -12.53 | 0.136 | 0.008 | -0.018 | 0.014 | -0.135 | 0.008 | 174.6 | 0.8 | 172.2 | 5.7 | 878 |
| -0.71 | 42.01 | 0.303 | 0.013 | -0.006 | 0.004 | 0.303 | 0.014 | 1.0 | 0.2 | 1.0 | 0.8 | – |
| -0.64 | -45.59 | 0.370 | 0.015 | 0.014 | 0.010 | -0.370 | 0.015 | 179.2 | 0.2 | 182.2 | 1.6 | – |
| -0.38 | -43.40 | 0.373 | 0.009 | -0.011 | 0.007 | -0.373 | 0.009 | 179.5 | 0.2 | 178.3 | 1.0 | – |
| – | – | – | – | – | – | – | – | – | – | – | – | -424 |
| – | – | – | – | – | – | – | – | – | – | – | – | -832 |
| – | – | – | – | – | – | – | – | – | – | – | – | -248 |
| – | – | – | – | – | – | – | – | – | – | – | – | 54 |
| – | – | – | – | – | – | – | – | – | – | – | – | -761 |
| – | – | – | – | – | – | – | – | – | – | – | – | 329 |
| – | – | – | – | – | – | – | – | – | – | – | – | 374 |

$^a$Extrapolated to 2004-01-29, or JD2453034.37.

Table 8.2: Radial velocity measurements of the three positions angles (325°, 62°, and 18°.9) of the nebula MWC 137. Measurements for the emission lines [N II] 6583.454Å and [S II] 6716.440Å together with the equivalent distances from the central star are listed. Errors on the radial velocity measurements were estimated based on the RMS of the wavelength calibration and are as follows: 0.89 km s$^{-1}$ ([S II]$_{62°,18°.9}$), 0.90 km s$^{-1}$ ([N II]$_{62°,18°.9}$ and [S II]$_{325°}$), 0.91 km s$^{-1}$ ([N II]$_{325°}$). The online published version of the table is available at https://iopscience.iop.org/article/10.3847/1538-3881/aa8df6/meta

| PA=325° | | | | PA=62° | | | | PA=18°.9 | | | |
|---|---|---|---|---|---|---|---|---|---|---|---|
| $v_{[S II]}$ km s$^{-1}$ | $d_{[S II]}$ ″ | $v_{[N II]}$ km s$^{-1}$ | $d_{[N II]}$ ″ | $v_{[S II]}$ km s$^{-1}$ | $d_{[S II]}$ ″ | $v_{[N II]}$ km s$^{-1}$ | $d_{[N II]}$ ″ | $v_{[S II]}$ km s$^{-1}$ | $d_{[S II]}$ ″ | $v_{[N II]}$ km s$^{-1}$ | $d_{[N II]}$ ″ |
| 5.23 | -29.63 | 4.15 | -30.08 | -0.28 | 2.89 | -0.42 | 2.85 | -10.62 | 43.93 | -11.10 | 40.16 |
| 11.03 | -29.42 | 5.93 | -29.87 | -1.62 | 3.10 | -1.06 | 3.06 | -10.53 | 43.72 | -11.10 | 39.95 |
| 8.27 | -29.21 | 9.75 | -29.66 | -0.06 | 3.31 | -1.19 | 3.27 | -15.04 | 43.51 | -8.51 | 39.74 |
| 9.83 | -29.00 | 4.38 | -29.45 | -1.13 | 3.52 | -0.92 | 3.48 | -18.87 | 43.30 | -8.96 | 39.53 |
| 5.86 | -28.79 | 2.97 | -29.24 | 1.01 | 3.73 | -1.42 | 3.69 | -16.87 | 43.09 | -12.92 | 39.32 |
| 9.16 | -28.58 | 5.38 | -29.03 | 2.98 | 3.94 | 0.63 | 3.90 | -14.81 | 42.88 | -11.15 | 39.11 |
| 9.20 | -28.37 | 5.79 | -28.82 | -2.25 | 4.15 | -0.01 | 4.11 | -10.84 | 42.67 | -11.87 | 38.90 |
| 5.68 | -28.16 | 7.48 | -28.61 | 0.66 | 4.36 | -1.37 | 4.32 | -15.48 | 42.46 | -12.01 | 38.69 |
| 7.91 | -27.95 | 6.47 | -28.40 | -0.95 | 4.57 | -0.83 | 4.53 | -14.37 | 42.25 | -12.10 | 38.48 |
| 7.06 | -27.74 | 8.11 | -28.19 | -0.73 | 4.78 | 0.54 | 4.74 | -15.39 | 42.04 | -12.42 | 38.27 |
| 6.88 | -27.53 | 6.70 | -27.98 | -1.62 | 4.99 | 0.36 | 4.95 | -15.04 | 41.83 | -12.01 | 38.06 |
| 6.84 | -27.32 | 4.70 | -27.77 | 0.08 | 5.20 | -1.19 | 5.16 | -17.09 | 41.62 | -12.06 | 37.85 |
| 9.52 | -27.11 | 6.52 | -27.56 | 0.21 | 5.41 | -0.46 | 5.37 | -16.60 | 41.41 | -11.83 | 37.64 |
| 7.33 | -26.90 | 7.11 | -27.35 | -0.37 | 5.62 | -0.74 | 5.58 | -12.40 | 41.20 | -12.28 | 37.43 |
| 9.47 | -26.69 | 6.61 | -27.14 | 0.75 | 5.83 | 1.59 | 5.79 | -13.21 | 40.99 | -11.51 | 37.22 |
| 8.67 | -26.48 | 6.66 | -26.93 | 2.80 | 6.04 | 1.45 | 6.00 | -14.19 | 40.78 | -11.37 | 37.01 |
| 10.14 | -26.27 | 6.52 | -26.72 | 1.46 | 6.25 | 2.27 | 6.21 | -12.49 | 40.57 | -9.83 | 36.80 |
| 11.48 | -26.06 | 7.43 | -26.51 | 4.27 | 6.46 | 2.72 | 6.42 | -12.36 | 40.36 | -10.96 | 36.59 |
| 11.17 | -25.85 | 7.61 | -26.30 | 0.34 | 6.67 | 2.27 | 6.63 | -12.31 | 40.15 | -10.83 | 36.38 |
| 11.75 | -25.64 | 8.71 | -26.09 | 4.23 | 6.88 | 3.00 | 6.84 | -11.55 | 39.94 | -10.24 | 36.17 |

Table 8.2: continued

| PA=325° | | | | PA=62° | | | | PA=18°.9 | | | |
|---|---|---|---|---|---|---|---|---|---|---|---|
| $v_{[\text{S\,II}]}$ km s$^{-1}$ | $d_{[\text{S\,II}]}$ " | $v_{[\text{N\,II}]}$ km s$^{-1}$ | $d_{[\text{N\,II}]}$ " | $v_{[\text{S\,II}]}$ km s$^{-1}$ | $d_{[\text{S\,II}]}$ " | $v_{[\text{N\,II}]}$ km s$^{-1}$ | $d_{[\text{N\,II}]}$ " | $v_{[\text{S\,II}]}$ km s$^{-1}$ | $d_{[\text{S\,II}]}$ " | $v_{[\text{N\,II}]}$ km s$^{-1}$ | $d_{[\text{N\,II}]}$ " |
| 13.35 | -25.43 | 10.25 | -25.88 | 5.03 | 7.09 | 3.18 | 7.05 | -10.66 | 39.73 | -11.24 | 35.96 |
| 12.86 | -25.22 | 9.39 | -25.67 | 4.36 | 7.30 | 4.77 | 7.26 | -12.36 | 39.52 | -12.06 | 35.75 |
| 13.71 | -25.01 | 11.03 | -25.46 | 6.32 | 7.51 | 2.95 | 7.47 | -12.27 | 39.31 | -11.92 | 35.54 |
| 14.78 | -24.80 | 12.67 | -25.25 | 2.13 | 7.72 | 4.09 | 7.68 | -12.40 | 39.10 | -11.97 | 35.33 |
| 15.59 | -24.59 | 10.85 | -25.04 | 5.03 | 7.93 | 4.82 | 7.89 | -13.47 | 38.89 | -11.97 | 35.12 |
| 13.49 | -24.38 | 11.48 | -24.83 | 3.69 | 8.14 | 3.18 | 8.10 | -14.01 | 38.68 | -12.47 | 34.91 |
| 15.45 | -24.17 | 12.94 | -24.62 | 3.51 | 8.35 | 4.04 | 8.31 | -14.90 | 38.47 | -13.29 | 34.70 |
| 13.67 | -23.96 | 12.80 | -24.41 | 3.29 | 8.56 | 3.09 | 8.52 | -14.14 | 38.26 | -14.11 | 34.49 |
| 12.55 | -23.75 | 13.35 | -24.20 | 4.05 | 8.77 | 3.36 | 8.73 | -12.98 | 38.05 | -13.65 | 34.28 |
| 13.53 | -23.54 | 13.26 | -23.99 | 6.15 | 8.98 | 2.72 | 8.94 | -14.19 | 37.84 | -13.97 | 34.07 |
| 12.15 | -23.33 | 11.57 | -23.78 | 0.03 | 9.19 | 3.09 | 9.15 | -11.69 | 37.63 | -14.15 | 33.86 |
| 13.93 | -23.12 | 11.39 | -23.57 | 3.56 | 9.40 | 1.40 | 9.36 | -12.36 | 37.42 | -14.65 | 33.65 |
| 12.19 | -22.91 | 11.98 | -23.36 | 3.16 | 9.61 | 4.45 | 9.57 | -11.73 | 37.21 | -15.34 | 33.44 |
| 12.69 | -22.70 | 11.39 | -23.15 | 4.45 | 9.82 | 2.36 | 9.78 | -12.27 | 37.00 | -15.29 | 33.23 |
| 12.06 | -22.49 | 10.39 | -22.94 | 2.31 | 10.03 | 4.18 | 9.99 | -10.17 | 36.79 | -15.93 | 33.02 |
| 10.90 | -22.28 | 12.85 | -22.73 | 2.22 | 10.24 | 2.81 | 10.20 | -11.60 | 36.58 | -15.29 | 32.81 |
| 11.12 | -22.07 | 11.16 | -22.52 | 2.66 | 10.45 | 2.63 | 10.41 | -10.08 | 36.37 | -15.06 | 32.60 |
| 9.78 | -21.86 | 9.48 | -22.31 | -2.07 | 10.66 | 2.27 | 10.62 | -11.60 | 36.16 | -15.15 | 32.39 |
| 8.71 | -21.65 | 11.21 | -22.10 | 2.04 | 10.87 | 1.36 | 10.83 | -12.36 | 35.95 | -14.38 | 32.18 |
| 7.91 | -21.44 | 9.02 | -21.89 | 3.11 | 11.08 | 1.54 | 11.04 | -11.73 | 35.74 | -13.74 | 31.97 |
| 11.44 | -21.23 | 9.98 | -21.68 | 2.93 | 11.29 | 1.99 | 11.25 | -12.09 | 35.53 | -12.51 | 31.76 |
| 10.68 | -21.02 | 6.66 | -21.47 | 0.97 | 11.50 | 0.54 | 11.46 | -11.91 | 35.32 | -13.47 | 31.55 |
| 13.71 | -20.81 | 9.71 | -21.26 | 2.22 | 11.71 | 2.72 | 11.67 | -12.04 | 35.11 | -13.29 | 31.34 |
| 11.48 | -20.60 | 7.75 | -21.05 | 3.74 | 11.92 | 3.54 | 11.88 | -12.36 | 34.90 | -12.24 | 31.13 |

181

Table 8.2: continued

| PA=325° | | | | PA=62° | | | | PA=18°.9 | | | |
|---|---|---|---|---|---|---|---|---|---|---|---|
| $v_{[SII]}$ km s⁻¹ | $d_{[SII]}$ ″ | $v_{[NII]}$ km s⁻¹ | $d_{[NII]}$ ″ | $v_{[SII]}$ km s⁻¹ | $d_{[SII]}$ ″ | $v_{[NII]}$ km s⁻¹ | $d_{[NII]}$ ″ | $v_{[SII]}$ km s⁻¹ | $d_{[SII]}$ ″ | $v_{[NII]}$ km s⁻¹ | $d_{[NII]}$ ″ |
| 9.25 | -20.39 | 9.94 | -20.84 | 4.09 | 12.13 | -0.10 | 12.09 | -13.70 | 34.69 | -12.51 | 30.92 |
| 8.76 | -20.18 | 9.25 | -20.63 | 1.86 | 12.34 | 1.45 | 12.30 | -14.45 | 34.48 | -13.29 | 30.71 |
| 12.42 | -19.97 | 11.44 | -20.42 | 0.70 | 12.55 | 0.17 | 12.51 | -13.79 | 34.27 | -13.61 | 30.50 |
| 12.95 | -19.76 | 11.44 | -20.21 | -1.13 | 12.76 | 0.58 | 12.72 | -13.92 | 34.06 | -13.61 | 30.29 |
| 10.23 | -19.55 | 10.57 | -20.00 | 3.16 | 12.97 | -1.47 | 12.93 | -14.68 | 33.85 | -12.74 | 30.08 |
| 10.90 | -19.34 | 10.25 | -19.79 | 0.66 | 13.18 | -3.20 | 13.14 | -14.28 | 33.64 | -12.47 | 29.87 |
| 8.53 | -19.13 | 11.80 | -19.58 | 3.16 | 13.39 | 1.18 | 13.35 | -14.72 | 33.43 | -13.10 | 29.66 |
| 10.32 | -18.92 | 11.57 | -19.37 | 0.08 | 13.60 | 2.95 | 13.56 | -14.77 | 33.22 | -11.42 | 29.45 |
| 10.86 | -18.71 | 11.21 | -19.16 | 0.25 | 13.81 | -1.83 | 13.77 | -13.65 | 33.01 | -10.96 | 29.24 |
| 10.05 | -18.50 | 12.26 | -18.95 | 2.31 | 14.02 | -0.96 | 13.98 | -14.23 | 32.80 | -10.24 | 29.03 |
| 11.12 | -18.29 | 10.44 | -18.74 | 2.44 | 14.23 | -2.56 | 14.19 | -13.12 | 32.59 | -10.74 | 28.82 |
| 9.78 | -18.08 | 11.71 | -18.53 | -0.42 | 14.44 | -3.01 | 14.40 | -13.70 | 32.38 | -10.01 | 28.61 |
| 12.69 | -17.87 | 14.03 | -18.32 | -4.48 | 14.65 | -2.47 | 14.61 | -14.14 | 32.17 | -9.87 | 28.40 |
| 12.37 | -17.66 | 12.03 | -18.11 | 1.59 | 14.86 | -4.02 | 14.82 | -13.61 | 31.96 | -10.78 | 28.19 |
| 11.12 | -17.45 | 13.67 | -17.90 | -4.43 | 15.07 | -4.02 | 15.03 | -13.70 | 31.75 | -10.42 | 27.98 |
| 15.85 | -17.24 | 13.35 | -17.69 | -2.96 | 15.28 | -6.88 | 15.24 | -12.98 | 31.54 | -10.10 | 27.77 |
| 13.09 | -17.03 | 14.49 | -17.48 | -8.49 | 15.49 | -6.61 | 15.45 | -14.59 | 31.33 | -10.24 | 27.56 |
| 13.67 | -16.82 | 12.76 | -17.27 | -4.79 | 15.70 | -8.11 | 15.66 | -12.98 | 31.12 | -9.51 | 27.35 |
| 13.93 | -16.61 | 14.31 | -17.06 | -10.24 | 15.91 | -9.94 | 15.87 | -12.40 | 30.91 | -10.10 | 27.14 |
| 13.27 | -16.40 | 13.26 | -16.85 | -5.99 | 16.12 | -11.67 | 16.08 | -11.78 | 30.70 | -10.28 | 26.93 |
| 15.85 | -16.19 | 15.86 | -16.64 | -6.26 | 16.33 | -11.98 | 16.29 | -12.49 | 30.49 | -9.60 | 26.72 |
| 16.88 | -15.98 | 15.31 | -16.43 | -9.39 | 16.54 | -10.98 | 16.50 | -12.94 | 30.28 | -12.51 | 25.67 |
| 11.03 | -15.77 | 16.58 | -16.22 | -9.65 | 16.75 | -13.49 | 16.71 | -11.37 | 30.07 | -10.24 | 25.46 |
| 11.17 | -15.56 | 15.72 | -16.01 | -16.04 | 16.96 | -13.53 | 16.92 | -11.02 | 29.86 | -10.74 | 25.25 |

Table 8.2: continued

| PA=325° | | | | PA=62° | | | | PA=18°.9 | | | |
|---|---|---|---|---|---|---|---|---|---|---|---|
| $v_{[SII]}$ km s$^{-1}$ | $d_{[SII]}$ ″ | $v_{[NII]}$ km s$^{-1}$ | $d_{[NII]}$ ″ | $v_{[SII]}$ km s$^{-1}$ | $d_{[SII]}$ ″ | $v_{[NII]}$ km s$^{-1}$ | $d_{[NII]}$ ″ | $v_{[SII]}$ km s$^{-1}$ | $d_{[SII]}$ ″ | $v_{[NII]}$ km s$^{-1}$ | $d_{[NII]}$ ″ |
| 10.77 | -15.35 | 17.04 | -15.80 | -12.87 | 17.17 | -16.08 | 17.13 | -10.97 | 29.65 | -13.33 | 25.04 |
| 14.02 | -15.14 | 15.81 | -15.59 | -14.12 | 17.38 | -15.58 | 17.34 | -11.78 | 29.44 | -13.79 | 24.83 |
| 15.10 | -14.93 | 16.04 | -15.38 | -14.61 | 17.59 | -16.81 | 17.55 | -12.04 | 29.23 | -13.01 | 24.62 |
| 10.32 | -14.72 | 9.71 | -15.17 | -15.19 | 17.80 | -15.54 | 17.76 | -11.60 | 29.02 | -14.06 | 24.41 |
| 3.09 | -14.51 | 15.17 | -14.96 | -16.98 | 18.01 | -16.63 | 17.97 | -11.11 | 28.81 | -15.56 | 24.20 |
| 7.46 | -14.30 | 11.48 | -14.75 | -14.92 | 18.22 | -15.86 | 18.18 | -10.71 | 28.60 | -14.06 | 23.99 |
| 3.27 | -14.09 | 11.16 | -14.54 | -14.16 | 18.43 | -14.17 | 18.39 | -11.51 | 28.39 | -13.61 | 23.78 |
| 9.11 | -13.88 | 11.76 | -14.33 | -14.48 | 18.64 | -15.22 | 18.60 | -10.97 | 28.18 | -15.02 | 23.57 |
| 7.60 | -13.67 | 9.07 | -14.12 | -11.35 | 18.85 | -13.17 | 18.81 | -10.53 | 27.97 | -13.47 | 23.36 |
| 7.69 | -13.46 | 9.02 | -13.91 | -9.25 | 19.06 | -13.17 | 19.02 | -10.57 | 27.76 | -14.88 | 23.15 |
| 11.30 | -13.25 | 7.66 | -13.70 | -12.24 | 19.27 | -13.53 | 19.23 | -11.11 | 27.55 | -13.70 | 22.94 |
| 10.01 | -13.04 | 6.88 | -13.49 | -9.43 | 19.48 | -11.94 | 19.44 | -8.83 | 27.34 | -14.38 | 22.73 |
| 9.29 | -12.83 | 8.52 | -13.28 | -12.20 | 19.69 | -10.71 | 19.65 | -9.50 | 27.13 | -14.79 | 22.52 |
| 4.43 | -12.62 | 9.98 | -13.07 | -8.85 | 19.90 | -11.21 | 19.86 | -9.41 | 26.92 | -13.10 | 22.31 |
| 7.06 | -12.41 | 6.38 | -12.86 | -8.49 | 20.11 | -10.80 | 20.07 | -9.81 | 26.71 | -13.15 | 22.10 |
| 6.66 | -12.20 | 6.25 | -12.65 | -7.20 | 20.32 | -8.43 | 20.28 | -8.34 | 26.50 | -12.69 | 21.89 |
| 5.36 | -11.99 | 4.61 | -12.44 | -8.32 | 20.53 | -11.48 | 20.49 | -6.51 | 26.29 | -12.88 | 21.68 |
| 4.52 | -11.78 | 6.06 | -12.23 | -10.06 | 20.74 | -10.21 | 20.70 | -2.49 | 26.08 | -11.97 | 21.47 |
| 5.81 | -11.57 | 4.11 | -12.02 | -8.90 | 20.95 | -8.98 | 20.91 | -8.96 | 25.87 | -10.55 | 21.26 |
| -0.30 | -11.36 | 3.51 | -11.81 | -8.94 | 21.16 | -7.43 | 21.12 | -13.65 | 25.66 | -10.83 | 21.05 |
| 3.49 | -11.15 | 2.15 | -11.60 | -7.78 | 21.37 | -8.39 | 21.33 | -14.14 | 25.45 | -9.69 | 20.84 |
| 2.64 | -10.94 | 2.38 | -11.39 | -7.65 | 21.58 | -10.16 | 21.54 | -15.70 | 25.24 | -7.82 | 20.63 |
| 3.76 | -10.73 | 2.69 | -11.18 | -9.57 | 21.79 | -9.16 | 21.75 | -13.12 | 25.03 | -9.01 | 20.42 |
| 0.23 | -10.52 | 0.24 | -10.97 | -11.93 | 22.00 | -9.94 | 21.96 | -15.70 | 24.82 | -10.55 | 20.21 |

Table 8.2: continued

| PA=325° | | | | PA=62° | | | | PA=18°.9 | | | |
|---|---|---|---|---|---|---|---|---|---|---|---|
| $v_{\text{[SII]}}$ km s⁻¹ | $d_{\text{[SII]}}$ ″ | $v_{\text{[NII]}}$ km s⁻¹ | $d_{\text{[NII]}}$ ″ | $v_{\text{[SII]}}$ km s⁻¹ | $d_{\text{[SII]}}$ ″ | $v_{\text{[NII]}}$ km s⁻¹ | $d_{\text{[NII]}}$ ″ | $v_{\text{[SII]}}$ km s⁻¹ | $d_{\text{[SII]}}$ ″ | $v_{\text{[NII]}}$ km s⁻¹ | $d_{\text{[NII]}}$ ″ |
| 5.19 | -10.31 | 3.51 | -10.76 | -9.79 | 22.21 | -8.89 | 22.17 | -12.18 | 24.61 | -9.42 | 20.00 |
| 4.61 | -10.10 | 1.74 | -10.55 | -7.51 | 22.42 | -10.21 | 22.38 | -7.31 | 24.40 | -9.87 | 19.79 |
| 1.53 | -9.89 | 0.74 | -10.34 | -10.82 | 22.63 | -10.53 | 22.59 | -14.63 | 24.19 | -8.10 | 19.58 |
| 0.54 | -9.68 | 2.10 | -10.13 | -9.52 | 22.84 | -10.98 | 22.80 | -10.66 | 23.98 | -9.60 | 19.37 |
| 2.24 | -9.47 | 1.97 | -9.92 | -10.46 | 23.05 | -10.85 | 23.01 | -11.02 | 23.77 | -10.37 | 19.16 |
| 3.71 | -9.26 | 0.42 | -9.71 | -9.65 | 23.26 | -9.34 | 23.22 | -12.62 | 23.56 | -8.73 | 18.95 |
| 2.24 | -9.05 | 2.69 | -9.50 | -6.98 | 23.47 | -8.25 | 23.43 | -13.65 | 23.35 | -7.91 | 18.74 |
| 4.34 | -8.84 | 1.15 | -9.29 | -8.32 | 23.68 | -8.89 | 23.64 | -13.61 | 23.14 | -6.55 | 18.53 |
| 4.07 | -8.63 | 1.28 | -9.08 | -7.69 | 23.89 | -9.80 | 23.85 | -12.67 | 22.93 | -8.55 | 18.32 |
| 5.36 | -8.42 | -0.13 | -8.87 | -8.81 | 24.10 | -9.34 | 24.06 | -11.06 | 22.72 | -6.87 | 18.11 |
| -0.35 | -8.21 | 2.10 | -8.66 | -7.65 | 24.31 | -8.02 | 24.27 | -9.28 | 22.51 | -5.32 | 17.90 |
| 1.44 | -8.00 | 3.06 | -8.45 | -6.89 | 24.52 | -7.52 | 25.32 | -9.68 | 22.30 | -5.68 | 17.69 |
| 1.75 | -7.79 | 0.42 | -8.24 | -7.33 | 24.73 | -7.98 | 25.53 | -8.16 | 22.09 | -5.00 | 17.48 |
| 1.44 | -7.58 | 1.65 | -8.03 | -8.36 | 24.94 | -7.80 | 25.74 | -10.39 | 21.88 | -8.05 | 17.27 |
| 2.24 | -7.37 | 1.47 | -7.82 | -5.77 | 25.15 | -8.75 | 25.95 | -10.66 | 21.67 | -5.36 | 17.06 |
| 2.28 | -7.16 | 1.24 | -7.61 | -7.20 | 25.36 | -8.75 | 26.16 | -7.76 | 21.46 | -5.73 | 16.85 |
| 2.64 | -6.95 | 2.24 | -7.40 | -8.41 | 25.57 | -9.71 | 26.37 | -10.97 | 21.25 | -6.41 | 16.64 |
| 2.33 | -6.74 | 2.15 | -7.19 | -7.47 | 25.78 | -8.89 | 26.58 | -7.80 | 21.04 | -5.36 | 16.43 |
| -1.20 | -6.53 | 1.19 | -6.98 | -9.61 | 25.99 | -9.62 | 26.79 | -6.69 | 20.83 | -5.50 | 16.22 |
| 2.60 | -6.32 | 1.56 | -6.77 | -8.32 | 26.20 | -9.62 | 27.00 | -7.58 | 20.62 | -6.18 | 16.01 |
| 1.48 | -6.11 | 1.47 | -6.56 | -9.39 | 26.41 | -9.66 | 27.21 | -6.29 | 20.41 | -7.59 | 15.80 |
| 1.17 | -5.90 | 1.97 | -6.35 | -10.06 | 26.62 | -8.11 | 27.42 | -9.37 | 20.20 | -5.36 | 15.59 |
| 3.67 | -5.69 | 1.06 | -6.14 | -9.61 | 26.83 | -8.02 | 27.63 | -6.11 | 19.99 | -4.27 | 15.38 |
| 2.37 | -5.48 | 2.28 | -5.93 | -10.10 | 27.04 | -4.84 | 27.84 | -8.07 | 19.78 | -4.13 | 15.17 |

Table 8.2: continued

| PA=325° | | | | PA=62° | | | | PA=18°.9 | | | |
|---|---|---|---|---|---|---|---|---|---|---|---|
| $v_{[SII]}$ km s⁻¹ | $d_{[SII]}$ " | $v_{[NII]}$ km s⁻¹ | $d_{[NII]}$ " | $v_{[SII]}$ km s⁻¹ | $d_{[SII]}$ " | $v_{[NII]}$ km s⁻¹ | $d_{[NII]}$ " | $v_{[SII]}$ km s⁻¹ | $d_{[SII]}$ " | $v_{[NII]}$ km s⁻¹ | $d_{[NII]}$ " |
| 4.96 | -5.27 | 1.51 | -5.72 | -5.91 | 27.25 | -2.33 | 28.05 | -10.44 | 19.57 | -5.27 | 14.96 |
| 0.14 | -5.06 | 2.28 | -5.51 | -7.78 | 27.46 | -5.84 | 28.26 | -6.60 | 19.36 | -4.36 | 14.75 |
| 3.31 | -4.85 | 1.51 | -5.30 | -9.61 | 27.67 | -2.65 | 28.47 | -5.44 | 19.15 | -2.68 | 14.54 |
| 1.04 | -4.64 | 2.47 | -5.09 | -10.15 | 27.88 | -2.47 | 28.68 | -7.05 | 18.94 | -4.00 | 14.33 |
| 2.28 | -4.43 | 2.47 | -4.88 | -6.98 | 28.09 | -3.24 | 28.89 | -6.60 | 18.73 | -4.27 | 14.12 |
| 1.39 | -4.22 | 2.24 | -4.67 | -3.32 | 28.30 | 2.59 | 29.10 | -8.83 | 18.52 | -4.95 | 13.91 |
| -0.62 | -4.01 | 2.10 | -4.46 | -5.86 | 28.51 | -3.52 | 29.31 | -8.70 | 18.31 | -3.27 | 13.70 |
| 2.46 | -3.80 | -0.36 | -4.25 | -7.02 | 28.72 | 1.59 | 29.52 | -9.59 | 18.10 | -4.13 | 13.49 |
| -0.93 | -3.59 | 0.65 | -4.04 | -3.27 | 28.93 | -10.57 | 29.73 | -7.27 | 17.89 | -3.45 | 13.28 |
| 1.17 | -3.38 | 0.10 | -3.83 | -5.91 | 29.14 | -3.06 | 29.94 | -5.04 | 17.68 | -2.95 | 13.07 |
| 1.08 | -3.17 | 0.96 | -3.62 | -9.83 | 29.35 | -4.20 | 30.15 | -5.80 | 17.47 | -2.27 | 12.86 |
| 3.31 | -2.96 | 0.28 | -3.41 | -2.74 | 29.56 | 0.17 | 30.36 | -6.24 | 17.26 | -3.27 | 12.65 |
| 3.98 | -2.75 | -0.58 | -3.20 | -6.71 | 29.77 | -0.56 | 26.55 | -5.39 | 17.05 | -4.59 | 12.44 |
| 2.33 | -2.54 | 1.37 | -2.99 | -4.92 | 29.98 | -2.38 | 26.34 | -3.21 | 16.84 | -4.04 | 12.23 |
| 3.80 | -2.33 | -2.41 | -2.78 | -5.64 | 30.19 | 1.22 | 26.13 | -5.26 | 16.63 | -3.36 | 12.02 |
| 1.93 | -2.12 | -0.81 | -2.57 | 0.39 | 30.40 | 1.31 | 25.92 | -5.39 | 16.42 | -3.27 | 11.81 |
| 3.09 | -1.91 | 0.69 | -2.36 | -0.64 | 30.61 | -0.87 | 25.71 | -3.43 | 16.21 | -4.13 | 11.60 |
| -0.39 | 2.71 | -0.40 | -1.94 | -3.85 | 30.82 | 0.13 | 25.50 | -5.13 | 16.00 | -4.91 | 11.39 |
| -0.48 | 2.92 | -1.54 | 2.68 | 1.77 | 31.03 | -3.01 | 25.29 | -4.14 | 15.79 | -4.63 | 11.18 |
| -1.20 | 3.13 | -0.90 | 2.89 | -3.00 | 31.24 | -1.24 | 25.08 | -1.91 | 15.58 | -3.31 | 10.97 |
| -1.38 | 3.34 | -0.04 | 3.10 | 8.24 | 31.45 | 0.49 | 24.87 | -3.79 | 15.37 | -3.50 | 10.76 |
| 0.05 | 3.55 | -0.49 | 3.31 | 3.02 | 31.66 | 0.67 | 24.66 | -3.74 | 15.16 | -4.22 | 10.55 |
| -0.17 | 3.76 | -1.27 | 3.52 | 0.03 | 31.87 | 0.90 | 24.45 | -2.80 | 14.95 | -3.45 | 10.34 |
| 1.04 | 3.97 | 1.78 | 3.73 | -4.70 | 32.08 | -1.83 | 24.24 | -4.59 | 14.74 | -2.95 | 10.13 |

185

Table 8.2: continued

| PA=325° | | | | PA=62° | | | | PA=18°.9 | | | |
|---|---|---|---|---|---|---|---|---|---|---|---|
| $v_{\rm [SII]}$ km s$^{-1}$ | $d_{\rm [SII]}$ " | $v_{\rm [NII]}$ km s$^{-1}$ | $d_{\rm [NII]}$ " | $v_{\rm [SII]}$ km s$^{-1}$ | $d_{\rm [SII]}$ " | $v_{\rm [NII]}$ km s$^{-1}$ | $d_{\rm [NII]}$ " | $v_{\rm [SII]}$ km s$^{-1}$ | $d_{\rm [SII]}$ " | $v_{\rm [NII]}$ km s$^{-1}$ | $d_{\rm [NII]}$ " |
| 0.19 | 4.18 | 0.28 | 3.94 | -6.53 | 32.29 | -1.60 | -24.03 | -6.20 | 14.53 | -3.68 | 9.92 |
| -1.38 | 4.39 | 0.37 | 4.15 | 0.43 | -26.51 | -1.19 | -23.82 | -3.47 | 14.32 | -3.59 | 9.71 |
| 1.04 | 4.60 | 1.06 | 4.36 | -2.69 | -26.30 | -0.92 | -23.61 | -1.47 | 14.11 | -3.45 | 9.50 |
| 0.14 | 4.81 | 1.47 | 4.57 | -2.02 | -26.09 | -1.24 | -23.40 | 0.32 | 13.90 | -3.63 | 9.29 |
| 4.29 | 5.02 | 2.51 | 4.78 | -4.03 | -25.88 | -2.01 | -23.19 | -2.67 | 13.69 | -2.86 | 9.08 |
| 1.53 | 5.23 | 2.88 | 4.99 | -3.72 | -25.67 | -0.42 | -22.98 | -6.42 | 13.48 | -3.36 | 8.87 |
| 3.85 | 5.44 | 1.24 | 5.20 | -2.42 | -25.46 | -2.47 | -22.77 | -1.91 | 13.27 | -3.50 | 8.66 |
| -1.29 | 5.65 | 2.97 | 5.41 | -3.00 | -25.25 | -2.15 | -22.56 | -1.82 | 13.06 | -3.54 | 8.45 |
| 2.02 | 5.86 | 4.56 | 5.62 | -5.15 | -25.04 | -1.19 | -22.35 | -3.03 | 12.85 | -4.68 | 8.24 |
| 3.89 | 6.07 | 4.29 | 5.83 | -2.91 | -24.83 | -0.42 | -22.14 | -2.36 | 12.64 | -6.23 | 8.03 |
| 3.89 | 6.28 | 2.51 | 6.04 | -3.67 | -24.62 | 0.17 | -21.93 | -3.97 | 12.43 | -5.91 | 7.82 |
| 4.87 | 6.49 | 4.70 | 6.25 | -1.53 | -24.41 | 2.04 | -21.72 | -3.21 | 12.22 | -5.64 | 7.61 |
| 3.45 | 6.70 | 5.61 | 6.46 | 0.61 | -24.20 | 0.90 | -21.51 | -3.21 | 12.01 | -6.77 | 7.40 |
| 9.43 | 6.91 | 8.07 | 6.67 | -4.34 | -23.99 | 0.99 | -21.30 | -4.01 | 11.80 | -6.46 | 7.19 |
| 11.03 | 7.12 | 5.52 | 6.88 | -4.12 | -23.78 | 2.54 | -21.09 | -6.11 | 11.59 | -6.09 | 6.98 |
| 9.16 | 7.33 | 11.39 | 7.09 | -2.60 | -23.57 | 3.95 | -20.88 | -4.81 | 11.38 | -6.23 | 6.77 |
| 7.46 | 7.54 | 10.21 | 7.30 | -4.21 | -23.36 | 3.09 | -20.67 | -4.90 | 11.17 | -4.86 | 6.56 |
| 12.51 | 7.75 | 10.12 | 7.51 | -1.67 | -23.15 | 3.82 | -20.46 | -3.16 | 10.96 | -5.50 | 6.35 |
| 7.37 | 7.96 | 9.07 | 7.72 | -1.80 | -22.94 | 2.86 | -20.25 | -3.30 | 10.75 | -5.41 | 6.14 |
| 7.33 | 8.17 | 10.21 | 7.93 | -2.91 | -22.73 | 4.09 | -20.04 | -3.39 | 10.54 | -4.54 | 5.93 |
| 7.06 | 8.38 | 9.16 | 8.14 | 0.03 | -22.52 | 5.32 | -19.83 | -3.12 | 10.33 | -4.77 | 5.72 |
| 8.76 | 8.59 | 9.30 | 8.35 | 0.12 | -22.31 | 4.77 | -19.62 | -3.16 | 10.12 | -4.86 | 5.51 |
| 8.27 | 8.80 | 7.25 | 8.56 | -0.68 | -22.10 | 5.27 | -19.41 | -3.52 | 9.91 | -4.50 | 5.30 |
| 5.45 | 9.01 | 6.70 | 8.77 | -0.77 | -21.89 | 5.14 | -19.20 | -0.93 | 9.70 | -4.86 | 5.09 |

Table 8.2: continued

| PA=325° | | | | PA=62° | | | | PA=18°.9 | | | |
|---|---|---|---|---|---|---|---|---|---|---|---|
| $v_{\rm[SII]}$ km s⁻¹ | $d_{\rm[SII]}$ ″ | $v_{\rm[NII]}$ km s⁻¹ | $d_{\rm[NII]}$ ″ | $v_{\rm[SII]}$ km s⁻¹ | $d_{\rm[SII]}$ ″ | $v_{\rm[NII]}$ km s⁻¹ | $d_{\rm[NII]}$ ″ | $v_{\rm[SII]}$ km s⁻¹ | $d_{\rm[SII]}$ ″ | $v_{\rm[NII]}$ km s⁻¹ | $d_{\rm[NII]}$ ″ |
| 5.81 | 9.22 | 4.88 | 8.98 | 0.39 | -21.68 | 6.14 | -18.99 | -2.72 | 9.49 | -3.95 | 4.88 |
| 7.06 | 9.43 | 4.11 | 9.19 | 0.66 | -21.47 | 6.91 | -18.78 | -2.98 | 9.28 | -4.73 | 4.67 |
| 7.78 | 9.64 | 4.65 | 9.40 | 2.75 | -21.26 | 7.05 | -18.57 | -4.46 | 9.07 | -5.09 | 4.46 |
| 1.30 | 9.85 | 5.65 | 9.61 | 2.22 | -21.05 | 7.32 | -18.36 | -6.51 | 8.86 | -5.45 | 4.25 |
| 6.12 | 10.06 | 2.38 | 9.82 | 3.29 | -20.84 | 6.37 | -18.15 | -5.13 | 8.65 | -4.82 | 4.04 |
| 5.45 | 10.27 | 3.70 | 10.03 | 3.38 | -20.63 | 5.36 | -17.94 | -5.48 | 8.44 | -4.95 | 3.83 |
| 1.97 | 10.48 | 3.51 | 10.24 | 3.20 | -20.42 | 6.59 | -17.73 | -5.35 | 8.23 | -5.00 | 3.62 |
| 5.23 | 10.69 | 4.33 | 10.45 | 3.65 | -20.21 | 4.50 | -17.52 | -4.81 | 8.02 | -5.18 | 3.41 |
| 5.72 | 10.90 | 4.06 | 10.66 | 3.74 | -20.00 | 3.18 | -17.31 | -5.71 | 7.81 | -4.91 | 3.20 |
| 6.17 | 11.11 | 3.20 | 10.87 | 5.39 | -19.79 | 0.31 | -17.10 | -6.60 | 7.60 | -4.32 | 2.99 |
| 1.97 | 11.32 | 4.11 | 11.08 | 4.58 | -19.58 | 1.59 | -16.89 | -6.60 | 7.39 | -4.32 | 2.78 |
| 6.61 | 11.53 | 4.15 | 11.29 | 5.16 | -19.37 | 1.63 | -16.68 | -5.53 | 7.18 | -6.05 | 2.57 |
| 10.27 | 11.74 | 1.88 | 11.50 | 6.37 | -19.16 | 2.31 | -16.47 | -7.09 | 6.97 | -5.18 | 2.36 |
| 0.28 | 11.95 | 1.19 | 11.71 | 4.45 | -18.95 | 0.67 | -16.26 | -4.81 | 6.76 | -6.91 | -1.84 |
| 4.70 | 12.16 | 2.69 | 11.92 | 5.39 | -18.74 | 1.54 | -16.05 | -4.14 | 6.55 | -7.23 | -2.05 |
| 4.20 | 12.37 | 2.19 | 12.13 | 6.95 | -18.53 | 0.45 | -15.84 | -3.16 | 6.34 | -7.05 | -2.26 |
| 1.84 | 12.58 | 2.56 | 12.34 | 5.79 | -18.32 | 0.72 | -15.63 | -4.68 | 6.13 | -6.96 | -2.47 |
| -1.87 | 12.79 | -0.54 | 12.55 | 8.51 | -18.11 | -1.01 | -15.42 | -3.70 | 5.92 | -6.46 | -2.68 |
| -2.85 | 13.00 | -4.18 | 12.76 | 5.88 | -17.90 | 1.45 | -15.21 | -3.92 | 5.71 | -7.18 | -2.89 |
| -2.45 | 13.21 | -2.36 | 12.97 | 4.05 | -17.69 | -0.56 | -15.00 | -3.52 | 5.50 | -7.09 | -3.10 |
| -10.88 | 13.42 | -5.64 | 13.18 | 6.06 | -17.48 | 0.58 | -14.79 | -4.05 | 5.29 | -7.82 | -3.31 |
| -7.40 | 13.63 | -7.64 | 13.39 | 1.28 | -17.27 | -0.15 | -14.58 | -3.74 | 5.08 | -8.05 | -3.52 |
| -7.09 | 13.84 | -13.84 | 13.60 | 0.52 | -17.06 | -0.74 | -14.37 | -3.70 | 4.87 | -7.91 | -3.73 |
| -13.25 | 14.05 | -11.01 | 13.81 | 0.57 | -16.85 | -0.51 | -14.16 | -3.61 | 4.66 | -8.60 | -3.94 |

Table 8.2: continued

| PA=325° | | | | PA=62° | | | | PA=18°.9 | | | |
|---|---|---|---|---|---|---|---|---|---|---|---|
| $v_{[S\,II]}$ km s⁻¹ | $d_{[S\,II]}$ " | $v_{[N\,II]}$ km s⁻¹ | $d_{[N\,II]}$ " | $v_{[S\,II]}$ km s⁻¹ | $d_{[S\,II]}$ " | $v_{[N\,II]}$ km s⁻¹ | $d_{[N\,II]}$ " | $v_{[S\,II]}$ km s⁻¹ | $d_{[S\,II]}$ " | $v_{[N\,II]}$ km s⁻¹ | $d_{[N\,II]}$ " |
| -7.04 | 14.26 | -13.43 | 14.02 | 1.99 | -16.64 | -1.47 | -13.95 | -2.67 | 4.45 | -9.10 | -4.15 |
| -8.47 | 14.47 | -10.83 | 14.23 | 1.91 | -16.43 | 0.17 | -13.74 | -3.65 | 4.24 | -10.01 | -4.36 |
| -8.25 | 14.68 | -9.33 | 14.44 | 0.57 | -16.22 | 1.22 | -13.53 | -4.41 | 4.03 | -9.78 | -4.57 |
| -10.84 | 14.89 | -13.24 | 14.65 | 2.22 | -16.01 | -0.01 | -13.32 | -4.14 | 3.82 | -10.01 | -4.78 |
| -3.47 | 15.10 | -11.15 | 14.86 | 1.19 | -15.80 | 0.54 | -13.11 | -3.21 | 3.61 | -9.46 | -4.99 |
| -9.54 | 15.31 | -12.83 | 15.07 | 1.15 | -15.59 | 1.81 | -12.90 | -3.79 | 3.40 | -8.10 | -5.20 |
| -14.23 | 15.52 | -11.60 | 15.28 | 4.00 | -15.38 | 2.27 | -12.69 | -3.39 | 3.19 | -8.96 | -5.41 |
| -12.49 | 15.73 | -9.87 | 15.49 | 0.83 | -15.17 | 2.22 | -12.48 | -3.83 | 2.98 | -8.55 | -5.62 |
| -4.90 | 15.94 | -9.87 | 15.70 | 1.99 | -14.96 | 1.45 | -12.27 | -4.32 | 2.77 | -8.10 | -5.83 |
| -10.88 | 16.15 | -14.43 | 15.91 | 2.40 | -14.75 | 1.95 | -12.06 | -6.15 | 2.56 | -7.32 | -6.04 |
| -14.94 | 16.36 | -10.24 | 16.12 | -0.64 | -14.54 | 0.67 | -11.85 | -7.18 | 2.35 | -7.09 | -6.25 |
| -14.90 | 16.57 | -14.70 | 16.33 | 0.34 | -14.33 | 0.08 | -11.64 | -7.18 | 2.14 | -5.59 | -6.46 |
| -14.59 | 16.78 | -16.02 | 16.54 | 1.91 | -14.12 | 3.09 | -11.43 | -8.65 | 1.93 | -5.32 | -6.67 |
| -17.76 | 16.99 | -14.34 | 16.75 | 1.24 | -13.91 | 0.40 | -11.22 | -7.71 | 1.72 | -4.91 | -6.88 |
| -22.67 | 17.20 | -23.12 | 16.96 | 1.73 | -13.70 | -0.19 | -11.01 | -5.75 | 1.51 | -4.59 | -7.09 |
| -30.61 | 18.04 | -21.53 | 17.17 | 0.16 | -13.49 | -0.28 | -10.80 | -2.14 | -1.22 | -4.41 | -7.30 |
| -24.41 | 18.25 | -28.86 | 17.38 | 2.44 | -13.28 | 1.13 | -10.59 | -5.84 | -1.43 | -4.82 | -7.51 |
| -29.67 | 18.46 | -27.72 | 17.59 | 5.16 | -13.07 | 0.04 | -10.38 | -6.47 | -1.64 | -2.77 | -7.72 |
| -27.62 | 18.67 | -29.50 | 17.80 | 3.11 | -12.86 | -2.51 | -10.17 | -6.20 | -1.85 | -2.90 | -7.93 |
| -21.42 | 18.88 | -29.00 | 18.01 | 3.96 | -12.65 | -3.42 | -9.96 | -5.62 | -2.06 | -2.27 | -8.14 |
| -22.09 | 19.09 | -31.00 | 18.22 | 3.78 | -12.44 | -1.97 | -9.75 | -5.39 | -2.27 | -1.36 | -8.35 |
| -22.40 | 19.30 | -30.23 | 18.43 | 3.69 | -12.23 | -1.97 | -9.54 | -5.84 | -2.48 | -2.36 | -8.56 |
| -21.55 | 19.51 | -26.63 | 18.64 | 2.40 | -12.02 | -4.56 | -9.33 | -5.35 | -2.69 | -2.31 | -8.77 |
| -17.67 | 19.72 | -29.45 | 18.85 | 1.46 | -11.81 | -2.83 | -9.12 | -7.18 | -2.90 | -1.45 | -8.98 |

188

Table 8.2: continued

| PA=325° | | | | PA=62° | | | | PA=18°.9 | | | |
|---|---|---|---|---|---|---|---|---|---|---|---|
| $v_{[S\,II]}$ km s⁻¹ | $d_{[S\,II]}$ ″ | $v_{[N\,II]}$ km s⁻¹ | $d_{[N\,II]}$ ″ | $v_{[S\,II]}$ km s⁻¹ | $d_{[S\,II]}$ ″ | $v_{[N\,II]}$ km s⁻¹ | $d_{[N\,II]}$ ″ | $v_{[S\,II]}$ km s⁻¹ | $d_{[S\,II]}$ ″ | $v_{[N\,II]}$ km s⁻¹ | $d_{[N\,II]}$ ″ |
| -23.78 | 19.93 | -23.85 | 19.06 | -0.19 | -11.60 | -2.92 | -8.91 | -6.33 | -3.11 | -2.90 | -9.19 |
| -22.18 | 20.14 | -20.35 | 19.27 | 2.22 | -11.39 | -3.88 | -8.70 | -5.75 | -3.32 | -2.49 | -9.40 |
| -16.95 | 20.35 | -24.26 | 19.48 | -0.77 | -11.18 | -2.60 | -8.49 | -5.84 | -3.53 | -0.95 | -9.61 |
| -23.20 | 20.56 | -20.98 | 19.69 | 0.83 | -10.97 | -1.83 | -8.28 | -5.75 | -3.74 | -0.35 | -9.82 |
| -23.60 | 20.77 | -23.72 | 19.90 | 1.82 | -10.76 | -3.01 | -8.07 | -6.82 | -3.95 | -0.13 | -10.03 |
| -20.75 | 20.98 | -22.44 | 20.11 | 3.07 | -10.55 | -3.88 | -7.86 | -6.64 | -4.16 | 0.56 | -10.24 |
| -21.77 | 21.19 | -23.12 | 20.32 | 0.03 | -10.34 | -4.02 | -7.65 | -6.69 | -4.37 | 0.51 | -10.45 |
| -26.86 | 21.40 | -24.40 | 20.53 | 0.08 | -10.13 | -0.37 | -7.44 | -8.03 | -4.58 | 0.15 | -10.66 |
| -23.92 | 21.61 | -25.45 | 20.74 | -3.54 | -9.92 | -1.74 | -7.23 | -6.91 | -4.79 | 0.65 | -10.87 |
| -25.26 | 21.82 | -22.81 | 20.95 | -2.25 | -9.71 | -2.19 | -7.02 | -6.24 | -5.00 | 0.88 | -11.08 |
| -24.41 | 22.03 | -24.26 | 21.16 | -1.62 | -9.50 | -1.24 | -6.81 | -6.87 | -5.21 | 0.69 | -11.29 |
| -19.05 | 22.24 | -26.13 | 21.37 | -3.90 | -9.29 | -1.47 | -6.60 | -6.20 | -5.42 | 1.65 | -11.50 |
| -27.13 | 22.45 | -23.35 | 21.58 | -3.18 | -9.08 | -3.47 | -6.39 | -5.88 | -5.63 | 2.74 | -11.71 |
| -21.24 | 22.66 | -29.14 | 21.79 | -1.04 | -8.87 | -2.79 | -6.18 | -6.64 | -5.84 | 2.01 | -11.92 |
| -24.09 | 22.87 | -24.76 | 22.63 | -1.26 | -8.66 | -3.29 | -5.97 | -5.93 | -6.05 | 1.92 | -12.13 |
| -22.49 | 23.08 | -25.22 | 22.84 | -0.15 | -8.45 | -1.42 | -5.76 | -3.70 | -6.26 | 3.29 | -12.34 |
| -16.51 | 23.29 | -24.40 | 23.05 | -0.64 | -8.24 | -2.10 | -5.55 | -4.14 | -6.47 | 3.56 | -12.55 |
| -23.20 | 23.50 | -22.58 | 23.26 | -2.42 | -8.03 | -0.51 | -5.34 | -3.74 | -6.68 | 3.70 | -12.76 |
| -22.18 | 23.71 | -23.26 | 23.47 | 0.75 | -7.82 | -1.42 | -5.13 | -3.21 | -6.89 | 4.84 | -12.97 |
| -22.18 | 23.92 | -25.95 | 23.68 | -2.02 | -7.61 | 0.36 | -4.92 | -2.00 | -7.10 | 4.97 | -13.18 |
| -24.54 | 24.13 | -24.49 | 23.89 | -3.00 | -6.56 | 0.13 | -4.71 | -2.09 | -7.31 | 5.79 | -13.39 |
| -23.87 | 24.34 | -24.22 | 24.10 | -1.35 | -6.35 | -0.60 | -4.50 | -2.14 | -7.52 | 5.61 | -13.60 |
| -25.34 | 24.55 | -27.63 | 24.31 | -0.68 | -6.14 | -0.92 | -4.29 | -2.49 | -7.73 | 6.52 | -13.81 |
| -24.14 | 24.76 | -28.13 | 24.52 | -1.93 | -5.93 | -1.65 | -4.08 | -1.33 | -7.94 | 7.71 | -14.02 |

| PA=325° | | | | PA=62° | | | | PA=18°.9 | | | |
|---|---|---|---|---|---|---|---|---|---|---|---|
| $v_{\rm [S\,II]}$ km s⁻¹ | $d_{\rm [S\,II]}$ ″ | $v_{\rm [N\,II]}$ km s⁻¹ | $d_{\rm [N\,II]}$ ″ | $v_{\rm [S\,II]}$ km s⁻¹ | $d_{\rm [S\,II]}$ ″ | $v_{\rm [N\,II]}$ km s⁻¹ | $d_{\rm [N\,II]}$ ″ | $v_{\rm [S\,II]}$ km s⁻¹ | $d_{\rm [S\,II]}$ ″ | $v_{\rm [N\,II]}$ km s⁻¹ | $d_{\rm [N\,II]}$ ″ |
| -28.47 | 24.97 | -29.91 | 24.73 | 0.25 | -5.72 | -0.87 | -3.87 | -0.39 | -8.15 | 9.16 | -14.23 |
| -33.38 | 25.18 | -30.00 | 24.94 | -0.37 | -5.51 | -0.42 | -3.66 | -1.15 | -8.36 | 8.48 | -14.44 |
| -30.25 | 25.39 | -34.83 | 25.15 | 0.43 | -5.30 | -2.24 | -3.45 | -0.04 | -8.57 | 7.52 | -14.65 |
| -31.46 | 25.60 | -36.38 | 25.36 | -0.77 | -5.09 | -1.42 | -3.24 | -1.64 | -8.78 | 8.57 | -14.86 |
| -32.84 | 25.81 | -36.69 | 25.57 | -0.73 | -4.88 | -1.97 | -3.03 | -0.26 | -8.99 | 8.43 | -15.07 |
| -31.86 | 26.02 | -36.60 | 25.78 | 1.10 | -4.67 | -3.29 | -2.82 | -0.89 | -9.20 | 8.16 | -15.28 |
| -32.62 | 26.23 | -37.97 | 25.99 | 1.82 | -4.46 | -2.79 | -2.61 | -1.02 | -9.41 | 7.75 | -15.49 |
| -33.65 | 26.44 | -40.79 | 26.20 | -0.50 | -4.25 | -1.42 | -2.40 | -0.39 | -9.62 | 6.93 | -15.70 |
| -30.61 | 26.65 | -38.79 | 26.41 | 1.19 | -4.04 | – | – | -0.31 | -9.83 | 6.93 | -15.91 |
| -29.63 | 26.86 | -38.56 | 26.62 | -0.33 | -3.83 | – | – | 1.30 | -10.04 | 6.02 | -16.12 |
| -35.16 | 27.07 | -39.47 | 26.83 | 1.06 | -3.62 | – | – | 1.30 | -10.25 | 5.02 | -16.33 |
| -32.31 | 27.28 | -36.33 | 27.04 | 1.77 | -3.41 | – | – | -0.89 | -10.46 | 5.02 | -16.54 |
| -32.04 | 27.49 | -36.38 | 27.25 | 0.48 | -3.20 | – | – | -0.89 | -10.67 | 6.57 | -16.75 |
| -31.28 | 27.70 | -37.38 | 27.46 | -1.58 | -2.99 | – | – | -0.62 | -10.88 | 5.70 | -16.96 |
| -32.35 | 27.91 | -37.24 | 27.67 | -1.84 | -2.78 | – | – | -0.22 | -11.09 | 3.84 | -17.17 |
| -32.66 | 28.12 | -37.51 | 27.88 | -2.65 | -2.57 | – | – | -0.57 | -11.30 | 2.06 | -17.38 |
| -28.96 | 28.33 | -36.29 | 28.09 | -2.96 | -2.36 | – | – | 1.52 | -11.51 | 4.79 | -17.59 |
| -33.24 | 28.54 | -35.60 | 28.30 | | | | | 2.15 | -11.72 | 3.02 | -17.80 |
| -31.01 | 28.75 | -34.96 | 28.51 | | | | | 2.24 | -11.93 | 0.47 | -18.01 |
| -29.99 | 28.96 | -36.56 | 28.72 | | | | | 3.44 | -12.14 | 0.42 | -18.22 |
| -30.08 | 29.17 | -35.69 | 28.93 | | | | | 3.09 | -12.35 | -0.85 | -18.43 |
| -28.20 | 29.38 | -35.01 | 29.14 | | | | | 1.52 | -12.56 | -1.45 | -18.64 |
| -29.50 | 29.59 | -35.10 | 29.35 | | | | | 4.74 | -12.77 | -1.54 | -18.85 |
| -29.90 | 29.80 | -34.87 | 29.56 | | | | | 4.87 | -12.98 | -3.68 | -19.06 |

Table 8.2: continued

| PA=325° | | | | PA=62° | | | | PA=18°.9 | | | |
|---|---|---|---|---|---|---|---|---|---|---|---|
| $v_{[\text{S\,II}]}$ km s$^{-1}$ | $d_{[\text{S\,II}]}$ ″ | $v_{[\text{N\,II}]}$ km s$^{-1}$ | $d_{[\text{N\,II}]}$ ″ | $v_{[\text{S\,II}]}$ km s$^{-1}$ | $d_{[\text{S\,II}]}$ ″ | $v_{[\text{N\,II}]}$ km s$^{-1}$ | $d_{[\text{N\,II}]}$ ″ | $v_{[\text{S\,II}]}$ km s$^{-1}$ | $d_{[\text{S\,II}]}$ ″ | $v_{[\text{N\,II}]}$ km s$^{-1}$ | $d_{[\text{N\,II}]}$ ″ |
| -27.58 | 30.01 | -33.05 | 29.77 | – | – | – | – | 4.38 | -13.19 | -3.13 | -19.27 |
| -27.17 | 30.22 | -32.96 | 29.98 | – | – | – | – | 4.87 | -13.40 | -1.04 | -19.48 |
| -27.31 | 30.43 | -31.96 | 30.19 | – | – | – | – | 8.13 | -13.61 | -0.26 | -19.69 |
| -22.76 | 30.64 | -32.10 | 30.40 | – | – | – | – | 10.05 | -13.82 | -0.99 | -19.90 |
| -26.86 | 30.85 | -31.64 | 30.61 | – | – | – | – | 10.63 | -14.03 | -0.67 | -20.11 |
| -26.55 | 31.06 | -29.77 | 30.82 | – | – | – | – | 11.12 | -14.24 | 0.01 | -20.32 |
| -24.99 | 31.27 | -29.14 | 31.03 | – | – | – | – | 9.25 | -14.45 | 0.01 | -20.53 |
| -26.91 | 31.48 | -29.96 | 31.24 | – | – | – | – | 9.56 | -14.66 | 0.33 | -20.74 |
| -27.80 | 31.69 | -28.82 | 31.45 | – | – | – | – | 7.95 | -14.87 | 0.83 | -20.95 |
| -24.18 | 31.90 | -29.04 | 31.66 | – | – | – | – | 7.37 | -15.08 | 0.56 | -21.16 |
| -23.74 | 32.11 | -28.59 | 31.87 | – | – | – | – | 7.64 | -15.29 | -0.04 | -21.37 |
| -24.90 | 32.32 | -28.63 | 32.08 | – | – | – | – | 5.45 | -15.50 | -0.49 | -21.58 |
| -22.44 | 32.53 | -28.95 | 32.29 | – | – | – | – | 5.10 | -15.71 | -0.26 | -21.79 |
| -23.60 | 32.74 | -28.82 | 32.50 | – | – | – | – | 2.28 | -15.92 | -0.40 | -22.00 |
| -23.11 | 32.95 | -29.32 | 32.71 | – | – | – | – | 5.23 | -16.13 | -0.85 | -22.21 |
| -23.65 | 33.16 | -28.63 | 32.92 | – | – | – | – | 4.87 | -16.34 | -0.31 | -22.42 |
| -21.95 | 33.37 | -26.54 | 33.13 | – | – | – | – | 6.08 | -16.55 | -0.17 | -22.63 |
| -22.04 | 33.58 | -28.95 | 33.34 | – | – | – | – | 2.91 | -16.76 | 0.42 | -22.84 |
| -19.90 | 33.79 | -27.22 | 33.55 | – | – | – | – | 5.94 | -16.97 | 0.69 | -23.05 |
| -22.26 | 34.00 | -26.77 | 33.76 | – | – | – | – | 9.56 | -17.18 | -0.22 | -23.26 |
| -17.67 | 34.21 | -27.04 | 33.97 | – | – | – | – | 6.93 | -17.39 | -0.13 | -23.47 |
| -20.03 | 34.42 | -26.08 | 34.18 | – | – | – | – | 4.25 | -17.60 | -0.95 | -23.68 |
| -18.52 | 34.63 | -22.40 | 34.39 | – | – | – | – | 3.13 | -17.81 | -1.40 | -23.89 |
| -13.87 | 34.84 | -22.40 | 34.60 | – | – | – | – | 3.71 | -18.02 | -1.45 | -24.10 |

Table 8.2: continued

| PA=325° | | | | PA=62° | | | | PA=18°.9 | | | |
|---|---|---|---|---|---|---|---|---|---|---|---|
| $v_{\rm [S\,II]}$ km s⁻¹ | $d_{\rm [S\,II]}$ " | $v_{\rm [N\,II]}$ km s⁻¹ | $d_{\rm [N\,II]}$ " | $v_{\rm [S\,II]}$ km s⁻¹ | $d_{\rm [S\,II]}$ " | $v_{\rm [N\,II]}$ km s⁻¹ | $d_{\rm [N\,II]}$ " | $v_{\rm [S\,II]}$ km s⁻¹ | $d_{\rm [S\,II]}$ " | $v_{\rm [N\,II]}$ km s⁻¹ | $d_{\rm [N\,II]}$ " |
| -11.69 | 35.05 | -27.04 | 34.81 | — | — | — | — | 3.80 | -18.23 | -0.90 | -24.31 |
| -13.07 | 35.26 | -22.12 | 35.02 | — | — | — | — | 3.09 | -18.44 | -1.13 | -24.52 |
| -12.85 | 35.47 | -19.44 | 35.23 | — | — | — | — | -2.27 | -18.65 | -1.67 | -24.73 |
| -11.86 | 35.68 | -16.25 | 35.44 | — | — | — | — | -2.27 | -18.86 | -2.31 | -24.94 |
| -9.05 | 35.89 | -14.20 | 35.65 | — | — | — | — | -2.27 | -19.07 | -2.90 | -25.15 |
| -11.15 | 36.10 | -15.02 | 35.86 | — | — | — | — | -2.27 | -19.28 | -3.18 | -25.36 |
| -8.07 | 36.31 | -19.44 | 36.07 | — | — | — | — | -2.31 | -19.49 | -3.54 | -25.57 |
| -12.94 | 36.52 | -13.24 | 36.28 | — | — | — | — | -1.51 | -19.70 | -4.45 | -25.78 |
| -6.91 | 36.73 | -13.97 | 36.49 | — | — | — | — | 0.72 | -19.91 | -5.41 | -25.99 |
| -4.54 | 36.94 | -16.57 | 36.70 | — | — | — | — | 0.86 | -20.12 | -4.41 | -26.20 |
| -1.06 | 37.15 | -12.01 | 36.91 | — | — | — | — | -0.17 | -20.33 | -5.14 | -26.41 |
| -2.67 | 37.36 | -14.25 | 42.79 | — | — | — | — | 0.45 | -20.54 | -4.82 | -26.62 |
| -2.71 | 37.57 | -18.25 | 43.00 | — | — | — | — | 1.75 | -20.75 | -4.77 | -26.83 |
| -14.19 | 37.78 | -14.38 | 43.21 | — | — | — | — | 1.30 | -20.96 | -5.27 | -27.04 |
| -1.91 | 37.99 | -12.20 | 43.42 | — | — | — | — | 1.26 | -21.17 | -4.18 | -27.25 |
| -4.54 | 38.20 | -14.15 | 43.63 | — | — | — | — | 1.35 | -21.38 | -4.18 | -27.46 |
| -1.46 | 38.41 | -13.20 | 43.84 | — | — | — | — | 2.24 | -21.59 | -3.36 | -27.67 |
| -0.30 | 38.62 | -15.38 | 44.05 | — | — | — | — | 1.88 | -21.80 | -4.18 | -27.88 |
| -6.37 | 38.83 | -17.48 | 44.26 | — | — | — | — | 1.30 | -22.01 | -3.91 | -28.09 |
| -14.81 | 39.04 | -14.79 | 44.47 | — | — | — | — | 0.77 | -22.22 | -2.86 | -28.30 |
| -9.50 | 39.25 | -11.56 | 44.68 | — | — | — | — | -0.22 | -22.43 | -2.36 | -28.51 |
| 2.37 | 39.46 | -16.07 | 44.89 | — | — | — | — | 0.72 | -22.64 | -2.31 | -28.72 |
| 0.19 | 39.67 | -14.15 | 45.10 | — | — | — | — | 1.61 | -22.85 | -2.27 | -28.93 |
| -2.71 | 42.61 | -12.42 | 45.31 | — | — | — | — | 2.51 | -23.06 | -2.59 | -29.14 |

Table 8.2: continued

| PA=325° | | | | PA=62° | | | | PA=18°.9 | | | |
|---|---|---|---|---|---|---|---|---|---|---|---|
| $v_{[SII]}$ km s$^{-1}$ | $d_{[SII]}$ " | $v_{[NII]}$ km s$^{-1}$ | $d_{[NII]}$ " | $v_{[SII]}$ km s$^{-1}$ | $d_{[SII]}$ " | $v_{[NII]}$ km s$^{-1}$ | $d_{[NII]}$ " | $v_{[SII]}$ km s$^{-1}$ | $d_{[SII]}$ " | $v_{[NII]}$ km s$^{-1}$ | $d_{[NII]}$ " |
| -3.47 | 42.82 | -13.11 | 45.52 | — | — | — | — | 1.57 | -23.27 | — | — |
| -1.91 | 43.03 | -11.24 | 45.73 | — | — | — | — | 0.54 | -23.48 | — | — |
| -1.46 | 43.24 | -12.51 | 45.94 | — | — | — | — | 0.86 | -23.69 | — | — |
| -5.04 | 43.45 | -13.84 | 46.15 | — | — | — | — | 0.54 | -23.90 | — | — |
| -9.59 | 43.66 | -8.92 | 46.36 | — | — | — | — | 0.54 | -24.11 | — | — |
| -7.67 | 43.87 | -13.15 | 46.57 | — | — | — | — | 1.08 | -24.32 | — | — |
| -8.25 | 44.08 | -9.01 | 46.78 | — | — | — | — | 0.59 | -24.53 | — | — |
| -4.45 | 44.29 | -13.15 | 46.99 | — | — | — | — | 0.23 | -24.74 | — | — |
| -11.86 | 44.50 | -10.24 | 47.20 | — | — | — | — | -0.39 | -24.95 | — | — |
| -4.23 | 44.71 | -13.33 | 47.41 | — | — | — | — | -1.96 | -25.16 | — | — |
| -6.20 | 44.92 | -9.92 | 47.62 | — | — | — | — | -2.49 | -25.37 | — | — |
| 0.01 | 45.13 | -14.84 | 47.83 | — | — | — | — | -2.76 | -25.58 | — | — |
| -5.84 | 45.34 | -13.02 | 48.04 | — | — | — | — | -4.01 | -25.79 | — | — |
| -3.03 | 45.55 | -13.20 | 48.25 | — | — | — | — | -3.92 | -26.00 | — | — |
| -9.77 | 45.76 | -15.79 | 48.46 | — | — | — | — | -3.39 | -26.21 | — | — |
| -6.33 | 45.97 | -13.97 | 48.67 | — | — | — | — | -4.28 | -26.42 | — | — |
| -4.63 | 46.18 | -11.51 | 48.88 | — | — | — | — | -3.97 | -26.63 | — | — |
| -6.33 | 46.39 | — | — | — | — | — | — | -3.65 | -26.84 | — | — |
| -6.46 | 46.60 | — | — | — | — | — | — | -3.97 | -27.05 | — | — |
| -4.86 | 46.81 | — | — | — | — | — | — | -2.18 | -27.26 | — | — |
| -7.22 | 47.02 | — | — | — | — | — | — | -3.12 | -27.47 | — | — |
| -7.45 | 47.23 | — | — | — | — | — | — | -3.83 | -27.68 | — | — |
| -8.61 | 47.44 | — | — | — | — | — | — | -3.30 | -27.89 | — | — |
| -7.00 | 47.65 | — | — | — | — | — | — | -3.21 | -28.10 | — | — |

Table 8.2: continued

| PA=325° | | | | PA=62° | | | | PA=18°.9 | | | |
| $v_{\rm [S II]}$ km s$^{-1}$ | $d_{\rm [S II]}$ ″ | $v_{\rm [N II]}$ km s$^{-1}$ | $d_{\rm [N II]}$ ″ | $v_{\rm [S II]}$ km s$^{-1}$ | $d_{\rm [S II]}$ ″ | $v_{\rm [N II]}$ km s$^{-1}$ | $d_{\rm [N II]}$ ″ | $v_{\rm [S II]}$ km s$^{-1}$ | $d_{\rm [S II]}$ ″ | $v_{\rm [N II]}$ km s$^{-1}$ | $d_{\rm [N II]}$ ″ |
|---|---|---|---|---|---|---|---|---|---|---|---|
| -7.67 | 47.86 | – | – | – | – | – | – | -2.22 | -28.31 | – | – |
| -5.93 | 48.07 | – | – | – | – | – | – | -1.24 | -28.52 | – | – |
| -8.56 | 48.28 | – | – | – | – | – | – | -2.14 | -28.73 | – | – |
| -5.57 | 48.49 | – | – | – | – | – | – | 0.10 | -28.94 | – | – |
| -5.53 | 48.70 | – | – | – | – | – | – | 2.69 | -29.15 | – | – |
| -3.12 | 48.91 | – | – | – | – | – | – | – | – | – | – |
| -1.91 | 49.12 | – | – | – | – | – | – | – | – | – | – |
| -6.11 | 49.33 | – | – | – | – | – | – | – | – | – | – |
| -1.60 | 49.54 | – | – | – | – | – | – | – | – | – | – |
| -6.78 | 49.75 | – | – | – | – | – | – | – | – | – | – |
| -4.81 | 49.96 | – | – | – | – | – | – | – | – | – | – |
| 1.17 | 50.17 | – | – | – | – | – | – | – | – | – | – |
| -4.81 | 50.38 | – | – | – | – | – | – | – | – | – | – |
| -1.60 | 50.59 | – | – | – | – | – | – | – | – | – | – |
| -7.45 | 50.80 | – | – | – | – | – | – | – | – | – | – |
| 0.99 | 51.01 | – | – | – | – | – | – | – | – | – | – |

# Curriculum vitae

**Personal data**

| | |
|---|---|
| Name | Tiina Liimets |
| Date and place of birth | 15 April 1982, Tallinn, Estonia |
| Citizenship | Estonian |
| Marital status | married |
| Current employment | Astronomical Institute of the Czech Academy of Sciences, researcher |
| Address | Astronomical Institute |
| | Fričova 298 |
| | 251 65 Ondřejov |
| | Czech Republic |
| Phone | (+372) 56 913161 |
| E-mail | tiina.liimets@ut.ee |

**Education**

| | |
|---|---|
| 1989 – 2000 | Tallinn Kuristiku High School |
| 2000 – 2005 | University of Tartu, undergraduate student, BSc (phyics) |
| 2005 – 2007 | University of Tartu, graduate student, MSc (astrophysics) |
| 2007 – 2019 | University of Tartu, PhD student |

**Employment**

| | |
|---|---|
| 2006 – 2010 | Tartu Observatory, engineer |
| 2007 – 2008 | Isaac Newton Group of Telescopes, student |
| 2009 – 2010 | Nordic Optical Telescope, student support astronomer |
| 2010 – 2015 | Tartu Observatory, researcher |
| 2015 – 2018 | Tartu Observatory, Maternity leave |
| 2016 – 2018 | Tartu Observatory, University of Tartu, researcher |
| 2018 – 2019 | Astronomical Institute, Academy of Sciences of the Czech Republic, researcher |
| 2018 – 2020 | Tartu Observatory, University of Tartu, engineer |

**Professional training**

| | |
|---|---|
| 31.07 – 11.08 2006 | Summer school "Life and Death of Stars", Utrecht, Netherlands. |

| 10.08 – 24.08 2008 | Nordic-Baltic Research Course "Observational Stellar Astrophysics", Moletai, Lithuania. |
| 08.06 – 18.06 2009 | Nordic-Baltic Optical/NIR and Radio Astronomy Summer School "Star Formation in the Milky Way and Nearby Galaxies",Turku, Finland. |
| 01.07 – 03.07 2009 | International Summer School "Future cosmic sky surveys and huge databases", Tartu, Estonia. |
| 25.01 – 28.01 2010 | "EuroVO-AIDA School 2010", Strasbourg, France. |

## Conferences attended

| 15.08 – 19.08 2005 | Workshop "Stellar Evolution at Low Mettallicity: Mass Loss, Explosions, Cosmology", Tartu, Estonia. |
| 16.05 – 19.05 2006 | Conference "The Nature of V838 Monocerotis and its Light Echo", La Palma, Spain. |
| | *Oral presentation*: "Photometry of V838 Mon and its light echo' |
| 18.06 – 22.06 2007 | Conference "Asymmetrical Planetary Nebulae IV", La Palma, Spain. |
| 02.09 – 03.09 2010 | Tuorla-Tartu annual meeting 2010 "Observational Cosmology", Tuorla, Finland. |
| 09.02 – 09.02 2011 | Concluding Seminar of the FP7 project EstSpacE (2008-2010) "Modern trends in space research", Tartu, Estonia. |
| | *Oral presentation*: "Possibilities for young space researchers through international networking" |
| 25.07 – 27.07 2011 | IAU symposium 283 "Planetary Nebulae: An Eye to the Future", Tenerife, Spain. |
| | *Poster*: "The growth of outflows from evolved stars: a live poster" |
| 04.02 – 08.02 2013 | Conference "Stella Novae: Past and Future Decades", Cape Town, South-Africa. |
| | *Oral presentation*: "Dynamical Study of the Nova Remnant of GK Per" |
| 08.07 – 12.07 2013 | European Week of Astronomy and Space Science 2013, Turku, Finland. |
| 18.08 – 23.08 2013 | Workshop on Symbiotic Stars, Binary Post-AGB and Related Objects, Wierzba, Poland. |
| | *Oral presentation*: "New Insights into the Jet of R Aquarii" |

| 25.09 – 27.09 2013 | Tartu-Tuorla Annual Meeting (Seminar on Cosmology), Villa Greete, Estonia. |
| | *Oral presentation*: "Structure and Kinematics of Circumstellar Matter in Real Time: GK Persei, R Aqr" |
| 10.09 – 15.09 2017 | Workshop "Massive Stars in Transition Phases", Tõravere, Estonia. |
| | *Oral presentation*: " Structure and Kinematics of Circumstellar Matter" |

**Language skills**

| Estonian | mother tongue |
| English | good |
| Spanish | beginner |
| Russian | beginner |

**Honours and stipends**

| 2006 | E. Öpik stipend (Tartu Observatory, Estonia) |
| 2009 | Honour, "Synnøve Irgens-Jensen Distinguished Research Student" (The Nordic Optical Telescope Scientific Association) |
| 2011 | Kristjan Jaak part-time study stipend for Ph.D. research in Instituto Astrofisica de Canarias, Spain (Estonian Ministry of Education and Research; Foundation Archimedes, Estonia) |
| 2012 | Honour, participation in a conference "62nd Lindau Nobel Laureate Meetin", 1-6 July, Lindau, Germany (Estonian Academy of Sciences; Foundation Lindau Nobel Laureate Meetings) |
| 2012 | Kristjan Jaak part-time study stipend for Ph.D. research in Instituto Astrofisica de Canarias, Spain (Estonian Ministry of Education and Research; Foundation Archimedes, Estonia) |
| 2012 | CWT Estonia stipend, participation in a conference "Stella Novae: Past and Future Decades", 4-8 February, 2013, Cape Town, South-Africa. (CWT Estonia; The University of Tartu Foundation, Estonia) |

| 2013 | DoRa stipend for Ph.D. research in Instituto Astrofisica de Canarias, Spain |
| | (Foundation Archimedes, Estonia) |
| 2013 | Stipend to attend a conference "Workshop on Symbiotic Stars, Binary Post-AGB and Related Objects", |
| | 18-23 August, Wierzba, Poland. |
| | (IUPAP Working Group on Women in Physics) |
| 2014 | Kristjan Jaak foreign visit stipend for Ph.D. research in Instituto Astrofisica de Canarias, Spain |
| | (Estonian Ministry of Education and Research; Foundation Archimedes, Estonia) |
| 2019 | Dora Pluss 1.1 travel grant for Ph.D. research in Centro Astrofisico en La Palma, Spain |
| | (Foundation Archimedes, Estonia) |

## Teaching experience

| 2010, 2013 | "General Astronomy: Observations", LOFY.04.058, lecturer, University of Tartu, Estonia |
| 2012 | "Seminar in Astrophysics", LOFY.04.027, organizer University of Tartu, Estonia |

## Supervised Bachelor's Thesis

| 2015 | Kristiina Verro, "Physical properties of the nova remnant Nova Persei 1901", University of Tartu, Estonia |

## Popularizing activities

| 2006 – present | Many popularizing talks for students, amateur astronomers, and wider public. |
| | Several interviews in radio, TV, and newspapers. |
| 2009 – 2012 | Board member of Estonian Astronomical Society. |
| 2011 – 2012 | Ambassador of the Year of Science in Estonia. |
| 2012 – 2014 | Chairman of Estonian Astronomical Society. |
| 2016 – 2018 | Board member of Estonian Astronomical Society. |

## Fields of research

Structure and kinematics of remnants around outbursting stars.

## Publications

# ELULOOKIRJELDUS

**Isikuandmed**

| | |
|---|---|
| Nimi | Tiina Liimets |
| Sünniaeg ja -koht | 15. Aprill 1982, Tallinn, Eesti |
| Kodakondsus | eesti |
| Perekonnaseis | abielus |
| Praegune töökoht | Tšehhi Vabariigi Teaduste Akadeemia Astronoomia Instituut, teadur |
| Address | Astronomical Institute |
| | Fričova 298 |
| | 251 65 Ondřejov |
| | Czech Republic |
| Phone | (+372) 56 913161 |
| E-mail | tiina.liimets@ut.ee |

**Haridustee**

| | |
|---|---|
| 1989 – 2000 | Tallinna Kuristiku Keskkool |
| 2000 – 2005 | Tartu Ülikool, üliõpilane, BSc (füüsika) |
| 2005 – 2007 | Tartu Ülikool, magistrant, MSc (astrofüüsika) |
| 2007 – 2019 | Tartu Ülikool, doktorant |

**Teenistuskäik**

| | |
|---|---|
| 2006 – 2010 | Tartu Observatoorium, insener |
| 2007 – 2008 | Isaac Newton Group of Telescopes, tudeng |
| 2009 – 2010 | Nordic Optical Telescope, tudeng-tugi-astronoom |
| 2010 – 2015 | Tartu Observatoorium, teadur |
| 2015 – 2018 | Tartu Observatoorium, lapsehoolduspuhkus |
| 2016 – 2018 | Tartu Observatoorium, Tartu Ülikool, teadur |
| 2018 – 2019 | Tšehhi Vabariigi Teaduste Akadeemia Astronoomia Instituut, teadur |
| 2018 – 2020 | Tartu Observatoorium, Tartu Ülikool, insener |

**Täiendkoolitus**

| | |
|---|---|
| 31.07 – 11.08 2006 | Suvekool "Life and Death of Stars", Utrecht, Holland. |
| 10.08 – 24.08 2008 | Vaatlusliku astronoomia suvekool "Observational Stellar Astrophysics", Moletai, Leedu. |
| 08.06 – 18.06 2009 | Optilise/lähisinfrapuna ja raadio astronoomia suvekool "Star Formation in the Milky Way and Nearby Galaxies", Turku, Soome. |
| 01.07 – 03.07 2009 | Suvekool "Future cosmic sky surveys and huge databases", Tartu, Eesti. |
| 25.01 – 28.01 2010 | Virtuaalse Observatooriumi talvekool "EuroVO-AIDA School 2010", Strasbourg, Prantsusmaa. |

**Külastatud konverentsid**

| | |
|---|---|
| 15.08 – 19.08 2005 | Konverents "Stellar Evolution at Low Mettallicity: Mass Loss, Explosions, Cosmology", Tartu, Eesti. |
| 16.05 – 19.05 2006 | Konverents "The Nature of V838 Monocerotis and its Light Echo", La Palma, Hispaania. *Suuline ettekanne*: "Photometry of V838 Mon and its light echo" |
| 18.06 – 22.06 2007 | Konverents "Asymmetrical Planetary Nebulae IV", La Palma, Hispaania. |
| 02.09 – 03.09 2010 | Konverents "Tuorla-Tartu annual meeting 2010. Observational Cosmology", Tuorla, Soome. |
| 09.02 – 09.02 2011 | Seminar "Kosmose uurimise uudsed trendid", Tartu, Eesti. *Suuline ettekanne*: "Possibilities for young space researchers through international networking" |
| 25.07 – 27.07 2011 | IAU sümpoosion 283 "Planetary Nebulae: An Eye to the Future", Tenerife, Hispaania. *Poster*: "The growth of outflows from evolved stars: a live poster" |
| 04.02 – 08.02 2013 | Konverents "Stella Novae: Past and Future Decades", Kapplinn, Lõuna-Aafrika Vabariik. *Suuline ettekanne*: "Dynamical Study of the Nova Remnant of GK Per" |
| 08.07 – 12.07 2013 | Konverents "European Week of Astronomy and Space Science 2013", Turku, Soome. |
| 18.08 – 23.08 2013 | Konverents "Workshop on Symbiotic Stars, Binary Post-AGB and Related Objects", Wierzba, Poola. *Suuline ettekanne*: "New Insights into the Jet of R Aqr" |

25.09 – 27.09 2013     Konverents "Tartu-Tuorla Annual Meeting (Seminar on Cosmology)", Villa Greete, Eesti.
*Suuline ettekanne*: "Structure and Kinematics of Circumstellar Matter in Real Time: GK Persei, R Aqr"

## Keelteoskus

| | |
|---|---|
| eesti keel | emakeel |
| inglise keel | hea |
| hispaania keel | algaja |
| vene keel | algaja |

## Autasud ja stipendiumid

| | |
|---|---|
| 2006 | E. Öpik stipendium (Tartu Observatoorium) |
| 2009 | Autasu, "Synnøve Irgens-Jensen väljapaistev noorteadlane" (The Nordic Optical Telescope Scientific Association) |
| 2011 | Kristjan Jaagu osalise õppe stipendium doktoritöö teadustööks Kanaari Astrofüüsikainstituudis, Hispaania (Haridus- ja Teadusministeerium; Sihtasutus Archimedes, Eesti) |
| 2012 | Autasu, osalemine konverentsil "62nd Lindau Nobel Laureate Meeting", 1-6 juuli, Lindau, Saksamaa (Eesti Teaduste Akadeemia; Foundation Lindau Nobel Laureate Meetings) |
| 2012 | Kristjan Jaagu osalise õppe stipendium doktoritöö teadustööks Kanaari Astrofüüsikainstituudis, Hispaania (Haridus- ja Teadusministeerium; Sihtasutus Archimedes, Eesti) |
| 2012 | CWT Estonia reisistipendium osalemaks konverentsil "Stella Novae: Past and Future Decades", 4-8 veebruar, 2013, Kapplinn, Lõuna-Aafrika Vabariik. (Tartu Ülikooli Sihtasutus, CWT Estonia) |
| 2013 | DoRa stipendium doktoritöö teadustööks Kanaari Astrofüüsikainstituudis, Hispaania (Sihtasutus Archimedes, Eesti) |
| 2013 | Stipendium osalemaks konverentsil "Workshop on Symbiotic Stars, Binary Post-AGB and Related Objects", 18-23 august, Wierzba, Poola. (IUPAP Working Group on Women in Physics) |
| 2014 | Kristjan Jaagu välissõidu stipendium doktoritöö teadustööks Kanaari Astrofüüsikainstituudis, Hispaania (Haridus- ja Teadusministeerium; Sihtasutus Archimedes, Eesti) |
| 2019 | Dora Pluss 1.1 stipendium doktoritöö teadustööks La Palma Astrofüüsikakeskuses, Hispaania (Sihtasutus Archimedes, Eesti) |

## Õpetamine Ülikoolis

| | |
|---|---|
| 2010, 2013 | "Praktiline astronoomia", LOFY.04.058, õppejõud, Tartu Ülikool, Eesti |
| 2012 | "Astrofüüsika seminar", LOFY.04.027, juhataja Tartu Ülikool, Eesti |

## Juhendatud bakalaureusetöö

| | |
|---|---|
| 2015 | Kristiina Verro, "Physical properties of the nova remnant Nova Persei 1901", Tartu Ülikool, Eesti |

## Populaarteaduslik tegevus

| | |
|---|---|
| 2006 – ... | Mimed populariseerivad loengud kooliõpilastele, astronoomiahuvilistele ja muule publikule. Mitmed intervjuud raadios, televisioonis ja ajakirjanduses. |
| 2009 – 2012 | Eesti Astronoomia Seltsi juhatuse liige. |
| 2011 – 2012 | Eesti Teadusaasta saadik. |
| 2012 – 2014 | Eesti Astronoomia Seltsi esimees. |
| 2014 – 2018 | Eesti Astronoomia Seltsi juhatuse liige. |

## Peamised uurimissuunad

Tähtedelt väljapaisatud jäänukite struktuuri ja kinemaatika uurimine.

# DISSERTATIONES ASTRONOMIAE
# UNIVERSITATIS TARTUENSIS

1. **Tõnu Viik.** Numerical realizations of analytical methods in theory of radiative transfer. Tartu, 1991.
2. **Enn Saar.** Geometry of the large scale structure of the Universe. Tartu, 1991.
3. **Maret Einasto.** Morphological and luminosity segregation of galaxies. Tartu, 1991.
4. **Urmas Haud.** Dark Matter in galaxies. Tartu, 1991.
5. **Eugene A. Ustinov.** Inverse problems of radiative transfer in sounding of planetary atmospheres. Tartu, 1992.
6. **Peeter Tenjes.** Models of regular galaxies. Tartu, 1993.
7. **Ivar Suisalu.** Simulation of the evolution of large scale structure elements with adaptive multigrid method. Tartu, 1995.
8. **Teimuraz Shvelidze.** Automated quantitative spectral classification of stars by means of objective prism spectra: the method and applications. Tartu, 1999.
9. **Jelena Gerškevitš.** Formation and evolution of binary systems with compact objects. Tartu, 2002.
10. **Ivan Suhhonenko**. Large-scale motions in the universe. Tartu, 2003.
11. **Antti Tamm.** Structure of distant disk galaxies. Tartu, 2006.
12. **Vladislav-Veniamin Pustynski.** Modeling the reflection effect in pre-cataclysmic binary systems. Tartu, 2007.
13. **Anna Aret.** Evolutionary separation of mercury isotopes in atmospheres of chemically peculiar stars. Tartu, 2009.
14. **Mari Burmeister.** Characteristics of the hot components of symbiotic stars. Tartu, 2010.
15. **Elmo Tempel.** Tracing galaxy evolution by their present-day luminosity function. Tartu, 2011.
16. **Anti Hirv.** Estimation of time delays from light curves of gravitationally lensed quasars. Tartu, 2011.
17. **Rain Kipper**. Galaxy modelling: dynamical methods and applications. Tartu, 2016, 134 p.
18. **Lauri Juhan Liivamägi**. Properties and spatial distribution of galaxy superclusters. Tartu, 2017, 185 p.
19. **Jaan Laur**. Variability survey of massive stars in Milky Way star clusters. Tartu, 2017, 183 p.
20. **Boris Zhivkov Deshev.** On the coevolution of galaxies and their host clusters. Tartu, 2019, 199 p.